%% file: 0-MNRAS_MN-20-4868-MJ_R1.tex
\documentclass[fleqn,usenatbib]{mnrasboskri}

\usepackage[T1]{fontenc}
\DeclareRobustCommand{\VAN}[3]{#2}
\let\VANthebibliography\thebibliography
\def\thebibliography{\DeclareRobustCommand{\VAN}[3]{##3}\VANthebibliography}

\usepackage{graphicx}	
\usepackage{amsmath}	
\usepackage{amssymb}	

\usepackage{newtxtext,newtxmath}
\usepackage{tikz}
\usepackage{comment}
\usepackage{siunitx}
\DeclareSIUnit\arcsecond{arcsecond}
\DeclareSIUnit\arcminute{arcminute}
\usetikzlibrary{calc,shapes.geometric}
\usepackage[utf8x]{inputenc}
\usepackage{overpic}
\usetikzlibrary{positioning}
\usepackage{subfigure}
\newcommand{\figref}[1]{{Fig. \ref{#1}}}
\usepackage{url}
\usepackage{soul}
\usepackage{booktabs}
\newcommand{\corr}[1]{#1}

\newcommand{\tabref}[1]{{Table \ref{#1}}}
\newcommand{\gmleft}{~\textquoteleft}
\newcommand{\gmright}{\textquoteright~}
\newcommand{\gmrightd}{\textquoteright.}
\newcommand{\gmrightc}{\textquoteright,}
\usepackage{textpos}

\renewcommand{\d}{\mathrm{d}}
\renewcommand{\SIrange}[3]{\SI{#1}{}$-$\SI{#2}{#3}}
\usepackage{array,multirow,makecell}
\newcolumntype{R}[1]{>{\raggedleft\arraybackslash }b{#1}}
\newcolumntype{L}[1]{>{\raggedright\arraybackslash }b{#1}}
\newcolumntype{C}[1]{>{\centering\arraybackslash }b{#1}}
\usetikzlibrary{3d}
\usetikzlibrary{arrows}
\usetikzlibrary{decorations.pathmorphing}
\pgfdeclaredecoration{random y steps}{start}
{%
	\state{start}[width=+0pt,next state=step,persistent precomputation=\pgfdecoratepathhascornerstrue]{}%
	\state{step}[auto end on length=1.5\pgfdecorationsegmentlength,
	auto corner on length=1.5\pgfdecorationsegmentlength,
	width=+\pgfdecorationsegmentlength]
	{
		\pgfpathlineto{
			\pgfpointadd
			{\pgfpoint{\pgfdecorationsegmentlength}{0pt}}
			{\pgfpoint{0pt}{rand*\pgfdecorationsegmentamplitude}}
		}
	}%
	\state{final}
	{}%
}%

\title[Off-axis fringe tracking and sky coverage]{Potential and sky coverage for off-axis fringe tracking in optical long baseline interferometry}

\input{1-authors}

 \date{Accepted 2021 May 20. Received 2021 May 18; in original form 2020 December 1}

\pubyear{2021}
\volume{00}
\journal{\begin{tabular}{p{\textwidth}}\vspace*{0.013cm}
		MNRAS \textbf{00,} 1 (2021)\hfill \url{https://doi.org/10.1093/mnras/stab1505} \\
Advance Access publication 2021 May 26
\end{tabular}}
\usepackage{hyperref}
\usepackage{orcidlink}
\usepackage[letterspace=150]{microtype}
\begin{document}
\label{firstpage}

\pagerange{1}
\maketitle

\begin{abstract}
\input{2-abstract}

\end{abstract}

\begin{textblock}{15}(-1.3,-10.28)
	\includegraphics[width=21cm]{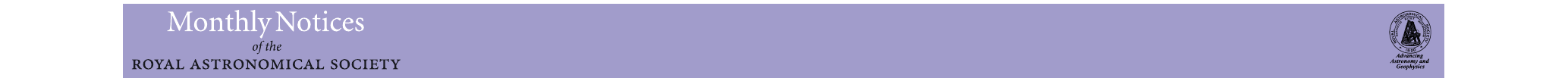}
\end{textblock}
\vspace*{-0.5cm}
\input{3-introduction}
\input{4-FT}
\input{4-suite-new}
\input{5-conclusion}

\input{6-Acknowledgements}


\input{7-Data-Availability}



\bibliographystyle{mnras}
\bibliography{8-Skybib} 






\bsp	
\label{lastpage}
\end{document}

%% file: 1-authors.tex
\author[A. Boskri et al.]{
Abdelkarim Boskri$^{\orcidlink{0000-0002-7370-9737}}$,$\color{blue}{^{1,2}}$\thanks{E-mail:\href{mailto:abdelkarim.boskri@ced.uca.ac.ma}{abdelkarim.boskri@ced.uca.ac.ma} (AB); \href{mailto:romain.petrov@unice.fr}{romain.petrov@unice.fr} (RGP)}
Romain G. Petrov,$\color{blue}{^{2,\star}}$
Thami El Halkouj,$\color{blue}{^{1}}$
Massinissa Hadjara,$\color{blue}{^{3,4,5}}$
\newauthor
James Leftley,$\color{blue}{^{2}}$
Zouhair  Benkhaldoun$^{\orcidlink{0000-0001-6285-9847}}$,$\color{blue}{^{1}}$
Pierre Cruzalèbes,$\color{blue}{^{2}}$
Aziz Ziad$\color{blue}{^{2}}$ and
Marcel Carbillet$\color{blue}{^{2}}$
\\
$^{1}$LPHEA Laboratory, Oukaimeden Observatory, Cadi Ayyad University/FSSM, BP 2390, Marrakesh, Morocco\\
$^{2}$Observatoire de la Côte d’Azur, Université de la Côte d’Azur,  CNRS, Laboratoire Lagrange, Parc Valrose, Bât. H. Fizeau, F-06108 Nice, France\\
$^{3}$Chinese Academy of Sciences South America Center for Astronomy, National Astronomical Observatories, CAS, Beijing 100101, China\\
$^{4}$Departamento de Astronomía, Universidad de Chile, Casilla 36-D, Santiago 7550000, Chile\\
$^{5}$Centre de Recherche en Astronomie, Astrophysique et Géophysique (CRAAG), Route de l’Observatoire, B.P. 63, Bouzareah, 16340, Alger, Algeria
}

%% file: 2-abstract.tex
The spectacular results provided by the second-generation \textit{VLTI} instruments \textit{GRAVITY} and \textit{MATISSE} on active galactic nuclei (AGN) trigger and justify a strong increase in the sensitivity limit of optical interferometers. A key component of such an upgrade is off-axis fringe tracking. To evaluate its potential and limitations, we describe and analyse its error budget including fringe sensing precision and temporal, angular and chromatic perturbations of the piston. The global tracking error is \corr{computed} using standard seeing parameters for different sites, seeing conditions and telescope sizes for the current \textit{GRAVITY} Fringe Tracker (GFT) and a new concept of Hierarchical Fringe Tracker. Then, it is combined with a large catalogue of guide star candidates from \textit{Gaia}  to produce sky coverage maps that give the probability to find a usable off-axis guide star in any part of the observable sky. These maps can be used to set the specifications of the system, check its sensitivity to seeing conditions, and evaluate the feasibility of science programs. We check the availability of guide stars and the tracking accuracy for a large set of $ 15\,799 $ Quasars to confirm the feasibility of a large program on Broad Line Regions in the $K$ band with the GFT and show how it can be extended to the $L$, $M$, and $N$ bands. Another set of 331 well-characterized nearby AGNs shows the high potential of \textit{MATISSE} for imaging and characterization of the dust torus in the $N$ band under off-axis tracking on \corr{both} \textit{Unit Telescopes} and \textit{Auxiliary Telescopes}.

%% file: 3-introduction.tex
\section{Introduction}
The second-generation \textit{VLTI} general user focal instruments, \textit{GRAVITY} \citep{2017A&A...602A..94G} and \textit{MATISSE} \citep{2014Msngr.157....5L}, which are now in operation, represent a major turning point for Optical
Long Baseline Interferometry. Both instruments combine either the four \textit{VLTI} \textit{UTs}
(\SI{8}{\meter} \textquoteleft \textit{Unit Telescopes}\textquoteright) or the four \textit{VLTI} \textit{ATs} (1.8 m \textquoteleft \textit{Auxiliary Telescopes}\textquoteright) and allow
spectro-interferometric imaging with the resolution of a \SI{130}{\meter} telescope in their respective
spectral domains: the $ K $ band (2\textendash\SI{2.4}{\micro\meter}) for \textit{GRAVITY} and the $ L $, $ M $, and $ N $ bands
(3 \textendash \SI{13}{\micro\meter}) for \textit{MATISSE}. \textit{GRAVITY} has been operational since 2016 and has presented a set of
breakthrough results on active galactic nuclei \citep[AGN; ][]{2018Natur.563..657G,2020A&A...634A...1G}, extrasolar planets \citep{2020A&A...633A.110G}, Young Stellar Objects (YSOs), and our Galactic Centre. \textit{MATISSE} saw its first light on sky in 2018 and is now publishing its first results on YSO \citep*{2020MNRAS.497.2540M,Varga2020}, exozodiacal light \citep{2020MNRAS.499L..47K}, AGNs, and Stellar Physics. Spectacular results are obtained on a small set of targets that illustrate the potential of the \textit{VLTI} for the study of protoplanetary discs and the co-evolution between supermassive black holes (SMBH) and their host galaxy. However, to be real game-changers such results have to be repeated on a large number of targets with a sufficient range of different physical conditions. Then the \textit{VLTI} meets its sensitivity limit. In spite of breakthrough sensitivities with \textit{GRAVITY} that allows routine high-quality fringe tracking up to $ \mathrm{K}=10.5 $ with \textit{UTs}, as well as with \textit{MATISSE} that can observe 20 mJy sources in $ L $, observations with the \textit{UTs} are limited to a few dozen bright AGNs and even fewer for which we can obtain the size and kinematics of the Broad Line Regions (BLR). That strongly limits the number of objects in which we can measure the mass of the SMBH, the distance to the AGN \citep{2020NatAs...4..517W}, or for which we can obtain detailed thermal images of the very complex dust distribution, which feeds the SMBH but also impacts {star formation near the galactic centres}. In the case of protoplanetary discs, we need detailed images of fine structures at high spectral resolution in quite resolved discs. This requires the use of the relocatable \textit{VLTI} \textit{ATs}. With the current sensitivity limits, this is again restricted to a small number of targets with a limited range of central star masses.

This situation has triggered the  \textit{GRAVITY+} proposal \citep{2019vltt.confE..30E} and a new generation fringe tracker (FT) proposal \citep{2019vltt.confE..37P} intended to drastically improve the sensitivity limit of the \textit{VLTI}. \textit{GRAVITY} contains two instruments, a high-performance FT (the \textit{GRAVITY} fringe tracker [GFT], the best FT in operation) that can work on an off-axis star and a science instrument that can reach a magnitude $\mathrm{K}=17$ \citep{2017Msngr.170...10G} using long exposure integration when the GFT allows. The GFT can also be used to stabilize the fringes for \textit{MATISSE} in the so-called GRA4MAT (GRAVITY for \textit{MATISSE}) mode. The current main limitation both for \textit{GRAVITY} and \textit{MATISSE} is due to the aging adaptive optics (AO) that feed the \textit{VLTI} beams on the \textit{UTs}.

This sets a strong $R$ band limit to all \textit{VLTI} programs and has a strong impact on the performances of the GFT on \textit{UTs}, a point that will be briefly discussed in this paper in Section \ref{Max-sent-PHDsens}. Two other limitations to the off-axis fringe tracking are the very limited (\SI{8}{\arcsecond}) separation currently possible between the off-axis tracking star
and the science target and the relatively high level of vibrations in the \textit{VLTI}, mainly with the \textit{UTs}. The main goal of \textit{\textit{GRAVITY}+} is to radically improve the \textit{UT} AO used for the \textit{VLTI} by replacing them with a new system with more
actuators coupled to a Laser Guide Star. The second important goal is to improve the off-axis fringe tracking capability by increasing the possible separation between the off-axis guide star and the science target. \textit{\textit{GRAVITY}+} will also work on the active damping of vibrations and on other features that can increase the GFT sensitivity.

Off-axis fringe tracking immediately raises the sky coverage issue. The sky coverage is the probability to find a usable off-axis guide star close enough to science target anywhere on the sky. It is limited by the sensitivity of the FT and by the angular variations of the atmospheric Optical Path Difference (OPD), which is called the atmospheric piston when it is averaged over each full individual aperture. The difference between the piston measured on the guide
star and the piston that affects the science source is called the anisopistonic error and the acceptable angle between guide star and observed target is called the isopistonic angle, by analogy with the isoplanatic angle that affects higher-order AO or wide-field high angular resolution observations through the atmosphere. Estimates of the sky coverage are often based on a fringe tracking limiting magnitude combined with an isopistonic angle. The sky coverage is then the probability to find a guide star brighter than the limiting magnitude within the isopistonic radius. Here we use a more complete definition based on an analysis of the full-error budget for off-axis tracking that includes the fringe sensing precision, the servo loop delay and error, the anisopistonic error, and the chromatic piston error that appears when we are fringe tracking at a certain wavelength and observe in another. We consider the interaction between all these terms, although with some simplifications, to establish a more realistic model of the off-axis fringe tracking precision as
a function of the angular separation. We develop analytical computations of the main error terms with parameters directly related to the standard seeing parameters that are measured at \textit{Paranal}. Then we use a large catalogue of more than \SI{8e8}{} guide star candidates, extracted from the $2^{nd}$ \textit{Gaia} data release, to evaluate the sky coverage for different telescope sizes and seeing conditions for observations with \textit{GRAVITY} ($K$ band), \textit{MATISSE} ($L$, $M$, and $N$ bands) and with a potential new instrument in the $J$ band.

We consider two FTs. The GFT, as it is described in \cite{2019A&A...624A..99L}, is used to validate our fringe sensing estimates and to predict the performance of \textit{\textit{GRAVITY}} with the \textit{\textit{GRAVITY}+} upgrade but with its unchanged internal FT. We consider also a new generation FT with a broader spectral domain, a slightly increased transmission and an innovative Hierarchical Fringe Tracking (HFT) concept \citep{2014SPIE.9146E..2PP,2016SPIE.9907E..1FP,2019vltt.confE..37P} to evaluate the gain that could be expected from such an additional upgrade.

To be more specific than with general sky coverage maps, we consider with more detail two possible observing programs on AGNs, after summarizing their goals and methods. We define a large list of more than 15 000 Quasars extracted from the Vista Hemisphere Survey and we use it to evaluate the number of targets accessible to a large BLR program on \textit{\textit{GRAVITY}}. A shorter list of 331 well-studied nearby AGNs, set up by \cite{2020MNRAS.494.1784A}, is used to investigate the potential of \textit{MATISSE} in the \textit{\textit{GRAVITY}+} context for detailed studies of AGN dust tori.

Our work on sky coverage and isopistonic errors started well before the \textit{\textit{GRAVITY}+} project. We first investigated the issue in the context of optical interferometry in Antarctica \citep{2006SPIE.6268E..12E}, which triggered the parametric analysis of the anisopistonic error that we use here \citep{2008A&A...477..337E}. Our isopistonic angle estimates are based on on-site testing campaigns with the Generalized Seeing Monitor (GSM) including two campaigns at \textit{Paranal} (1998 and 2007) and one campaign at Dome C. We have combined our analytic computation of the anisopistonic error with the statistics obtained on all these sites on the Fried parameter, the coherence time, the isoplanatic angle, and the outer scale, to produce the set of
isopistonic angle statistics of \cite{2016MNRAS.458.4044Z} that is used in the present paper.

In the following, we first present the components of the error budget of an off-axis FT (Section~\ref{sec:FTB}), then we discuss the measurement errors on phase delay (Section~\ref{sec:phasedelay}) and group delay (Section~\ref{sec:groupdelay}), first for the GFT (Section~\ref{sec:GFT}). Then we briefly present the HFT  (Section~\ref{sec:HFT}) and its estimated performances. We discuss the anisopistonic error  (Section~\ref{sec:Anis-error}), the servo loop error (Section~\ref{sec:serv-loop}), and the optimum frame time set by its combination with the sensing error (Section~\ref{Optimum-time}). A note about the chromatic OPD error (Section~\ref{Chrom-OPD}) closes our estimation of the off-axis error analysis. Then we compute sky coverage maps for \textit{VLTI} \textit{UTs} (Section~\ref{sec:VLTI-UTS}) and \textit{VLTI} \textit{ATs} (Section~\ref{sec:VLTI-ATS}) after a general presentation of our sky coverage computation methodology (Section~\ref{sec:Methodology}). As an example of the application of our methodology, we investigate the potential of off-axis tracking for interferometric observations of AGNs, briefly presented in Section~\ref{AGN-Examp} and we consider the potential of \textit{\textit{GRAVITY}+} for a large program on BLR  (Section~\ref{Interf-BLR}) and the potential of \textit{MATISSE} for several types of measures on dust torus, including imaging (Section~\ref{Img-BLR}). We finally briefly consider the impact of our analysis on the feasibility of the Planet Formation Imager \citep[PFI;][]{2018ExA....46..517M} that is the new generation interferometric project intended for high-resolution images of planetary formation down to the Hill sphere of giant planets. We consider the main goal of PFI that is YSOs observed in the $ N $ band and its capacity to observe AGNs both in a \textit{Paranal} site and at Dome C.

%% file: 4-FT.tex
\section{Off-axis fringe tracking in optical interferometry}
An optical interferometer produces a Young fringe pattern for each one of its baselines. These patterns, called interferograms, can be merged in 
\textquoteleft global\textquoteright ~ beam combiners or analysed separately in \textquoteleft pairwise\textquoteright ~ beam combiners. For each baseline $ij$ between the apertures $T_i$ and $T_j$, the Fourier transform of the interferograms yields the measured complex coherent flux 
\begin{eqnarray}
C_{i j}(\lambda)&=&\sqrt{n_{* i}(\lambda) n_{* j}(\lambda)} V_{_{* i j}}(\lambda) e^{i \varphi_{_{* ij}}(\lambda)} V_{_{I i j}}(\lambda) e^{i \varphi_{_{I i j}}(\lambda)}\nonumber\\
&=&V_{C i j}(\lambda) e^{i \varphi_{_{C ij}}(\lambda)},
\end{eqnarray}
where $n_{*i}(\lambda)$ is flux received from $T_i$ in the interferogram, $V_{* i j}(\lambda) e^{i \varphi_{_{* ij}}(\lambda)}$  is the complex visibility of the source and  $V_{I i j}(\lambda) e^{i \varphi_{_{I i j}}(\lambda)}$ is the interferometer complex visibility. The atmospheric turbulence and the interferometer vibrations introduce rapid variations of the OPD between the two beams, which affect the interferometer phase
\begin{equation}
\varphi_{_{I i j}}(\lambda, t)=\varphi_{_{S i j}}(\lambda)+\frac{2 \pi}{\lambda} p_{i j}(t),
\end{equation}
where $\varphi_{_{I i j}}(\lambda, t)$ is a fixed (or slowly varying) instrumental phase and $p_{i j}(t)$ is the instantaneous value of the OPD between $T_i$ and $T_j$, called the piston. The typical time needed to observe piston variations larger than $\frac{\lambda}{10}$ is called the piston coherence time $\mathcal{T}_{pC}(\lambda)$. The atmospheric turbulence sets seeing dependent piston coherence time values ranging typically from \SI{15}{\milli\second} in the $J$ band at \SI{1.25}{\micro\meter} to \SI{200}{\milli\second} in the $ N $ band at \SI{11}{\micro\meter}. These typical values are from the \textit{AMBER} \citep{2007A&A...464....1P}, \textit{MIDI} \citep{2004A&A...423..537L}, and \textit{MATISSE} \citep{2020SPIE11446E..0LP} experience on the \textit{VLTI}. The relevant values for the $K$ band used in this paper are discussed in Sections \ref{section-loop-error} and \ref{Optimum-time}. If the piston is not corrected, the average  $\left\langle e^{i \varphi_{_{* ij}}}\right\rangle $  rapidly tends to zero for exposures longer than $\mathcal{T}_{pC}(\lambda)$. Then, it is only possible to record short exposures and to compute piston independent quantities such as the squared modulus of the coherent flux $\left| C_{ij} \right| ^2$. This limits the sensitivity of interferometric instruments, for example to K<10 and N<7 on \textit{UTs} for \textit{AMBER} and \textit{MIDI}.

The device used to measure and correct in real-time the interferometric pistons $p_{ij}(t)$ is called a FT that is the AO system specific to a long-baseline interferometer. As there is no practical way to measure the contribution of the atmosphere to $p_{ij}(t)$ on artificial sources, the piston must be deduced from the coherent flux phase $\varphi_{_{I i j}}(t)$ measured on the observed source itself (on-axis fringe tracking) or on a nearby guide star (off-axis fringe tracking).

In practice, we measure the variation of the coherent flux phase with time: 
\begin{eqnarray}
\varphi_{_{C i j}}(\lambda, t+\d t)-\varphi_{_{C i j}}(\lambda, t)=&&\hspace*{-5mm}\arg \left[C_{i j}(\lambda, t+\d t)\cdot C^\ast_{i j}(\lambda, t)\right] \nonumber\\ \simeq&& \hspace*{-5mm}\frac{2 \pi}{\lambda}\left[p_{i j}(t+\d t)-p_{i j}(t)\right],
\end{eqnarray}

if we assume that all other phase terms vary much slower than the piston. We actively correct this measured variation of the piston to maintain it constant. This variation of the phase with time is called \textit{phase delay}. It is directly extracted from the phase of the measured coherent flux.
The phase delay is defined modulo $2\pi$ and measures the piston with a $k\lambda$ ambiguity. The absolute value of the piston can be deduced from the variation of the phase with wavelength $\lambda$ or wave number $\sigma=\frac{1}{\lambda}$,
\begin{eqnarray}
d \varphi_{_{C i j}}(\sigma, t)&=&\arg \left[C_{i j}(\sigma+\d \sigma, t) \cdot C_{i j}^{*}(\sigma, t)\right]\nonumber\\&=&\left[\d \varphi_{_{* i j}}(\sigma)+\d \varphi_{_{S i j}}(\sigma)-2 \pi p_{i j}(t)\right] \d \sigma.
\end{eqnarray}

The slope of the coherent flux phase as a function of the wavenumber  $\frac{\d\varphi_{_{C i j}}(\sigma,t)}{\d\sigma}$ is called the\textit{ group delay}. It merges the absolute piston $p_{ij}(t)$ with instrumental and source terms. It could be used for fringe tracking, but it is usually not accurate enough to control the piston within a fraction of a wavelength at high frequency. However, it is a critical component of the robustness of a FT as it is used to acquire and reacquire the fringes when the tracking is lost and to maintain the fringes near the centre of the coherence length. To make an analogy with the optics of single large apertures, the \textit{group delay} tracking is the \textit{active optics} system that allows setting the working point of the interferometer, while the \textit{phase delay} tracking is the AO system that allows correcting the effects of atmospheric turbulence and to make long exposures for high sensitivity observations.

The most efficient, state of the art, FT in operation is the GFT 
\citep{2019A&A...624A..99L}. It can currently track up to magnitudes $\mathrm{K}=9.5$ with the \SI{1.8}{\meter} diameter \textit{ATs} and $\mathrm{K}=10.5$ with the \SI{8}{\meter} diameter \textit{UTs}. It can be used off-axis with so far a small separation between the science source and the guide star. Then it allows making long exposure observations on science targets as faint as $\mathrm{K}\simeq$17 \citep{2017Msngr.170...10G}. The \textit{GRAVITY+} proposal for \textquoteleft sky faint sources \textit{VLTI} observations\textquoteright  \citep{2019vltt.confE..30E} includes plans to increase the GFT sensitivity on \textit{UTs} and allowing off-axis tracking with much larger separations is a part of it that already went through its first tests.

Several authors \citep*{2012SPIE.8445E..1LM,2019vltt.confE..34I} have proposed alternative FT designs. Below we will consider more in detail the Hierarchical Fringe Tracker \citep[HFT;][]{2016SPIE.9907E..1FP,2019vltt.confE..37P} that we have designed to be intrinsically more sensitive than the GFT. In the next subsection,  we discuss first the parameters and the computation of the sensitivity of a FT.

\subsection{Error budget of off-axis fringe tracker and sky coverage \label{sec:FTB}}

The quality of fringe tracking will depend on :

\begin{itemize}
	\setlength{\itemindent}{0.8cm}
	\item[(i)] The precision of the phase delay measurement (variance $\sigma_D^2$) that, for a given design, is a function of the the number of detected photons collected from the guide source in each  frame with Detector Integration Time  $\mathcal{T}_{F}$.
	
	\item[(ii)] The error introduced by the servo loop (variance $\sigma_L^2$) that applies an OPD correction sometime after its measurement. It is dominated by the seeing dependent temporal variation of the atmospheric piston over the FT period $\mathcal{T}_{L}$, if the telescope and interferometer mechanical vibrations are properly damped.
	
	\item[(iii)] The anisopistonic error (variance $\sigma_A^2$) appears because the guide star and the observed target are not seen through the same atmospheric turbulence. For given seeing and site conditions, it depends on the angular separation between the two sources.
	
	\item[(iv)] The chromatic phase delay difference (variance $\sigma_H^2$) between the wavelength used for guiding and for the observation that appears between near and mid-infrared (IR) mainly because of the variations of the differences in water vapor columns. This term is relevant only when FT and scientific observations are performed in different spectral bands.
\end{itemize}

If we have a specification $N_{spec}$ to reduce the fringe tracking residual (variance $\sigma_T^2$) below $\lambda_{o}/N_{spec}$ at the wavelength of observation $\lambda_{o}$, it writes:
\begin{equation}
\sigma_{T}^{2}\left(\lambda_{o}\right)=\sigma_{D}^{2}\left(\lambda_{o}\right)+\sigma_{L}^{2}\left(\lambda_{o}\right)+\sigma_{A}^{2}\left(\lambda_{o}\right)<\left(\frac{\lambda_{o}}{ N_{s p e c}}\right)^{2}.
\label{equ:ST:obs}
\end{equation}

If we are tracking at a wavelength $\lambda_{t}\neq\lambda_{o}$ we have a specification on the fringe tracking quality and one of the final tracking residual,
\begin{equation}
\sigma_{F}^{2}\left(\lambda_{t}\right)=\sigma_{D}^{2}\left(\lambda_{t}\right)+\sigma_{L}^{2}\left(\lambda_{t}\right)<\left(\dfrac{\lambda_{t}}{N_{t r a}}\right)^{2},
\label{eq:6}
\end{equation}
to ensure that the loop remains closed and that we do not have fringe jumps that would destroy the information at $\lambda_{o}$,
\begin{equation}
\sigma_{T}^{2}\left(\lambda_{o}\right)=\sigma_{F}^{2}\left(\lambda_{t}\right)+\sigma_{H}^{2}\left(\lambda_{t}, \lambda_{o}\right)+\sigma_{A}^{2}\left(\lambda_{o}\right)<\left(\frac{\lambda_{o}}{ N_{s p e c}}\right)^{2},
\label{equ:ST:obs:tra}
\end{equation}
where $\sigma_{H}^{2}(\lambda_{t}, \lambda_{o})$ is the variance of the phase delay difference between $\lambda_{t}$  and $\lambda_{o}$ on the guide star and $\sigma_{A}^2(\lambda_{o})$ is the variance of the anisopistonic error at the wavelength $\lambda_{o}$. 

The sky coverage near coordinates ($\alpha$, $\delta$) is the probability to find at least one usable guide star that complies with the conditions set by equations (\ref{equ:ST:obs}  or \ref{equ:ST:obs:tra}) for any science target  in that region of the sky. This definition is more restrictive and more accurate than the often used \textquoteleft the probability to find a guide star closer than the Isopistonic angle $\theta_p$ and brighter than the fringe tracking magnitude limit $M_{lim}$\textquoteright. Indeed, if the definition of the Isopistonic angle is $\sigma_{A}(\theta_p)=(\frac{\lambda_{o}}{N_{spec}})^2$, a guide star at the distance $\theta_p$ is usable only if $\sigma_{D}^2 + \sigma_{L}^2 =0$ and a guide star of magnitude $M_{lim}$ such as  $\sigma_{D}^2(M_l) + \sigma_{L}^2=(\frac{\lambda_{o}}{N_{spec}})^2$ is usable only if the distance to the target is zero. Furthermore, we have to consider the variations of seeing and hence $\sigma_{D}^2$, $\sigma_{L}^2$, and $\sigma_{A}^2$ with the minimum zenith distance of the target.

\subsection{Phase delay measurement\label{sec:phasedelay}}

The phase delay in each frame is the phase $\varphi_{_{C i j}}(\lambda)$  of the complex coherent flux $C_{i j}(\lambda)$:
\begin{equation}
\varphi_{_{C i j}}(\lambda)=\arg \left\{\frac{\mathfrak{I}\left[C_{i j}(\lambda)\right]}{\mathfrak{R}\left[C_{i j}(\lambda)\right]}\right\},
\label{equ:phase-var}
\end{equation}
where $\mathfrak{R}$ and $\mathfrak{I}$ note the real and imaginary parts of the complex coherent flux. 

The variance of the coherent flux is 
\begin{equation}
\operatorname{Var}\left[C_{i j}(\lambda)\right]=\operatorname{Var}\left(\mathfrak{I}\left[C_{i j}(\lambda)\right]\right)+\operatorname{Var}\left(\mathfrak{R}\left[C_{i j}(\lambda)\right]\right).
\label{equ:phase}
\end{equation}

Assuming that the real and imaginary parts of the coherent flux are independent variables with the same variance, which would be the case for an ideal beam combiner and has been found to be a good approximation in real instruments, yields:
\begin{equation}
\operatorname{Var}\left(\Re\left[C_{i j}(\lambda)\right]\right)=\operatorname{Var}\left(\Im\left[C_{i j}(\lambda)\right]\right)=\operatorname{Var}\left[C_{i j}(\lambda)\right] / 2.
\end{equation}

Many authors \citep{1989ASIC..274..249P,2006MNRAS.367..825V,2008SPIE.7013E..1BH} have computed the variance of the coherent flux in the presence of detector noise and source plus background photon noise. For a pairwise beam combiner using single mode spatial filters, this variance is given by:
\begin{equation}
\operatorname{Var}\left[C_{i j}(\lambda)\right]=\frac{n_{* i}+n_{* j}+n_{B i}+n_{B j}}{N_{P} N_{\lambda}} \mathcal{T}_{F} \alpha^{2}+N_{X} \sigma_{R}^{2},
\end{equation}

where $n_{* i}$ is the number of detected photons/s received from the telescope $T_i$ in the FT spectral band, $n_{Bi}$  is the number of detected background photons/s in the beam i, in the same spectral band, at the entrance of the cold optics, $\mathcal{T}_{F}$ is the frame time, $N_P$ is the number of fringe tracking baselines involving each telescope, $N_\lambda$ is the number of spectral channels, $N_X$ is the number of pixels per spectral channel, $\sigma_{R}^2$ is the variance of the detector read-out noise, and $\alpha$ is the photon noise amplification factor observed in intensified detectors. Equation \eqref{equ:phase} simply expresses the fact that the coherent flux variance is the sum of the variance of the source and background photon noise and the detector read-out noise on all used pixels.

When we are close to the centre of the coherence length (i.e. the fringes have been acquired and centred by the group delay tracker), we can coherently add the different spectral channels (i.e. use an average of the interferograms over the available bandpass) and the variance of the broadband coherent flux becomes:
\begin{equation}
\operatorname{Var}\left(C_{i j L}\right)=\frac{n_{* i}+n_{* j}+n_{B i}+n_{B j}}{N_{P}} \mathcal{T}_{F} \alpha^{2}+N_{\lambda} N_{X} \sigma_{R}^{2}.
\end{equation}

The average coherent flux is given by
\begin{equation}
\left\langle C_{i j L}\right\rangle=\frac{\mathcal{T}_{F} \sqrt{n_{* i} n_{* j}}}{N_{P}} V_{i j},
\end{equation}
where $V_{i j}$ is the visibility measured on the fringe tracking target. In single-mode instruments, the variations of $V_{ij}$ are strongly dominated by changes in the residual piston jitter:
\begin{equation}
V_{I i j}=\exp \left[-2\left(\frac{\pi}{\lambda}\right)^{2}\left(\sigma_{V i j}^{2}+\sigma_{T i j}^{2}\right)\right],
\end{equation}
with $\sigma_{V}^2$ and $\sigma_{T}^2$ are the residual variances of the vibrational and atmospheric piston jitters.

When the coherent flux signal-to-noise ratio (SNR) is high, and the phase delay is small, equations ( \ref{equ:phase-var} and \ref{equ:phase} ) yield \cite*{1986JOSAA...3..634P},
\begin{eqnarray}
\varphi_{C i j L}&=&\frac{\Im C_{i j L}}{\Re C_{i j L}} \quad ,
\text { and } \nonumber\\
\quad \sigma_{\varphi_{C i j L}}&=&\frac{\sqrt{\operatorname{Var}\left(\Im C_{i j L}\right)}}{\left\langle\Re C_{i j L}\right\rangle}=\frac{1}{\sqrt{2} S N R_{C_{i j L}}}.
\end{eqnarray}

We define the ratio:
\begin{eqnarray}
Z_{i j L}&=&\frac{\sqrt{\operatorname{Var}\left(\Im C_{i j L}\right)}}{\left\langle\Re C_{i j L}\right\rangle}\\\nonumber
&=&\frac{\sqrt{\left(n_{* i}+n_{* j}+n_{B i}+n_{B j}\right) \mathcal{T}_{F} \alpha^{2}+N_{X} N_{P} N_{\lambda} \sigma_{R}^{2}}}{\mathcal{T}_{F} \sqrt{2 n_{* i} n_{* j} V_{i j}}} \sqrt{N_{P}}.
\end{eqnarray} 

We assume that the two beams see comparable backgrounds 
$n_{B i}\simeq n_{B j}\simeq n_{B}$. To express the unbalance between beams, we write $n_{* i}=n_{*}(1-\gamma)$ and $n_{* j}=n_{*}(1+\gamma)$ where $n_{*}$ is the average source flux and $\gamma=\frac{n_{* j}-n_{*} i}{n_{* j}+n_{* i}}$ describes the source flux ratio between the two beams. 

Then
\begin{eqnarray}
Z_{i j L}&=&\frac{\sqrt{\left(n_{*}+n_{B}\right) \mathcal{T}_{F} \alpha^{2}+\frac{N_{X} N_{P} N_{\lambda}}{2} \sigma_{R}^{2}}}{\mathcal{T}_{F} n_{*} V_{i j} \sqrt{1-\gamma^{2}}} \sqrt{N_{P}}\nonumber\\&=&\frac{1}{\sqrt{2} S N R_{C_{i j L}}}.
\label{equ:fx-ratio}
\end{eqnarray}

The flux unbalance is equivalent to a change in the instrument visibility, negligible for small flux differences $(\gamma \sim 0)$, and dramatic when one beam is lost $(\gamma=1)$. Note that if the collected flux varies rapidly within the frame time, the correct expression for $\gamma$ is given by an average of its instantaneous values:
\begin{equation}
\gamma(t)=\frac{1}{\mathcal{T}_{F}} \int_{t}^{t+\mathcal{T}_{F}} \frac{n_{* j}\left(t^{\prime}\right)-n_{* i}\left(t^{\prime}\right)}{n_{* j}\left(t^{\prime}\right)+n_{* i}\left(t^{\prime}\right)} \d t^{\prime}.
\end{equation}

Micro cuts in the injection efficiency at different instants in each beam can result in strong SNR drops even if the average fluxes $\left\langle n_{* i}(t)\right\rangle$ and $\left\langle n_{* j}(t)\right\rangle$ are comparable.

The approximation $\sigma_{\varphi_{C i j L}}=\frac{\sqrt{\operatorname{Var}\left(\Im C_{i j L}\right)}}{\left\langle\Re C_{i j L}\right\rangle}=Z_{i j L}$ is considered in the literature as acceptable up to $Z_{i j L}=1$ rad for phase precision estimates. We found that this underestimates the phase delay error at low SNR with possibly significant impact on the actual fringe tracking limiting magnitude. \cite{1965prvs.book.....P} gives an expression of the probability density of $\phi=\arg (\mathfrak{I} / \Re)$ as a function of the ratio $Z=\sqrt{\operatorname{Var}(\mathfrak{J})} /\langle\Re\rangle,$ when $\mathfrak{I}$ and $\Re$ are two independent normal variables with the same variance $\operatorname{Var}(\mathfrak{I})=\operatorname{Var}(\Re)$ and $<\mathfrak{I}>=0,$
\begin{equation}
f_{\phi}(\theta)=\begin{cases} 
\dfrac{e^{\frac{-1}{2 Z^{2}}}}{2 \pi}+\dfrac{\cos(\theta) e^{\frac{-\sin ^{2}(\theta)}{2 Z^{2}}}}{2 Z \sqrt{2 \pi}} \\ \times\left[1+2\mathrm{erf}\left(\frac{\cos(\theta)}{Z}\right)\right];&\text{if }|\theta| \leq min(\kappa Z,\pi) \\ 
0;	& \text{ else }
\end{cases},
\label{equ:paplois}
\end{equation}

	\begin{figure}
	\centering
	\includegraphics[width=\linewidth]{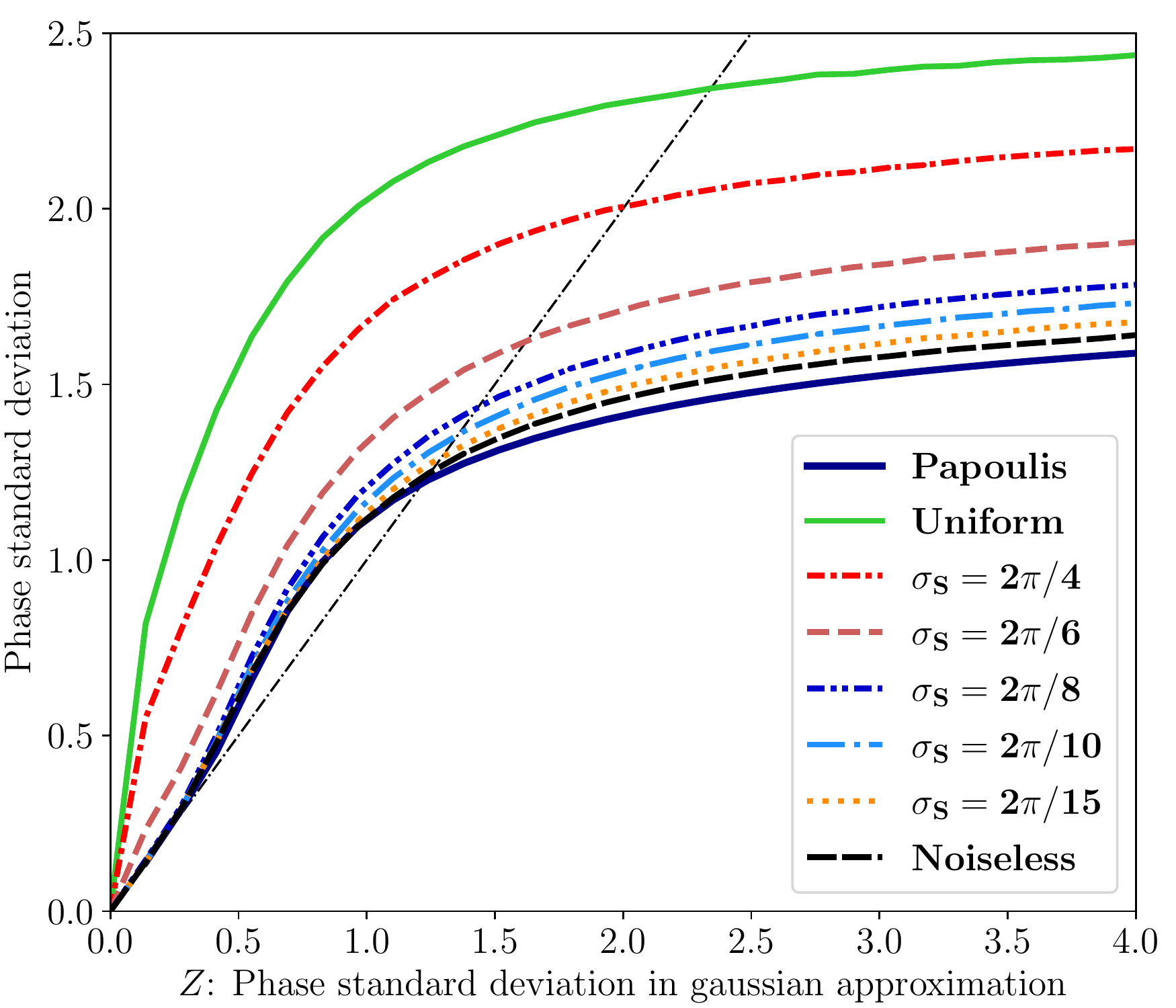}
	\caption{Standard deviation of the phase estimation error in one frame $\sigma_{\phi}$, as a function
		of $Z$ that is its value in the Gaussian approximation ($Z=1/\sqrt{2SNR_C}$), for different statistics
		of the \textquoteleft input\textquoteright ~ phase that has to be estimated. From top to bottom: input phases uniformly distributed between $\pm\pi$ (top, full green line), input phases with a normal distribution with standard deviations $\sigma_{S}=2\pi/4$, $\sigma_{S}=2\pi/6$, $\sigma_{S}=2\pi/8$ ; $\sigma_{S}=2\pi/10$ ; $\sigma_{S}=2\pi/15$, and finally $\sigma_{S}=0$ for the noiseless case (bottom, black dashed line). The thin straight line represents the Gaussian approximation.
	}
	\label{fig:1b}
\end{figure}

where we have introduced the  $min(\kappa Z,\pi)$ limit to eliminate the incorrect behaviour of that Papoulis function near $\pm\pi$. The parameter $\kappa=5$ has been selected from numerical tests to allow a good representation of the actual probability density.

When $Z \ll 1, f_{\phi}(\theta)$ is very close to a normal distribution with standard deviation $Z$, and when $Z>1, f_{\phi}(\theta)$ rapidly tends towards an uniform distribution between $-\pi$ and $+\pi$ with a standard deviation $\pi / \sqrt{3}$. The variance of $\arg (\mathfrak{J} / \Re)$ is given by:
\begin{equation}
\sigma_{\phi, Z}^{2}=\int_{-\pi}^{\pi} \theta^{2} f_{\phi, Z}(\theta) \d \theta,
\end{equation}

The standard deviation $\sigma_{\phi}$ is a correct estimation of the error on a phase very close to zero.
The function $\sigma_{\phi}(Z)$is plotted in  \figref{fig:1b} that shows that $\sigma_{\phi}(Z)\backsimeq Z$ looks like an acceptable approximation up to $Z\sim1.2$. 

\begin{figure}
		\centering
		\includegraphics[width=\linewidth]{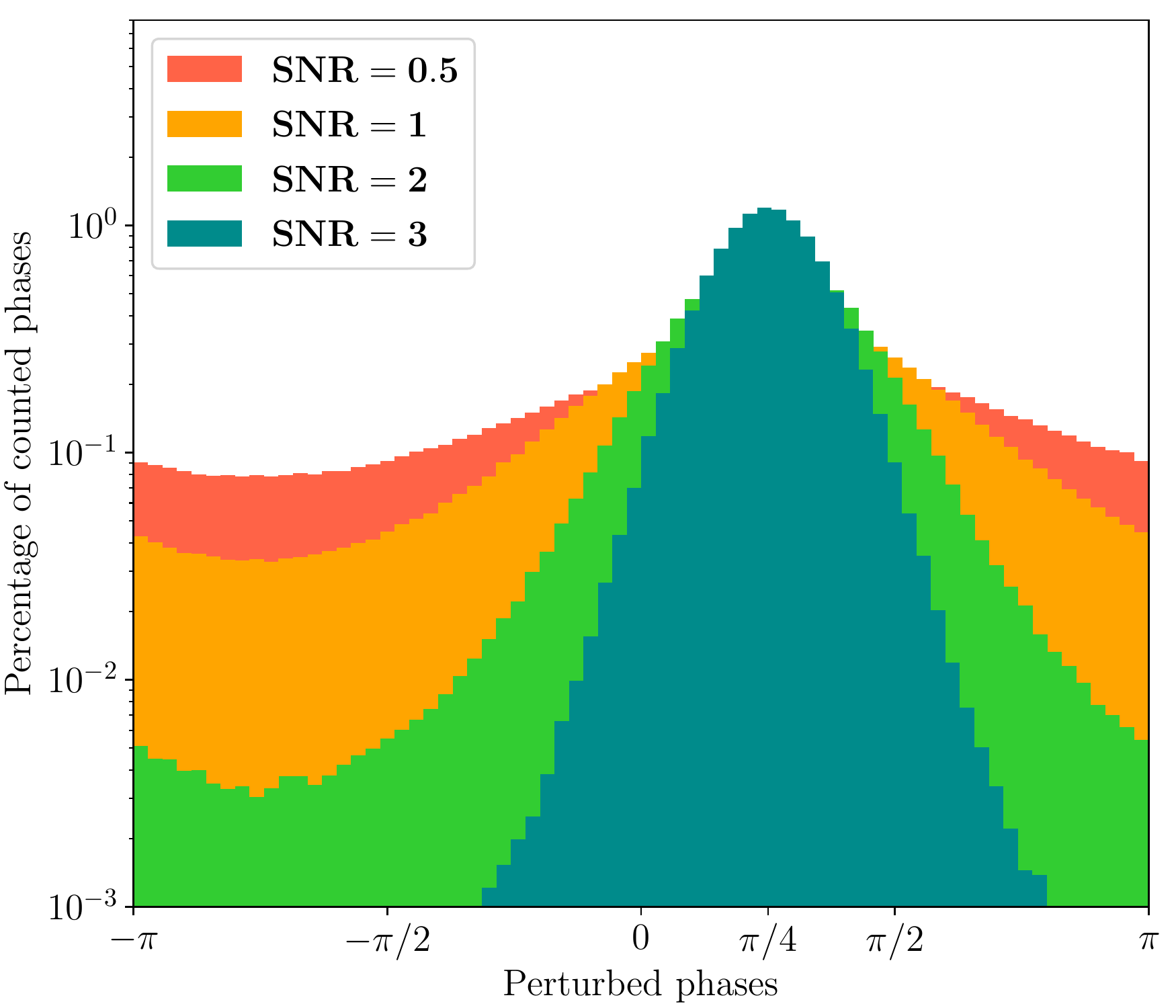}
		\caption{Histogram of the phase estimator of a coherent flux phase of \SI{45}{\degree}, for different
			values of the coherent flux SNR (from top to bottom): 0.5; 1; 2; and 3.}
		\label{fig:1a}
\end{figure}

However, when it is applied to the non-zero phase $\theta$ of a noisy coherent flux, the estimator
\begin{equation}
\theta^{\prime}=\arg \left[M \sin (\theta)+I_{N}, M \sin (\theta)+R_{N}\right],
\label{equ:theta_hist}
\end{equation}
is biased with a bias depending on $\theta$ and the coherent flux SNR. This is a mathematical bias entirely due to the non-linearity of the argument function for values that are far from zero. The probability density of the phase can be roughly divided in a narrow part centred around $\theta$ on top of broad part close to a uniform distribution between $\pm \pi$, as illustrated by  \figref{fig:1a}, with an average shifted towards zero in proportion to the relative importance of the broad part, i.e. of the coherent flux SNR. Therefore, the phase is systematically underestimated. This bias could be evaluated and explicitly corrected with a certain cost in phase delay precision. It can also be implicitly corrected by the optimization of the gain of the fringe tracking loop as a function of coherent flux SNR. A full discussion of this mathematical bias is beyond the scope of this paper and here we only evaluate its global effect on the fringe detection SNR.

We compute the variance of the estimator $\theta_{Z}^{\prime}$ in equation \eqref{equ:theta_hist} from a large number of values of $\theta^{\prime}$ for random values of $\theta, I_{N}$ and $R_{N} .$ The phase $\theta$ to be estimated is a centred Gaussian random variable with variance $\sigma_{S}^{2}.$  $I_{N}$ and $R_{N}$ are centred Gaussian random variables with variance 1 that represent the noise on the imaginary and real parts of the coherent flux. The parameter $M$ sets the value $Z=\frac{\sqrt{\operatorname{Var}(\mathfrak{I})}}{\langle\Re\rangle}=1 / M .$ We compute the standard deviation of the estimation error
\begin{equation}
\sigma_{\varphi, \sigma_{S}}^{2}(Z)=\left\langle\left(\theta_{Z}^{\prime}-\theta\right)^{2}\right\rangle,
\end{equation}

\figref{fig:1b} shows $\sigma_{\varphi, \sigma_{S}}=f(Z)$ for various values of $\sigma_{S} .$ The line $\sigma_{\varphi}(Z)=Z$ represents the Gaussian approximation of the phase delay error. For $\sigma_{S}=0$, we obtain the standard deviation of $f_{\phi}(\theta)$: $\sigma_{\varphi, 0}(Z)=\sigma_{\phi}$ discussed above. For $\sigma_{S} \lesssim 8.5$, we have $\sigma_{\varphi, \sigma_{S}}(Z)=Z$ for $Z \leqslant 0.25$ and then $\sigma_{\varphi, \sigma_{S}}(Z)-Z$ grows with $\sigma_{S} .$ For $\sigma_{S} \geq 8.5$, we have $\sigma_{\varphi, \sigma_{S}}(Z) \neq Z$ for all values of $Z$ which means that $\sigma_{\varphi}$ is not proportional to $1 / \sqrt{n_{*}}$ even at very high flux.

In a given frame, the fringe sensor has to estimate a phase delay $\theta\neq 0$ because the phase delay in the previous frame has been estimated with an error and corrected with an even larger error due to the variation of the atmospheric phase delay during the frame duration. The change in phase delay between measure and correction is what we have called the \textquoteleft servo loop\textquoteright ~ error. Its value is discussed in Section \ref{section-loop-error}. Here, we evaluate only its influence on the phase delay estimation error. 

For given observing conditions and a given frame time $\mathcal{T}_{F}$, the servo loop error has a constant standard deviation $\sigma_{L}$. It will trigger an estimation error $\sigma_{\varphi, \sigma_{L}}(Z)$ that will add quadratically to the servo loop error in the next frame. To estimate this effect, we have computed numerically the series:
\begin{equation}
\sigma_{\varphi, \sigma_{L}}(Z, N)=\sigma_{\varphi, \sigma_{N}} \text { with } \sigma_{N}^{2}=\sigma_{\varphi, \sigma_{L}}^{2}(Z, N-1)+\sigma_{L}^{2}.
\label{equ:sigmaphi}
\end{equation}

\begin{figure}
\centering
\includegraphics[width=\linewidth]{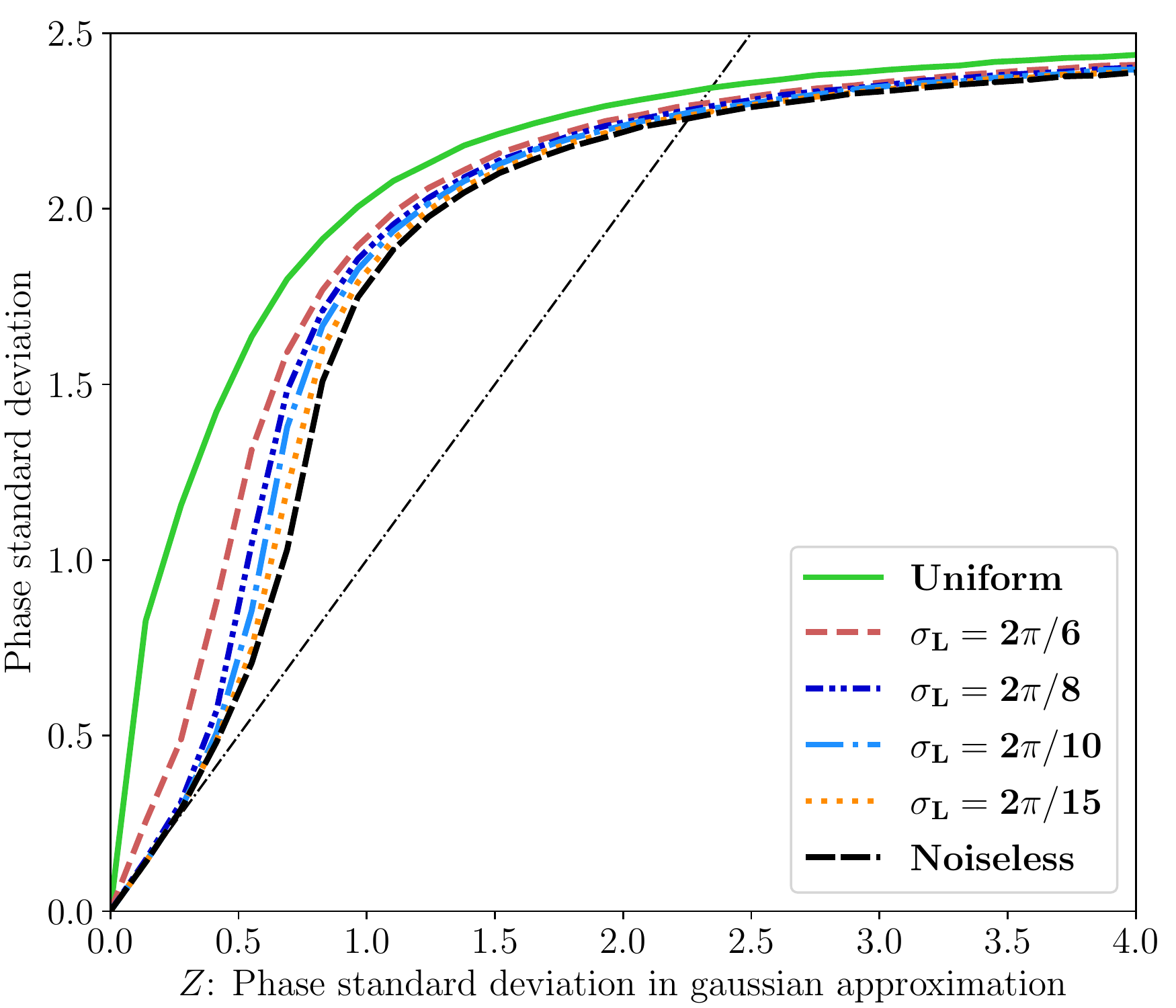}
\caption{Asymptotic standard deviation of the phase estimation error after a few frames, for different values of the servo loop standard deviation $\sigma_{L}$. From bottom to top: $\sigma_{L}=0, \frac{2\pi}{15} ; \frac{2\pi}{10} ; \frac{2\pi}{8} ;  \frac{2\pi}{6}$. The full green line represents the estimation error on a phase randomly	distributed between $\pm\pi$.}
\label{fig:1c}
\end{figure}

This series converges very rapidly, in less than 10 frames, towards an asymptotic value $\sigma_{final}(Z,\sigma_{L})$ that is independent of the starting point of the series. This final phase delay estimation error is plotted in  \figref{fig:1c}. We see that for servo loop errors larger than $\sim 2\pi/8$, $\sigma_{final}(Z,\sigma_{L})>Z$ for any value of $Z\lesssim2.5$. For smaller servo loop errors, the Gaussian approximation $\sigma_{final}(Z,\sigma_{L})=Z$ is valid up to $Z\sim 0.25$ (phase delay SNR>4) and increase much more rapidly with $Z$ after that value. 

A phase delay SNR=1 rad can be achieved for a coherent flux SNR~1 ($Z\sim 0.7$) on a single-phase estimator.

\subsection{Phase delay estimators combining all baselines}

For a four telescope interferometer, we can retrieve only three combined phase delays from the three piston differences:
\begin{equation}
\theta_{i j}=\frac{2 \pi}{\lambda}\left(P_{j}-P_{i}\right) \text{ with } i \in\{1,2,3\} \text{ and } j=i+1.
\end{equation}

We have a phase delay $b_{ij}$ and a phase delay measurement $m_{ij}$ for each baseline with:
\begin{equation}
\left\{\begin{array}{l}
	b_{12}=\theta_{12} \\
	b_{13}=\theta_{12}+\theta_{23} \\
	b_{14}=\theta_{12}+\theta_{23}+\theta_{34} \\
	b_{23}=\theta_{23} \\
	b_{24}=\theta_{23}+\theta_{34} \\
	b_{34}=\theta_{34}
	\end{array}\right.,
	\label{equ:pha-delay-mes}
\end{equation}
and, as in equation \eqref{equ:theta_hist}

\begin{equation}
\resizebox{0.91\linewidth}{!}{$m_{i j}=\arg \left[M_{i j} \sin \left(b_{i j}+\varphi_{n i j}\right)+I_{i j}, M_{i j} \cos \left(b_{i j}+\varphi_{n i j}\right)+R_{i j}\right]$},
\end{equation}
$\varphi_{n i j}$ is the noise on the phase resulting from the error on the correction of the previous frame, $R_{ij}$ and $I_{ij}$  are the noise on the real and imaginary parts of the coherent flux of mean modulus $M_{ij}$.

The estimators $\theta_{i j}^\prime$ of the phase delay are the parameters that minimize the sum of square differences:
\begin{equation}
\chi^{2}=\sum_{i=1}^{N_{T}-1} \sum_{j=i+1}^{N_{T}}\left(m_{i j}-b_{i j}^{\prime}\right)^{2},
\end{equation}
with $b^\prime_{ij}$ deduced from the $\theta_{lk}^\prime$ using equations \eqref{equ:pha-delay-mes}.
\begin{eqnarray}
\chi^{2}&=&\left(m_{12}-\theta_{12}^{\prime}\right)^{2}+\left(m_{13}-\theta_{12}^{\prime}-\theta_{23}^{\prime}\right)^{2}\nonumber\\
&+&\left(m_{14}-\theta_{12}^{\prime}-\theta_{23}^{\prime}-\theta_{34}^{\prime}\right)^{2}+\left(m_{23}-\theta_{23}^{\prime}\right)^{2}\nonumber\\
&+&\left(m_{24}-\theta_{23}^{\prime}-\theta_{34}^{\prime}\right)^{2}+\left(m_{34}-\theta_{34}^{\prime}\right)^{2}.
\end{eqnarray}

The three equations setting the minimum value of $\chi^2$ for each of the parameters
$$\dfrac{\d\chi^2}{\d\theta_{lk}^\prime}=0,$$

write
\begin{equation}
\left[\begin{array}{c}
\theta_{12}^{\prime} \\
\theta_{23}^{\prime} \\
\theta_{34}^{\prime}
\end{array}\right]\left[\begin{array}{lll}
3 & 2 & 1 \\
2 & 4 & 2 \\
1 & 2 & 3
\end{array}\right]=\left[\begin{array}{c}
m_{12}+m_{13}+m_{14} \\
m_{13}+m_{14}+m_{23}+m_{24} \\
m_{14}+m_{24}+m_{34}
\end{array}\right],
\end{equation}

hence, from the inversion of that system matrix
\begin{equation}
\resizebox{0.91\linewidth}{!}{$
\left[\begin{array}{c}
\theta_{12}^{\prime} \\
\theta_{23}^{\prime} \\
\theta_{34}^{\prime}
\end{array}\right]=\frac{1}{4}\left[\begin{array}{ccc}
2 & -1 & 0 \\
-1 & 2 & -1 \\
0 & -1 & 2
\end{array}\right]\left[\begin{array}{c}
m_{12}+m_{13}+m_{14} \\
m_{13}+m_{14}+m_{23}+m_{24} \\
m_{14}+m_{24}+m_{34}
\end{array}\right]$},
\end{equation}

or
\begin{equation}
\begin{array}{l}
\theta_{12}^{\prime}=\dfrac{2 m_{12}+\left(m_{13}-m_{23}\right)+\left(m_{14}-m_{24}\right)}{4} \\
\theta_{23}^{\prime}=\dfrac{2 m_{23}+\left(m_{13}-m_{12}\right)+\left(m_{24}-m_{34}\right)}{4} \\
\theta_{34}^{\prime}=\dfrac{2 m_{34}+\left(m_{14}-m_{13}\right)+\left(m_{24}-m_{23}\right)}{4}
\end{array},
\end{equation}

In each one of these equations, we combine three estimators of the phase delay. For example $m_{12}$, $(m_{13}-m_{23})$, and $(m_{14}-m_{24})$ are all estimators of $\theta_{12}$. When one of these estimators jumps by $2\pi$, the combined estimator $\theta_{34}^\prime$ will jump by $\pi/2$ or by $\pi$. To avoid this partial wrapping of the phase we write:
\begin{equation}
\begin{array}{l}
\theta_{12}^{\prime}=\left(\frac{2 m_{12}+\left(m_{13}-m_{23}+\pi\right)_{[2 \pi]}+\left(m_{14}-m_{24}+\pi\right)_{[2 \pi]}-2 \pi}{4}+\pi\right)_{[2 \pi]}-\pi \\
\theta_{23}^{\prime}=\left(\frac{2 m_{23}+\left(m_{13}-m_{12}+\pi\right)_{[2 \pi]}+\left(m_{24}-m_{34}+\pi\right)_{[2 \pi]}-2 \pi}{4}+\pi\right)_{[2 \pi]}-\pi \\
\theta_{12}^{\prime}=\left(\frac{2 m_{34}+\left(m_{14}-m_{13}+\pi\right)_{[2 \pi]}+\left(m_{24}-m_{23}+\pi\right)_{[2 \pi]}-2 \pi}{4}+\pi\right)_{[2 \pi]}-\pi
\end{array},
\end{equation}

The standard error on the phase delay estimation will then be
\begin{equation}
 \sigma_{\theta_{i j}}^{2}=\operatorname{Var}\left(\theta_{12}^{\prime}-\theta_{i j}\right).
\label{equ:var-def-phase}
\end{equation}

\begin{figure}
	\centering
	\includegraphics[width=\linewidth]{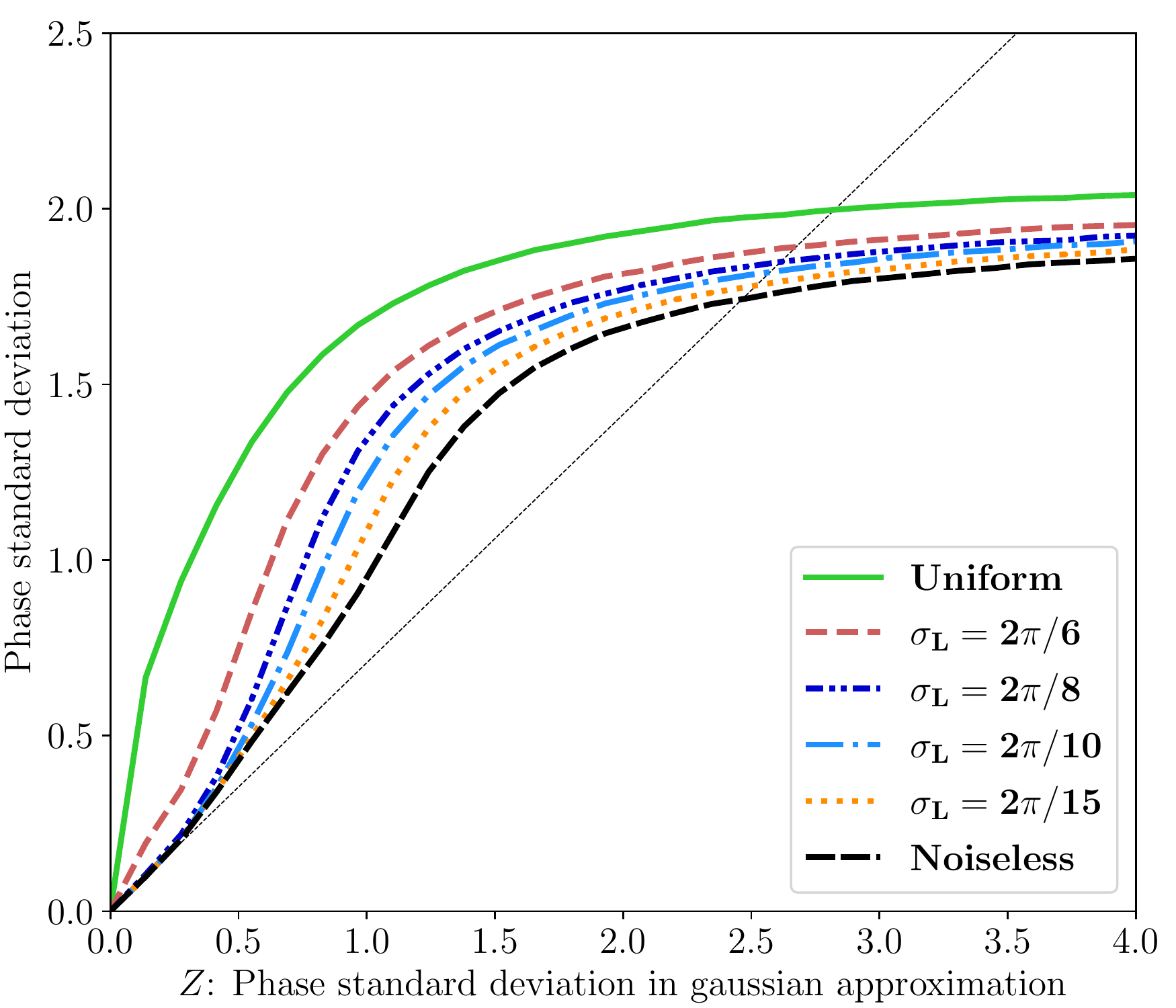}
	\caption{Standard deviation of the three independent phase delays computed from the six
		baselines as a function of $Z$ that is the Gaussian approximation for the phase delay error per
		baselines. The parameters are the same as in \figref{fig:1c}.}
	\label{fig:1d}
\end{figure}

As we have seen above for the phase delay estimator on a single baseline, $\sigma_{\theta_{i j}}^2$ depends on the coherent flux $SNR_C=\frac{M_{ij}}{\sqrt{\operatorname{Var}(I_{ij})}}=\frac{1}{Z}$ and from the variance of $\theta_{i j}$. We apply the same iterative simulation as for a single-phase delay and the result is given in  \figref{fig:1d}.

The final result is that the Gaussian approximation is not valid below $SNR_C<3$ $(Z=0.5)$. A phase delay estimation error of \SI{0.4}{\radian}, corresponding to $\frac{\lambda}{15}$, can be achieved for $SNR_C\sim2$ for $\sigma_{L}=\frac{2\pi}{15}$.

Equation \eqref{equ:var-def-phase} and the numerical computation described above give the standard deviation of the
phase delay estimator as a function of Z= $1/(\sqrt{2} SNTR_c)$ described in equation \eqref{equ:fx-ratio}. 
We perform a polynomial fit of the result given by equation \eqref{equ:var-def-phase} up to $\sigma_{\theta_{i j}}=1$:
\begin{equation}
\sigma_{\theta_{i j}}\left(Z_{i j L}\right)=Z_{i j L}+a_{1} Z_{i j L}^{3}+a_{5} Z_{i j L}^{5}+a_{7} Z_{i j L}^{7}+a_{9} Z_{i j L}^{9},
\label{equ:sigma_theta_poly}
\end{equation}
with $a_1$, $a_3$,  $a_5 $, $a_7$, and $a_9 $ are fixed coefficients (for a given FT) that describe
the bias introduced by the estimator argument applied to a noisy complex coherent flux. Combining equations (\ref{equ:fx-ratio} and \ref{equ:sigma_theta_poly}) yields the phase delay measurement error $\sigma_t$.

\subsection{Phase delay measurement in the GFT\label{sec:GFT}}

The GFT \citep{2019A&A...624A..99L} is a pairwise FT using an integrated optics beam combiner that splits the signal for each of the six baselines into four pixels that sample the phases $0$, $\pi/2$, $\pi$, and $3\pi/2$ in the so-called \textquoteleft ABCD\textquoteright ~ scheme. It is important to remember that integrated optics and single-mode fibers yield a spatial filtering of each telescope beam that reduces the dependence of the measured instrumental contrast and SNR to only two factors: piston jitters between beams, which affect the term $V_{Iij}$ of equation \eqref{equ:fx-ratio}, and fluctuations of the ratios of flux injected in each beam that set the term $\gamma$ in the same equation. Each output is dispersed over six spectral channels covering the $ K $ band. It uses a SAPHIRA detector \citep{2016SPIE.9909E..12F} with a very low read-out noise ($\sigma_{R}\sim0.7e^-$) and a noise amplification factor $\alpha\sim1.5$. The measured overall transmission of the \textit{VLTI} and the GFT, including the detector efficiency, is $\sim1\mathrm{\ per\ cent}$. The six phase delay measurements on each baseline are combined to obtain the three phase delays corresponding to the OPD difference between the four telescopes. For phase delay measurements, the interferograms in the six spectral channels are added after the correction of the known chromatic OPD. For group delay measurements, the spectral channels are processed separately and combined in a group delay estimator that uses 40 individual frames.

\begin{figure}
	\centering
	\includegraphics[width=\linewidth]{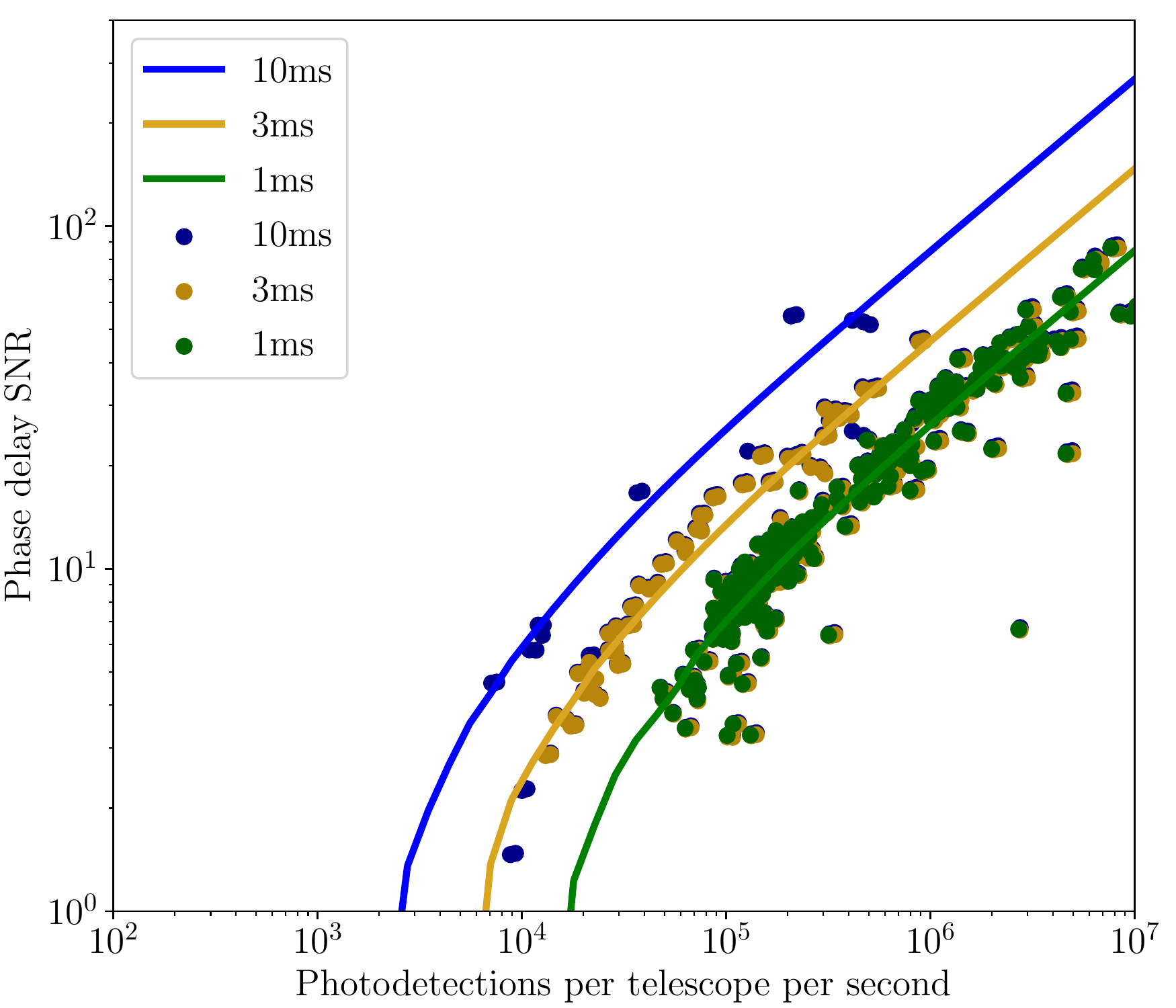}
	\caption{Phase delay  as {function of the number of detected photons per telescope per}
		second. The full lines are for frame times \SI{1}{\milli\second} (green), 3 ms (gold), and 10 ms (blue). The
		dots represent measured SNR values with the GFT copied from \protect\citep{2019A&A...624A..99L}.}
	\label{fig:lacours-data}
\end{figure}

Equation \eqref{equ:fx-ratio} allows us to express $Z_{ijL}$ as a function of the number $n_*$ per detected photons per telescope per second. Here, we assume that $Z_{ijL}-Z$ is the same for all baselines. Then we use equation \eqref{equ:var-def-phase} and the numerical computation described above to obtain the variance $\sigma_{\varphi_{\theta i j}}^2[Z(n_*)]$. With the parameters given in \tabref{tab:1} and for $V_I\sqrt{1-\gamma^2}\sim0.8$, which corresponds to the \textit{VLTI} with \textit{ATs} that has moderate vibrations and flux imbalances, we obtain the phase delay $SNR=1/\sigma_{\theta}$ as a function of $n_*$ shown in  \figref{fig:lacours-data}, where it has been plotted over the experimental data points copied from \cite{2019A&A...624A..99L}.
\begin{table}
	\centering
	\caption{Parameters for SNR and sky coverage estimate at the \textit{VLTI} and for the various FTs.}
	\small{
		\begin{tabular}{C{2.78cm}C{0.68cm}C{0.95cm}C{0.95cm}C{0.95cm}}
			\hline
			\multirow{2}{*}{}                                 & \multicolumn{2}{c}{GFT}  & GFT+     & HFT      \\ 
			& \textit{ATs}       & \textit{UTs}           & \textit{UTs}      & \textit{UTs}      \\  \hline
			\multicolumn{1}{m{2.78cm}}{Nbr. telescopes $N_T$}   
			& \multicolumn{4}{c}{4}       \\ 
			\multicolumn{1}{m{2.78cm}}{Detector}                    & \multicolumn{4}{c}{SAPHIRA}                    \\ 
			\multicolumn{1}{m{2.78cm}}{Read out noise $\sigma_{D}$}            & \multicolumn{4}{c}{0.7}                        \\ 
			\multicolumn{1}{m{2.78cm}}{Noise amplification $\alpha$}         & \multicolumn{4}{c}{1.5}                        \\ 
			\multicolumn{1}{m{2.78cm}}{Nbr. pixels $N_X$}                 &\multicolumn{4}{c}{4}  \\ 
			\multicolumn{1}{m{2.78cm}}{Seeing at \SI{0.5}{\micro\metre}}            & \multicolumn{4}{c}{\SI{0.5}{\arcsecond}}                 \\ 
			\multicolumn{1}{m{2.78cm}}{Coherence time 
				
				at \SI{0.5}{\micro\meter} }    & \multicolumn{4}{c}{\SI{7}{\milli\second}}                       \\ 
			\multicolumn{1}{p{2.78cm}}{{Atmospheric residual T}}      & \multicolumn{4}{c}{200}          \\ 
			\multicolumn{1}{m{2.78cm}}{Isoplanatic angle
				
				at \SI{0.5}{\micro\meter}} & \multicolumn{4}{c}{ \SI{2.6}{\arcsecond}}                 \\ 
			\multicolumn{1}{m{2.78cm}}{Outer Scale}                 & \multicolumn{4}{c}{\SI{25}{\meter}}                       \\
			
			\multicolumn{1}{m{2.78cm}}{Nbr. pairs $N_P$}               & 3         & 3             & 3        & 1        \\ 
			\multicolumn{1}{m{2.78cm}}{Nbr. spectral 
				
				channels $N_\lambda$}      & 6         & 6             & 6        & 1        \\ 
			\multicolumn{1}{m{2.78cm}}{Spectral band 
				
				phase delay}   & $ K $         & $ K $             & $ K $        & \SIrange{1.58}{2.3}{\micro\meter}     \\ 
			\multicolumn{1}{m{2.78cm}}{Spectral band
				
				group delay}   & $ K  $        & $ K $             & $ K $        &      \SIrange{1.1}{1.58}{\micro\meter}   \\ 
			\multicolumn{1}{m{2.78cm}}{Transmission}                & 1~per~cent       & 1~per~cent         & 1~per~cent      & 2~per~cent      \\ 
			\multicolumn{1}{m{2.78cm}}{\multirow{2}{*}{\vspace*{-0.5cm}Photons/s  ($K=0$, 	$S_{tr}$=$1$)}}      & \multirow{2}{*}{\vspace*{-0.5cm}\num{6e7}  }   & \multirow{2}{*}{\vspace*{-0.5cm}\num{1.2e9}}     &\multirow{2}{*}{\vspace*{-0.5cm}\num{1.2e9}}  &\vspace*{0.04cm}\num{2.9e9} $(\mathcal{G}{_{\tiny\mathbf{Delay}}})$\\
			\multicolumn{1}{m{2.78cm}}{}&&&&  \vspace*{0.04cm}  \num{4.4e9} $(\varphi_{_{\tiny\mathbf{Delay}}} )$       \\ 
			\multicolumn{1}{p{2.78cm}}{Strehl in $ K $}                 & 0.5       & \multicolumn{3}{c}{0.6 (LGS mode)} \\ 
			\multicolumn{1}{p{2.78cm}}{Strehl in $ H $}                 &     \textemdash    &      \textemdash         &    \textemdash      & 0.4      \\ 
			\multicolumn{1}{p{2.78cm}}{Strehl in $ J $}                 &   \textemdash        &     \textemdash          &   \textemdash       & 0.2      \\ 
			\multicolumn{1}{p{2.78cm}}{Vibration residual $\sigma_{V}$}        & 100       & 200           & 100      & 100      \\ 
			\multicolumn{1}{m{2.78cm}}{$V_I=\exp^{\left(-\frac{\pi^2(\sigma_{V}^2+\sigma_{T}^2)}{2\lambda^2}\right)}$}            & 0.81      & 0.69          & 0.81     & 0.75     \\ 
			\multicolumn{1}{m{2.78cm}}{$V_E=\sqrt{1-\gamma^2}$}                      & $\sim$1   & $\sim$0.16    & $\sim$1  & $\sim$1  \\ 
			\multicolumn{1}{m{2.78cm}}{Baseline}                    & G$_1$-J$_3$     & UT$_1$-UT$_4$       & UT$_1$-UT$_4$  & UT$_1$-UT$_4$  \\ \hline
	\end{tabular}}
	\label{tab:1}
\end{table}

When $n_*$ is converted in $ K $ band magnitude with \textit{ATs} using again the parameters in \tabref{tab:1}, the limiting magnitude that corresponds to a limit $SNR\sim3$, i.e. $5\cdot10^3$ detected photons/s, is $\mathrm{K}=9.45$, that is indeed the GFT magnitude offered to users. This shows that our analytical SNR is in good agreement with the experimental results of the GFT with \textit{ATs}.

 \subsection{Maximum sensitivity of phase delay sensing \label{Max-sent-PHDsens}}
Equation \eqref{equ:fx-ratio} and \tabref{tab:1} set the conditions to optimize the limiting sensitivity of a phase delay tracker. For a given detector technology, i.e. fixed readout noise $\sigma_R$ and noise amplification factor $\alpha$, the first step is to maximize $V_{ij}$ and $\sqrt{1-\gamma^2}$. Minimizing the interferometer vibrations will increase $V_{ij}$. An optimization of the AO, from the telescope to the control of image motion in the focal laboratory, will increase $\sqrt{1-\gamma^2}$. Then, we have to use the longest possible frame time $\mathcal{T}_F$ that is limited by the atmospheric turbulence as discussed in detail in section \ref{sec:serv-loop} but also by the vibrations of the interferometer. To maximize the collected flux $n_*$, we can increase the Strehl ratio of the AO and the transmission of the instrument. Increasing the transmission of the interferometer itself is quite difficult in an existing interferometer but will be a key design parameter in any new generation interferometer such as PFI.  We can also maximize the wavelength range used for phase delay tracking but in doing so we have to avoid increasing too much the background photon noise $n_B$ that grows very rapidly with wavelength after \SI{2.3}{\micro\meter}. Finally, we can work on the phase delay tracker itself, by reducing the triplet $N_X$, $N_\lambda$, and $N_P$, where $N_X$ is the number of pixels necessary to measure a phase delay, $N_\lambda$ is the number of spectral channels, and $N_P$ is the numbers of the unit phase delay sensors fed by each telescope.

Maximizing the AO efficiency and minimizing the vibrations of the \textit{VLTI} are two key elements of the \textit{GRAVITY+} plan. They would also be corner stones of any new interferometer such as PFI together with an optimization of its global transmission. In this paper, we focus on the optimization of the phase delay sensor that is described in the next section.

\subsection{Fringe sensing optimization}

Historically, the first alternative to pairwise FTs has been based on \textquoteleft global\textquoteright\ beam combiners like the ones used in \cite{2007A&A...464....1P} and \cite{2014Msngr.157....5L}. For such a system, equation \eqref{equ:fx-ratio} becomes: 
\begin{eqnarray}
Z_{i j G}&=&\frac{\sqrt{\operatorname{Var}\left(\mathfrak{\Im} C_{i j G}\right)}}{\left\langle\Re C_{i j G}\right\rangle}\nonumber\\&=&\frac{\sqrt{N_{T}\left({n}_{*}+{n}_{B}\right) {\mathcal{T}}_{F} {\alpha}^{2}+{N}_{X G} {N}_{\lambda} {\sigma}_{R}^{2}}}{\sqrt{{2}} {\mathcal{T}}_{{F}} {n}_{*} {V}_{i j} \sqrt{{1}-{\gamma}_{i j}^{2}}}.
\label{equ:ZIJG}
\end{eqnarray}
$N_T$ is the number of telescopes and $N_{XG}$ is the number of pixels necessary to sample a global interferogram. The minimum value for $N_{XG}$ is set by the number of baselines: 
\begin{equation}
N_{X G}=2 \mathcal{K} N_{T}\left(N_{T}-1\right),
\end{equation}
the term $\mathcal{K}$ depends on the specifications of the beam combiner. To measure properly all scientific measurable of an interferometer, we need $\mathcal{K}>2$ and for example in \textit{MATISSE} $\mathcal{K}>6$. For a system designed only to measure only phase delays in a fringe sensor, it is possible to set $\mathcal{K}=2$, at least to be able to use the standard data processing routines with fully evaluated noise propagation. \cite{2011A&A...530A.121B} claims that pairwise FT is better when the noise is dominated by detector read-out noise and  \cite{2014Msngr.157....5L} show that global beam combiners are better when the noise is dominated by background noise. Comparing equations (\ref{equ:fx-ratio} and \ref{equ:ZIJG}) at the sensitivity limit; when noise is dominated by the detector read-out yields: 
\begin{equation}
\left(\frac{Z_{i j G}}{Z_{i j L}}\right)^{2}=\frac{\mathcal{K} N_{T}}{2\left(N_{T}-1\right)}<1 \text { if } \mathcal{K}<\frac{2\left(N_{T}-1\right)}{N_{T}}.
\label{equ:read-out}
\end{equation}

This confirms an advantage to pairwise systems for $N_T=4$ and $\mathcal{K}\geq2$ when all baselines are used in the same way.

Many authors \citep[e.g.][]{2012SPIE.8445E..1LM,2016SPIE.9907E..1FP,2011A&A...530A.121B} have proposed systems where only a fraction of the $\frac{N_T(N_T-1)}{2}$ baselines are used to evaluate the $N_T-1$ phase delays, i.e. to reduce $N_P$ that is the number of pairwise unit sensors fed by a telescope. The ultimate solution seems to be the HFT that reduces $N_P$ to the minimum value of 1.

\subsection{ The HFT phase delay sensor\label{sec:HFT}}
 
 The HFT  \citep{2014SPIE.9146E..2PP,2016SPIE.9907E..1FP} splits the phase delay measurements between different levels. In the first level, we cophase pairs of apertures individually using all the available flux ($N_P=1$). Then we transmit the cophased flux to the next level as if it was produced by a single cophased aperture. If the transmission of the first level is larger than 50 per cent, the second level of the HFT that cophases pairs of pairs of telescopes, is fed by more photons than the first level and it has a higher sensitivity if the source is unresolved on all baselines, which is always the case with the \textit{VLTI} for the very faint fringe tracking targets considered in this paper. Then the cophased beams are again fed in the next level that cophases pairs of groups of telescopes and so on. As the phase delay sensing SNR increases in an HFT with the hierarchical level, the limiting performances are set by the first level that cophases two telescopes. \newline
 
 
 \indent An additional characteristic of the HFT is that the beam combination and the phase delay estimation at each level are performed by a modified ABCD. In an ABCD beam combiner, we have four outputs corresponding to 0, $\pi/2$, $\pi$, and $-\pi/2$ (or the equivalent $3\pi/4$, $-\pi/4$, $\pi/4$, and $3\pi/4$) phase delays between the two input beams. The maximum flux in one of these outputs is 47.5 per cent of the total input (without transmission losses). In the HFT, we use a beam combiner with three outputs at A:$-3\pi/4$, C:0, and B:$3\pi/4$. The central output is obtained by merging the $-\pi/4$ and $\pi/4$ outputs of an ABCD after introducing the appropriate fixed phase shift that cophases them whatever the piston between the two input beams, as illustrated in \figref{fig:ABCD-ABC}. 

\begin{figure}
	\centering
	\input{fig/full-shape-optics}
	\caption{Conceptual design of a four telescopes HFT, with two levels. Note that in this design
		the central outputs $_1^1C$, and $_2^1C$, of the first level are not measured directly but deduced from
		the seven outputs of the system.}
	\label{fig:shape-optics-full}
\end{figure}
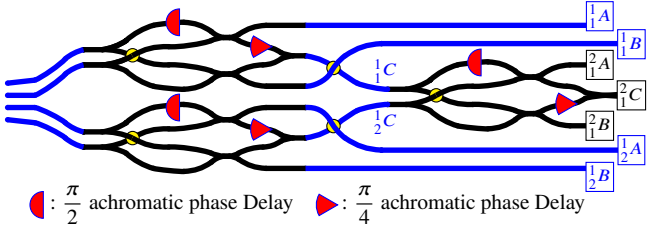

The estimator $\phi = \arg{(A-B,C-A-B)} $ yields the phase delay between the input beams independently of the source visibility and flux ratio between input beams \citep{2016SPIE.9907E..1FP}. When the two input beams are cophased and have the same intensity, the central output of this ACB beam combiner contains 80.8 per cent of the total input (without transmission losses). This beam can then be efficiently fed in an identical ACB beam combiner of the $2^\text{nd}$ level of the HFT as illustrated by \figref{fig:shape-optics-full}.

\figref{fig:ABCD-ABC} and \ref{fig:shape-optics-full} that illustrate the conceptual design of a four telescopes HFT also show the design of integrated optics components of an HFT that have been manufactured and successfully tested. There are also bulk optic solutions to implement these concepts. They can yield a much better transmission but have a more complex optomechanical design and  a loss in precision resulting from the lack of the perfect spatial filtering introduced by single mode fibers or integrated optics. However, this should have a limited impact on the phase delay accuracy, at least for the moderate tracking quality discussed here. We leave the discussion of this point for future work and in this paper we use only transmissions compatible with single mode filtering. The full description of the HFT and of its proof of concept from analytical computations, numerical simulations, and laboratory tests will be described in another paper (Boskri et al., in preparation). Here, we just have to note that the phase delay SNR of an HFT is set by equation \eqref{equ:fx-ratio} with $N_P=1$ and $N_X=4$.

\begin{figure}
	\centering
	\input{fig/ABCD-ABC}
	\caption{Conceptual design of the ACB beam combiner that is the basic unit of a
		HFT. The upper image shows the classical ABCD beam combiner used
		in PIONIER or \textit{GRAVITY} with its four outputs at $-\pi/2$, $\pi/2$, 0, and $-\pi$. The bottom image shows
		our modified beam combiner where the \textquoteleft $\pi/2$\textquoteright\ and \textquoteleft0\textquoteright\ outputs of the ABCD have been
		cophased by an appropriate phase shift and then merged.}
	\label{fig:ABCD-ABC}
\end{figure}
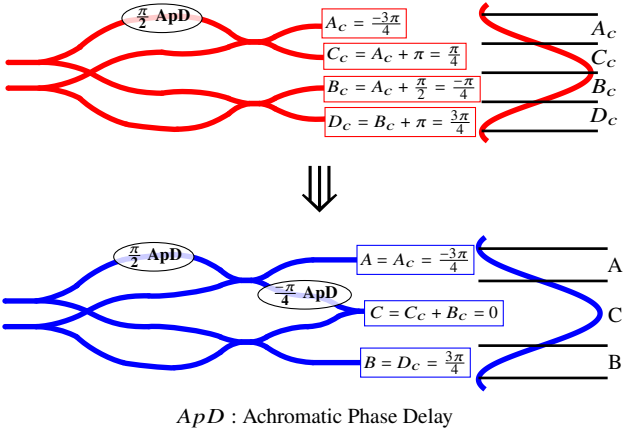

For a maximum efficiency of the HFT phase delay measurements, we also decided to use very broad-band measurements with $N_\lambda$. Equation \eqref{equ:fx-ratio} then becomes for the first level of the HFT:
\begin{equation}
Z_{i j H 1}=\frac{\sqrt{\left(n_{*}+n_{B}\right) \mathcal{T}_{F} \alpha^{2}+1.5 \sigma_{R}^{2}}}{\mathcal{T}_{F} n_{*} V_{i j} \sqrt{1-\gamma_{i j}^{2}}}.
\label{equ:ZIJH1}
\end{equation}

As our goal in this paper is to evaluate the sky coverage for the best fringe sensor that can be reasonably foreseen, we will also consider that the HFT uses 80 per cent of the full $ H $ and $ K $ spectral bands, which is acceptable for off-axis tracking where we do not need to feed a fraction of this bands into science instruments that work on-axis. We eliminate the wavelengths larger than \SI{2.3}{\micro\meter} because their contribution to SNR loss due to background noise exceeds the gain provided by their contribution to the source flux. We also reserve a fraction of the H band for the group delay sensor discussed in the next section.

Finally, we estimate that the overall efficiency of a new generation FT will be 2 per cent instead of the current 1 per cent value for the GFT. Beyond the gain provided by the fact that we do not include dispersive elements in our design, this 2 per cent value is quite conservative, as for example the overall transmission of the \textit{AMBER} \textit{VLTI} instrument, with its complex combination of internal dichroic plates to split and then recombine the $J$, $H$, and $K$ bands; optical fibers for spatial filtering and the high-resolution spectrograph, was of the order of 4 per cent in the $ K $ band.

The final numbers used in this paper to estimate the sensitivity and the sky-coverage with an HFT are given in \tabref{tab:1}. Note that we take into account the variation of Strehl ratio and instrument visibility as a function of wavelength as well as the average magnitude difference $\mathrm{H-K}=0.1$ and $\mathrm{J-K}=0.5$ for our (very large) selection of calibrators. The corresponding HFT phase delay SNR, computed from equation \eqref{equ:ZIJH1} combined with equation \eqref{equ:sigmaphi}, is displayed in \figref{fig:GFTphasedelay}.

\subsection{Group delay sensing in association with the HFT\label{sec:groupdelay}}

With $N_\lambda=1$, the HFT sketched above cannot control the group delay. As it is a very broadband system, the phase delay SNR and the global transmission in the central \textquoteleft C\textquoteright\ outputs will have a strong maximum at OPD=0 allowing to detect immediately that the central fringe has been lost, but we would not know in what direction to search to reacquire it and scan the baselines for that purpose is considered as too costly in observing time.

\begin{figure}
	\centering
	\includegraphics[width=\linewidth]{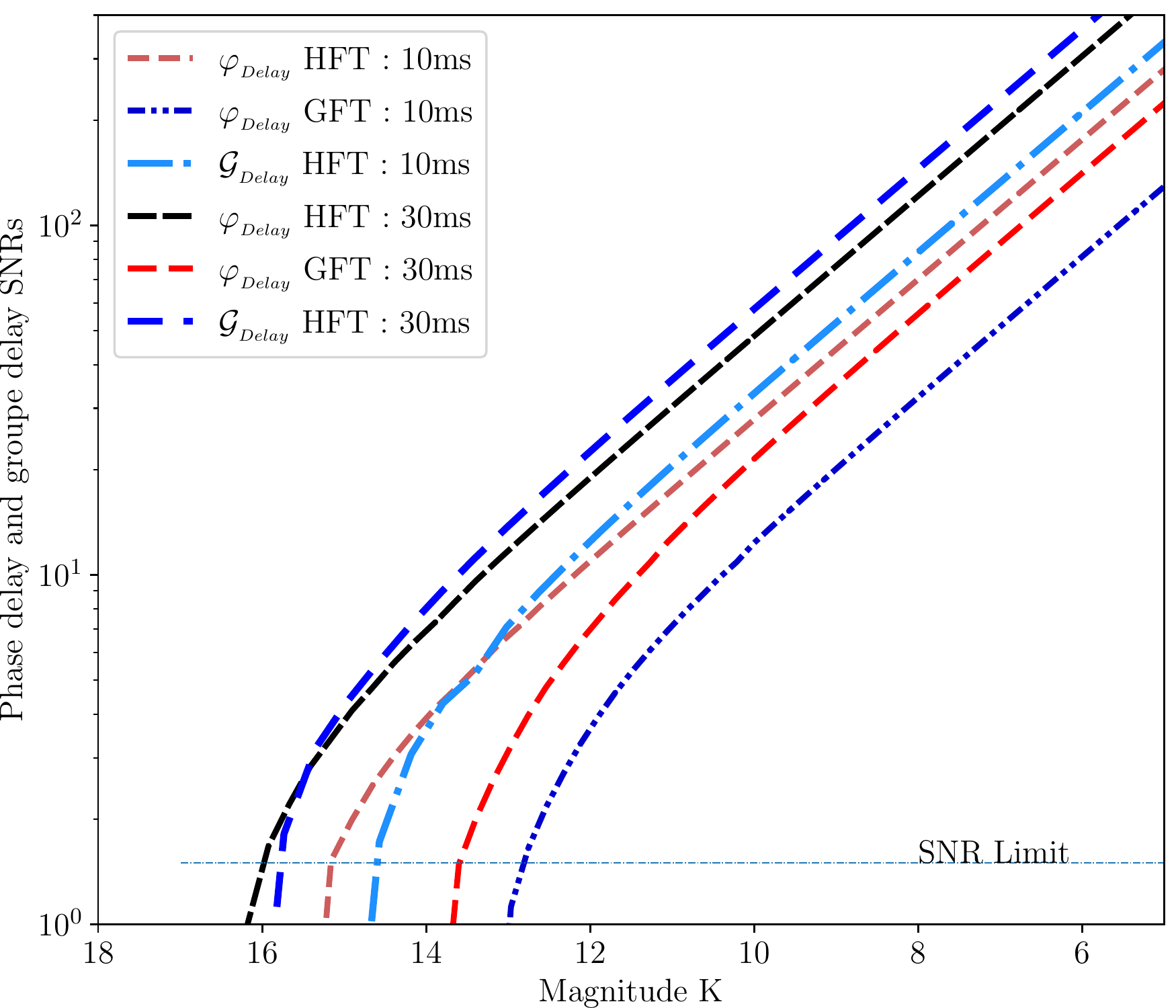}
	\caption{Phase delay and group delay SNR as a function of the $ K $ band magnitude for the HFT and GFT used with \textit{UTs} with the future \textit{GRAVITY+} AO.}
	\label{fig:GFTphasedelay}
\end{figure}

There are several options to associate a group delay sensor to the HFT. The original plan was to install a group delay sensor in the final output of the HFT (labelled $_1^2C$ in \figref{fig:shape-optics-full}). This works in principle but implies some design and data processing complexity. Another option is to use the J band and a small fraction of the H band to implement a classical group delay sensor based on the same dispersed fringe scheme as \textit{MATISSE}. This yields the parameters in \tabref{tab:1} for the group delay estimation that is plotted in \figref{fig:GFTphasedelay} for a \SI{1}{\hertz} Group Delay frequency. Another possibility is to disperse the A and B outputs of the HFT over 3 spectral channels covering the $ J $, $ H $, and $ K $ bands with a minor degradation of the phase delay performance. The phase delay SNR of such a $N_\lambda=3$ JHK HFT is also plotted in \figref{fig:GFTphasedelay}.

%% file: fig/ABCD-ABC.tex
\begin{tikzpicture}[scale=0.27,dot/.style ={circle,color=red, fill, minimum size=2pt,inner sep=0pt, outer sep=0pt}]
\node [dot] (0) at (-8.25, 1.25) {};
\node [dot] (1) at (-10.25, -0.75) {};
\node [dot] (2) at (-11.75, -0.75) {};
\node [dot] (3) at (-10.25, 0.5) {};
\node [dot] (5) at (-7.75, 0) {};
\node [dot] (6) at (-6, -2.5) {};
\node [dot] (8) at (-3.5, -2.75) {};
\node [dot] (11) at (-4.25, -1) {};
\node [dot] (13) at (0, -1.5) {};
\node [dot] (14) at (0.5, -1.5) {};
\node [dot] (15) at (3.75, -0.75) {};
\node [dot] (16) at (0.5, -1.5) {};
\node [dot] (17) at (3.75, -2.25) {};
\node [dot] (22) at (-11.75, 0.5) {};
\node [dot] (23) at (-8.25, -1.25) {};
\node [dot] (24) at (-1, -2) {};
\node [dot] (25) at (-1, -1.25) {};
\node [dot] (28) at (1.25, -1.25) {};
\node [dot] (29) at (1.25, -1.75) {};
\node [dot] (30) at (-7.75, 0) {};
\node [dot] (39) at (-8.25, 1.25) {};
\node [dot] (40) at (-7.75, 0) {};
\node [dot] (41) at (-6, 2.5) {};
\node [dot] (42) at (-3.5, 2.75) {};
\node [dot] (43) at (-4.25, 1) {};
\node [dot] (44) at (0, 1.5) {};
\node [dot] (45) at (0.5, 1.5) {};
\node [dot] (46) at (3.75, 0.75) {};
\node [dot] (47) at (0.5, 1.5) {};
\node [dot] (48) at (3.75, 2.25) {};
\node [dot] (49) at (-8.25, 1.25) {};
\node [dot] (50) at (-1, 2) {};
\node [dot] (51) at (-1, 1.25) {};
\node [dot] (52) at (1.25, 1.25) {};
\node [dot] (53) at (1.25, 1.75) {};
\node [dot] (54) at (-7.75, 0) {};
\draw [color=red,line width=2pt,bend right=15, looseness=0.75] (3.center) to (0.center);
\draw [color=red,line width=2pt,bend right=15, looseness=0.75] (1.center) to (5.center);
\draw [color=red,line width=2pt](2.center) to (1.center);
\draw [color=red,line width=2pt,bend right=15] (5.center) to (11.center);
\draw [color=red,line width=2pt,bend right=15, looseness=0.25] (6.center) to (8.center);
\draw [color=red,line width=2pt](13.center) to (14.center);
\draw [color=red,line width=2pt](16.center) to (13.center);
\draw [color=red,line width=2pt,bend left=15, looseness=0.75] (3.center) to (5.center);
\draw [color=red,line width=2pt](3.center) to (22.center);
\draw [color=red,line width=2pt,bend left=15, looseness=0.75] (1.center) to (23.center);
\draw [color=red,line width=2pt,bend right=15, looseness=0.75] (23.center) to (6.center);
\draw [color=red,line width=2pt,bend left=15, looseness=0.75] (11.center) to (25.center);
\draw [color=red,line width=2pt,bend right=15] (8.center) to (24.center);
\draw [color=red,line width=2pt,bend right=15, looseness=0.50] (25.center) to (13.center);
\draw [color=red,line width=2pt,bend left=15] (24.center) to (13.center);
\draw [color=red,line width=2pt,bend left=15, looseness=0.75] (16.center) to (29.center);
\draw [color=red,line width=2pt,bend right=15] (29.center) to (17.center);
\draw [color=red,line width=2pt,bend right=15, looseness=0.75] (16.center) to (28.center);
\draw [color=red,line width=2pt,bend left=15, looseness=0.75] (28.center) to (15.center);
\draw [color=red,line width=2pt,bend left=15] (40.center) to (43.center);
\draw [color=red,line width=2pt,bend left=15, looseness=0.25] (41.center) to (42.center);
\draw [color=red,line width=2pt](44.center) to (45.center);
\draw [color=red,line width=2pt](47.center) to (44.center);
\draw [color=red,line width=2pt,bend left=15, looseness=0.75] (49.center) to (41.center);
\draw [color=red,line width=2pt,bend right=15, looseness=0.75] (43.center) to (51.center);
\draw [color=red,line width=2pt,bend left=15] (42.center) to (50.center);
\draw [color=red,line width=2pt,bend left=15, looseness=0.50] (51.center) to (44.center);
\draw [color=red,line width=2pt,bend right=15] (50.center) to (44.center);
\draw [color=red,line width=2pt,bend right=15, looseness=0.75] (47.center) to (53.center);
\draw [color=red,line width=2pt,bend left=15] (53.center) to (48.center);
\draw [color=red,line width=2pt,bend left=15, looseness=0.75] (47.center) to (52.center);
\draw [color=red,line width=2pt,bend right=15, looseness=0.75] (52.center) to (46.center);
\node [draw=black,fill=white!50 ,ellipse,fill opacity=0.8,text opacity=1,minimum size=2pt,inner sep=0pt, outer sep=0pt] at (-4.25,2.65) {\scriptsize $\mathbf{\frac{\pi}{2}~{ApD}}$};
\node [draw=red,fill=white,minimum size=2pt,inner sep=1pt, outer sep=1pt] at (5.6,2.4){\scriptsize$A_c=\frac{-3\pi}{4}$};
\node [draw=red,fill=white,minimum size=2pt,inner sep=1.1pt, outer sep=1pt] at (7.1,0.8){\scriptsize$C_c=A_c+\pi=\frac{\pi}{4}$};
\node [draw=red,fill=white,minimum size=2pt,inner sep=1pt, outer sep=1pt]at (7.45, -0.8){\scriptsize$B_c=A_c+\frac{\pi}{2}=\frac{-\pi}{4}$};
\node [draw=red,fill=white,minimum size=2pt,inner sep=1pt, outer sep=1pt] at (7.36, -2.4){\scriptsize$D_c=B_c+\pi=\frac{3\pi}{4}$};
\node at (14.8, -0.){\begin{tikzpicture}[rotate=-90,scale=0.9]
	\draw[scale=0.1,smooth,samples=100,domain=4:24.,color=red,line width =2] plot(\x,{(1+sin(20.7*\x+160))*8+100}); 
	\draw[line width = 1] (0.55,10) -- (0.55,11.7);
	\draw[line width = 1] (0.975,10) -- (0.975,11.7);
	\draw[line width = 1] (1.4,10) -- (1.4,11.7);
	\draw[line width = 1] (1.825,10) -- (1.825,11.7);
	\draw[line width = 1] (2.25,10) -- (2.25,11.7);
	\node at ({(0.55+0.975)/2},11.8) {\small $A_c$};
	\node at ({(1.4+0.975)/2},11.8) {\small $C_c$};
	\node at ({(1.4+1.825)/2},11.8) {\small $B_c$};
	\node at ({(2.25+1.825)/2},11.8) {\small $D_c$};
	\end{tikzpicture}};
\end{tikzpicture}

\begin{tikzpicture}
\node[rotate=90] at (0, 0){\Huge $\Lleftarrow$ };
\end{tikzpicture}

\begin{tikzpicture}[scale=0.27,dot/.style ={circle, fill, color=blue, minimum size=2pt,inner sep=0pt, outer sep=0pt}]
\node [dot] (0) at (-8, 1.25) {};
\node [dot] (1) at (-10, -0.75) {};
\node [dot] (2) at (-11.5, -0.75) {};
\node [dot] (3) at (-10, 0.5) {};
\node [dot] (5) at (-7.5, 0) {};
\node [dot] (6) at (-5.75, -2.5) {};
\node [dot] (8) at (-3.25, -2.75) {};
\node [dot] (11) at (-4.5, -0.75) {};
\node [dot] (13) at (0, -1.5) {};
\node [dot] (14) at (0.5, -1.5) {};
\node [dot] (15) at (3.25, -0.75) {};
\node [dot] (16) at (0.5, -1.5) {};
\node [dot] (17) at (3.5, -2.5) {};
\node [dot] (22) at (-11.5, 0.5) {};
\node [dot] (23) at (-8, -1.25) {};
\node [dot] (24) at (-1, -2) {};
\node [dot] (25) at (-1, -1.25) {};
\node [dot] (28) at (1.25, -1.25) {};
\node [dot] (29) at (1.25, -1.75) {};
\node [dot] (30) at (-7.5, 0) {};
\node [dot] (39) at (-8, 1.25) {};
\node [dot] (40) at (-7.5, 0) {};
\node [dot] (41) at (-5.75, 2.5) {};
\node [dot] (42) at (-3.25, 2.75) {};
\node [dot] (43) at (-4.5, 0.75) {};
\node [dot] (44) at (0, 1.5) {};
\node [dot] (45) at (0.5, 1.5) {};
\node [dot] (46) at (3.25, 0.75) {};
\node [dot] (47) at (0.5, 1.5) {};
\node [dot] (48) at (3.5, 2.5) {};
\node [dot] (49) at (-8, 1.25) {};
\node [dot] (50) at (-1, 2) {};
\node [dot] (51) at (-1, 1.25) {};
\node [dot] (52) at (1.25, 1.25) {};
\node [dot] (53) at (1.25, 1.75) {};
\node [dot] (54) at (-7.5, 0) {};
\node [dot] (55) at (5.75, 0) {};
\node [dot] (56) at (6.75, 0) {};
\node [dot] (57) at (4.75, 0.25) {};
\node [dot] (60) at (6.75, 2.5) {};
\node [dot] (61) at (6.75, -2.5) {};
\node [dot] (62) at (4.75, -0.25) {};
\draw [color=blue,line width=2pt,bend right=15, looseness=0.75] (3.center) to (0.center);
\draw [color=blue,line width=2pt,bend right=15, looseness=0.75] (1.center) to (5.center);
\draw [color=blue,line width=2pt](2.center) to (1.center);
\draw [color=blue,line width=2pt,bend right=15] (5.center) to (11.center);
\draw [color=blue,line width=2pt,bend right=15, looseness=0.25] (6.center) to (8.center);
\draw [color=blue,line width=2pt](13.center) to (14.center);
\draw [color=blue,line width=2pt](16.center) to (13.center);
\draw [color=blue,line width=2pt,bend left=15, looseness=0.75] (3.center) to (5.center);
\draw [color=blue,line width=2pt](3.center) to (22.center);
\draw [color=blue,line width=2pt,bend left=15, looseness=0.75] (1.center) to (23.center);
\draw [color=blue,line width=2pt,bend right=15, looseness=0.75] (23.center) to (6.center);
\draw [color=blue,line width=2pt,bend left=15, looseness=0.50] (11.center) to (25.center);
\draw [color=blue,line width=2pt,bend right=15] (8.center) to (24.center);
\draw [color=blue,line width=2pt,bend right=15, looseness=0.50] (25.center) to (13.center);
\draw [color=blue,line width=2pt,bend left=15] (24.center) to (13.center);
\draw [color=blue,line width=2pt,bend left=15, looseness=0.75] (16.center) to (29.center);
\draw [color=blue,line width=2pt,bend right=15] (29.center) to (17.center);
\draw [color=blue,line width=2pt,bend right=15, looseness=0.75] (16.center) to (28.center);
\draw [color=blue,line width=2pt,bend left=15, looseness=0.75] (28.center) to (15.center);
\draw [color=blue,line width=2pt,bend left=15] (40.center) to (43.center);
\draw [color=blue,line width=2pt,bend left=15, looseness=0.25] (41.center) to (42.center);
\draw [color=blue,line width=2pt](44.center) to (45.center);
\draw [color=blue,line width=2pt](47.center) to (44.center);
\draw [color=blue,line width=2pt,bend left=15, looseness=0.75] (49.center) to (41.center);
\draw [color=blue,line width=2pt,bend right=15, looseness=0.50] (43.center) to (51.center);
\draw [color=blue,line width=2pt,bend left=15] (42.center) to (50.center);
\draw [color=blue,line width=2pt,bend left=15, looseness=0.50] (51.center) to (44.center);
\draw [color=blue,line width=2pt,bend right=15] (50.center) to (44.center);
\draw [color=blue,line width=2pt,bend right=15, looseness=0.75] (47.center) to (53.center);
\draw [color=blue,line width=2pt,bend left=15] (53.center) to (48.center);
\draw [color=blue,line width=2pt,bend left=15, looseness=0.75] (47.center) to (52.center);
\draw [color=blue,line width=2pt,bend right=15, looseness=0.75] (52.center) to (46.center);
\draw [color=blue,line width=2pt,bend left=15] (46.center) to (57.center);
\draw [color=blue,line width=2pt,bend right=15, looseness=0.50] (57.center) to (55.center);
\draw [color=blue,line width=2pt](55.center) to (56.center);
\draw [color=blue,line width=2pt](48.center) to (60.center);
\draw [color=blue,line width=2pt](17.center) to (61.center);
\draw [color=blue,line width=2pt,bend right=15] (15.center) to (62.center);
\draw [color=blue,line width=2pt,bend left=15, looseness=0.50] (62.center) to (55.center);
\node [draw=black,fill=white!50 ,ellipse,fill opacity=0.8,text opacity=1,minimum size=2pt,inner sep=0pt, outer sep=0pt] at (-4.25,2.65) {\scriptsize $\mathbf{\frac{\pi}{2}~{ApD}}$};
\node [draw=black,fill=white!50 ,ellipse,fill opacity=0.8,text opacity=1,minimum size=2pt,inner sep=0pt, outer sep=0pt] at (3.1,0.8) {\scriptsize $\mathbf{\frac{-\pi}{4}~{ApD}}$};
\node [draw=blue,fill=white,fill=white,minimum size=7pt,inner sep=1pt, outer sep=0pt] at (8.5,2.5){\scriptsize$A=A_c=\frac{-3\pi}{4}$};
\node [draw=blue,fill=white,fill=white,minimum size=7pt,inner sep=2pt, outer sep=0pt]at (9.4, -0.1){\scriptsize$C=C_c+B_c=0$};
\node [draw=blue,fill=white,minimum size=7pt,inner sep=1pt, outer sep=0pt] at (8.5, -2.5){\scriptsize$B=D_c=\frac{3\pi}{4}$};
\node at (15.2, -0.1){\begin{tikzpicture}[rotate=-90,scale=1]
	\draw[scale=0.1,smooth,samples=100,domain=4:24.,color=blue,line width =2] plot(\x,{(1+sin(20.7*\x+160))*8+100}); 
	\draw[line width = 1] (0.55,10) -- (0.55,11.7);
	\draw[line width = 1] (0.975,10) -- (0.975,11.7);
	\draw[line width = 1] (1.825,10) -- (1.825,11.7);
	\draw[line width = 1] (2.25,10) -- (2.25,11.7);
	\node at ({(0.55+0.975)/2},11.8) {\small A};
	\node at ({(0.975+1.855)/2},11.8) {\small C};
	\node at ({(2.25+1.825)/2},11.8) {\small B};
	\end{tikzpicture}};
\end{tikzpicture}

\begin{tikzpicture}
\node at (-5, 0){\footnotesize $ApD$ : Achromatic Phase Delay};
\end{tikzpicture}

%% file: fig/full-shape-optics.tex
\begin{tikzpicture}[scale=0.16,dot/.style ={circle, fill, minimum size=2pt,inner sep=0pt, outer sep=0pt}]
\node [dot] (0) at (-7.75, 0.5) {};
\node [dot] (1) at (-9.75, -1.5) {};
\node [dot] (2) at (-11.25, -1.5) {};
\node [dot] (3) at (-9.75, -0.25) {};
\node [dot] (5) at (-7.25, -0.75) {};
\node [dot] (6) at (-5.5, -3.25) {};
\node [dot] (8) at (-3, -3.5) {};
\node [dot] (11) at (-4.25, -1.5) {};
\node [dot] (13) at (0.25, -2.25) {};
\node [dot] (14) at (0.75, -2.25) {};
\node [dot] (15) at (3.5, -1.5) {};
\node [dot] (16) at (0.75, -2.25) {};
\node [dot] (17) at (3.75, -3.25) {};
\node [dot] (22) at (-11.25, -0.25) {};
\node [dot] (23) at (-7.75, -2) {};
\node [dot] (24) at (-0.75, -2.75) {};
\node [dot] (25) at (-0.75, -2) {};
\node [dot] (28) at (1.5, -2) {};
\node [dot] (29) at (1.5, -2.5) {};
\node [dot] (30) at (-7.25, -0.75) {};
\node [dot] (39) at (-7.75, 0.5) {};
\node [dot] (40) at (-7.25, -0.75) {};
\node [dot] (41) at (-5.5, 1.75) {};
\node [dot] (42) at (-3, 2) {};
\node [dot] (43) at (-4.25, 0) {};
\node [dot] (44) at (0.25, 0.75) {};
\node [dot] (45) at (0.75, 0.75) {};
\node [dot] (46) at (3.5, 0) {};
\node [dot] (47) at (0.75, 0.75) {};
\node [dot] (48) at (3.75, 1.75) {};
\node [dot] (49) at (-7.75, 0.5) {};
\node [dot] (50) at (-0.75, 1.25) {};
\node [dot] (51) at (-0.75, 0.5) {};
\node [dot] (52) at (1.5, 0.5) {};
\node [dot] (53) at (1.5, 1) {};
\node [dot] (54) at (-7.25, -0.75) {};
\node [dot] (55) at (6, -0.75) {};
\node [dot] (56) at (7, -0.75) {};
\node [dot] (57) at (5, -0.5) {};
\node [dot] (60) at (7, 1.75) {};
\node [dot] (62) at (5, -1) {};
\node [dot] (63) at (-7.75, -6.25) {};
\node [dot] (64) at (-9.75, -8.25) {};
\node [dot] (65) at (-11.25, -8.25) {};
\node [dot] (66) at (-9.75, -7) {};
\node [dot] (67) at (-7.25, -7.5) {};
\node [dot] (68) at (-5.5, -10) {};
\node [dot] (69) at (-3, -10.25) {};
\node [dot] (70) at (-4.25, -8.25) {};
\node [dot] (71) at (0.25, -9) {};
\node [dot] (72) at (0.75, -9) {};
\node [dot] (73) at (3.5, -8.25) {};
\node [dot] (74) at (0.75, -9) {};
\node [dot] (75) at (3.75, -10) {};
\node [dot] (76) at (-11.25, -7) {};
\node [dot] (77) at (-7.75, -8.75) {};
\node [dot] (78) at (-0.75, -9.5) {};
\node [dot] (79) at (-0.75, -8.75) {};
\node [dot] (80) at (1.5, -8.75) {};
\node [dot] (81) at (1.5, -9.25) {};
\node [dot] (82) at (-7.25, -7.5) {};
\node [dot] (83) at (-7.75, -6.25) {};
\node [dot] (84) at (-7.25, -7.5) {};
\node [dot] (85) at (-5.5, -5) {};
\node [dot] (86) at (-3, -4.75) {};
\node [dot] (87) at (-4.25, -6.75) {};
\node [dot] (88) at (0.25, -6) {};
\node [dot] (89) at (0.75, -6) {};
\node [dot] (90) at (3.5, -6.75) {};
\node [dot] (91) at (0.75, -6) {};
\node [dot] (92) at (3.75, -5) {};
\node [dot] (93) at (-7.75, -6.25) {};
\node [dot] (94) at (-0.75, -5.5) {};
\node [dot] (95) at (-0.75, -6.25) {};
\node [dot] (96) at (1.5, -6.25) {};
\node [dot] (97) at (1.5, -5.75) {};
\node [dot] (98) at (-7.25, -7.5) {};
\node [dot] (99) at (6, -7.5) {};
\node [dot] (100) at (7, -7.5) {};
\node [dot] (101) at (5, -7.25) {};
\node [dot] (102) at (7, -5) {};
\node [dot] (103) at (7, -10) {};
\node [dot] (104) at (5, -7.75) {};
\node [dot] (105) at (17.25, -2.75) {};
\node [dot] (106) at (15.25, -4.75) {};
\node [dot] (108) at (15.25, -3.5) {};
\node [dot] (109) at (17.75, -4) {};
\node [dot] (110) at (19.5, -6.5) {};
\node [dot] (111) at (22, -6.75) {};
\node [dot] (112) at (20.75, -4.75) {};
\node [dot] (113) at (25.25, -5.5) {};
\node [dot] (114) at (25.75, -5.5) {};
\node [dot] (115) at (28.5, -4.75) {};
\node [dot] (116) at (25.75, -5.5) {};
\node [dot] (117) at (28.75, -6.5) {};
\node [dot] (118) at (13.75, -3.5) {};
\node [dot] (119) at (17.25, -5.25) {};
\node [dot] (120) at (24.25, -6) {};
\node [dot] (121) at (24.25, -5.25) {};
\node [dot] (122) at (26.5, -5.25) {};
\node [dot] (123) at (26.5, -5.75) {};
\node [dot] (124) at (17.75, -4) {};
\node [dot] (125) at (17.25, -2.75) {};
\node [dot] (126) at (17.75, -4) {};
\node [dot] (127) at (19.5, -1.5) {};
\node [dot] (128) at (22, -1.25) {};
\node [dot] (129) at (20.75, -3.25) {};
\node [dot] (130) at (25.25, -2.5) {};
\node [dot] (131) at (25.75, -2.5) {};
\node [dot] (132) at (28.5, -3.25) {};
\node [dot] (133) at (25.75, -2.5) {};
\node [dot] (134) at (28.75, -1.5) {};
\node [dot] (135) at (17.25, -2.75) {};
\node [dot] (136) at (24.25, -2) {};
\node [dot] (137) at (24.25, -2.75) {};
\node [dot] (138) at (26.5, -2.75) {};
\node [dot] (139) at (26.5, -2.25) {};
\node [dot] (140) at (17.75, -4) {};
\node [dot] (141) at (34, -4) {};
\node [dot] (142) at (34, -4) {};
\node [dot] (143) at (30, -3.75) {};
\node [dot] (144) at (32, -1.5) {};
\node [dot] (145) at (32, -6.5) {};
\node [dot] (146) at (30, -4.25) {};
\node [dot] (147) at (13.25, 0.25) {};
\node [dot] (167) at (7, -3.25) {};
\node [dot] (150) at (13.25, -8.5) {};
\node [dot] (153) at (7, -3.25) {};
\node [dot] (154) at (13.75, -4.75) {};
\node [dot] (159) at (7, -7.5) {};
\node [dot] (171) at (7, -5) {};
\node [dot] (183) at (-11.25, -1.5) {};
\node [dot] (195) at (-11.25, -8.25) {};
\node [dot] (196) at (-11.25, -0.25) {};
\node [dot] (200) at (-11.25, -0.25) {};
\node [dot] (205) at (-11.25, -8.25) {};
\node [dot] (206) at (-11.25, -7) {};
\node [dot] (210) at (-11.25, -7) {};
\node [dot] (211) at (7, -0.75) {};
\draw [line width=2pt](17.center) to (153.center);
\draw [line width=2pt,bend right=15, looseness=0.75] (3.center) to (0.center);
\draw [line width=2pt,bend right=15, looseness=0.75] (1.center) to (5.center);
\draw [line width=2pt](2.center) to (1.center);

\draw [line width=2pt,bend right=15] (5.center) to (11.center);

\draw [line width=2pt,bend right=15, looseness=0.25] (6.center) to (8.center);
\draw [line width=2pt](13.center) to (14.center);
\draw [line width=2pt](16.center) to (13.center);
\draw [line width=2pt,bend left=15, looseness=0.75] (3.center) to (5.center);
\draw [line width=2pt](3.center) to (22.center);
\node [circle,fill=yellow,draw=black,rotate=90,minimum size=5pt,inner sep=0pt, outer sep=0pt]  at (-7.223, -0.709752) {};
\draw [line width=2pt](17.center) to (153.center);
\draw [line width=2pt,bend right=15, looseness=0.75] (3.center) to (0.center);
\draw [line width=2pt,bend right=15, looseness=0.75] (1.center) to (5.center);
\draw [line width=2pt](2.center) to (1.center);
\draw [line width=2pt,bend left=15, looseness=0.75] (1.center) to (23.center);
\draw [line width=2pt,bend right=15, looseness=0.75] (23.center) to (6.center);
\draw [line width=2pt,bend left=15, looseness=0.50] (11.center) to (25.center);
\draw [line width=2pt,bend right=15] (8.center) to (24.center);
\draw [line width=2pt,bend right=15, looseness=0.50] (25.center) to (13.center);
\draw [line width=2pt,bend left=15] (24.center) to (13.center);
\draw [line width=2pt,bend left=15, looseness=0.75] (16.center) to (29.center);
\draw [line width=2pt,bend right=15] (29.center) to (17.center);
\draw [line width=2pt,bend right=15, looseness=0.75] (16.center) to (28.center);
\draw [line width=2pt,bend left=15, looseness=0.75] (28.center) to (15.center);
\draw [line width=2pt,bend left=15] (40.center) to (43.center);
\draw [line width=2pt,bend left=15, looseness=0.25] (41.center) to (42.center);
\draw [line width=2pt](44.center) to (45.center);
\draw [line width=2pt](47.center) to (44.center);
\draw [line width=2pt,bend left=15, looseness=0.75] (49.center) to (41.center);
\draw [line width=2pt,bend right=15, looseness=0.50] (43.center) to (51.center);
\draw [line width=2pt,bend left=15] (42.center) to (50.center);
\draw [line width=2pt,bend left=15, looseness=0.50] (51.center) to (44.center);
\draw [line width=2pt,bend right=15] (50.center) to (44.center);
\draw [line width=2pt,bend right=15, looseness=0.75] (47.center) to (53.center);
\draw [line width=2pt,bend left=15] (53.center) to (48.center);
\draw [line width=2pt,bend left=15, looseness=0.75] (47.center) to (52.center);
\draw [line width=2pt,bend right=15, looseness=0.75] (52.center) to (46.center);
\draw [line width=2pt,bend left=15] (46.center) to (57.center);
\draw [line width=2pt,bend right=15, looseness=0.50] (57.center) to (55.center);
\draw [line width=2pt](55.center) to (56.center);
\draw [line width=2pt](48.center) to (60.center);

\draw [line width=2pt,bend right=15] (15.center) to (62.center);
\draw [line width=2pt,bend left=15, looseness=0.50] (62.center) to (55.center);
\draw [line width=2pt,bend right=15, looseness=0.75] (66.center) to (63.center);
\draw [line width=2pt,bend right=15, looseness=0.75] (64.center) to (67.center);
\draw [line width=2pt](65.center) to (64.center);

\draw [line width=2pt,bend right=15, looseness=0.25] (68.center) to (69.center);

\draw [line width=2pt](71.center) to (72.center);
\draw [line width=2pt](74.center) to (71.center);

\draw [line width=2pt](66.center) to (76.center);
\draw [line width=2pt,bend left=15, looseness=0.75] (64.center) to (77.center);
\draw [line width=2pt,bend right=15, looseness=0.75] (77.center) to (68.center);
\draw [line width=2pt,bend left=15, looseness=0.50] (70.center) to (79.center);
\draw [line width=2pt,bend right=15] (69.center) to (78.center);
\draw [line width=2pt,bend right=15, looseness=0.50] (79.center) to (71.center);
\draw [line width=2pt,bend left=15] (78.center) to (71.center);

\draw [line width=2pt,bend left=15, looseness=0.75] (74.center) to (81.center);
\draw [line width=2pt,bend right=15] (81.center) to (75.center);
\draw [line width=2pt,bend right=15, looseness=0.75] (74.center) to (80.center);
\draw [line width=2pt,bend left=15, looseness=0.75] (80.center) to (73.center);
\draw [line width=2pt,bend left=15] (84.center) to (87.center);
\draw [line width=2pt,bend left=15, looseness=0.25] (85.center) to (86.center);
\draw [line width=2pt](88.center) to (89.center);
\draw [line width=2pt](91.center) to (88.center);
\node [circle,fill=yellow,draw=black,rotate=90,minimum size=5pt,inner sep=0pt, outer sep=0pt]  at (-7.223, -7.509752) {};
\draw [line width=2pt,bend right=15] (67.center) to (70.center);
\draw [line width=2pt,bend left=15, looseness=0.75] (66.center) to (67.center);
\draw [line width=2pt,bend left=15, looseness=0.75] (93.center) to (85.center);
\draw [line width=2pt,bend right=15, looseness=0.50] (87.center) to (95.center);
\draw [line width=2pt,bend left=15] (86.center) to (94.center);
\draw [line width=2pt,bend left=15, looseness=0.50] (95.center) to (88.center);
\draw [line width=2pt,bend right=15] (94.center) to (88.center);
\draw [line width=2pt,bend right=15, looseness=0.75] (91.center) to (97.center);
\draw [line width=2pt,bend left=15] (97.center) to (92.center);
\draw [line width=2pt,bend left=15, looseness=0.75] (91.center) to (96.center);
\draw [line width=2pt,bend right=15, looseness=0.75] (96.center) to (90.center);
\draw [line width=2pt,bend left=15] (90.center) to (101.center);
\draw [line width=2pt,bend right=15, looseness=0.50] (101.center) to (99.center);
\draw [line width=2pt](99.center) to (100.center);
\draw [line width=2pt](92.center) to (102.center);
\draw [line width=2pt](75.center) to (103.center);
\draw [line width=2pt,bend right=15] (73.center) to (104.center);
\draw [line width=2pt,bend left=15, looseness=0.50] (104.center) to (99.center);
\draw [line width=2pt,bend right=15] (109.center) to (112.center);
\draw [line width=2pt,bend right=15, looseness=0.25] (110.center) to (111.center);

\draw [line width=2pt,bend left=15, looseness=0.75] (108.center) to (109.center);
\node [circle,fill=yellow,draw=black,rotate=90,minimum size=5pt,inner sep=0pt, outer sep=0pt]  at (17.75, -4.) {};
\draw [line width=2pt,bend right=15, looseness=0.75] (108.center) to (105.center);
\draw [line width=2pt,bend right=15, looseness=0.75] (106.center) to (109.center);

\draw [line width=2pt](113.center) to (114.center);
\draw [line width=2pt](116.center) to (113.center);
\draw [line width=2pt](108.center) to (118.center);
\draw [line width=2pt,bend left=15, looseness=0.75] (106.center) to (119.center);
\draw [line width=2pt,bend right=15, looseness=0.75] (119.center) to (110.center);

\draw [line width=2pt,bend left=15, looseness=0.50] (112.center) to (121.center);
\draw [line width=2pt,bend right=15] (111.center) to (120.center);
\draw [line width=2pt,bend right=15, looseness=0.50] (121.center) to (113.center);
\draw [line width=2pt,bend left=15] (120.center) to (113.center);
\draw [line width=2pt,bend left=15, looseness=0.75] (116.center) to (123.center);
\draw [line width=2pt,bend right=15] (123.center) to (117.center);
\draw [line width=2pt,bend right=15, looseness=0.75] (116.center) to (122.center);
\draw [line width=2pt,bend left=15, looseness=0.75] (122.center) to (115.center);
\draw [line width=2pt,bend left=15] (126.center) to (129.center);
\draw [line width=2pt,bend left=15, looseness=0.25] (127.center) to (128.center);
\draw [line width=2pt](130.center) to (131.center);
\draw [line width=2pt](133.center) to (130.center);

\draw [line width=2pt,bend left=15, looseness=0.75] (135.center) to (127.center);
\draw [line width=2pt,bend right=15, looseness=0.50] (129.center) to (137.center);
\draw [line width=2pt,bend left=15] (128.center) to (136.center);
\draw [line width=2pt,bend left=15, looseness=0.50] (137.center) to (130.center);
\draw [line width=2pt,bend right=15] (136.center) to (130.center);
\draw [line width=2pt,bend right=15, looseness=0.75] (133.center) to (139.center);
\draw [line width=2pt,bend left=15] (139.center) to (134.center);
\draw [line width=2pt,bend left=15, looseness=0.75] (133.center) to (138.center);
\draw [line width=2pt,bend right=15, looseness=0.75] (138.center) to (132.center);
\draw [line width=2pt,bend left=15] (132.center) to (143.center);
\draw [line width=2pt,bend right=15, looseness=0.50] (143.center) to (141.center);
\draw [line width=2pt](141.center) to (142.center);
\draw [line width=2pt](134.center) to (144.center);
\draw [line width=2pt](117.center) to (145.center);
\draw [line width=2pt,bend right=15] (115.center) to (146.center);
\draw [line width=2pt,bend left=15, looseness=0.50] (146.center) to (141.center);
\draw [line width=2pt](106.center) to (154.center);
\node [dot,color=blue] (160) at (13.75, -4.75) {};
\node [dot,color=blue] (161) at (9.75, -6.25) {};
\node [dot,color=blue] (162) at (11.25, -5.25) {};
\draw [line width=2pt,color=blue,bend right=15] (159.center) to (161.center);
\draw [line width=2pt,color=blue,bend left=15, looseness=0.25] (161.center) to (162.center);
\draw [line width=2pt,color=blue,bend left=15, looseness=0.50] (162.center) to (160.center);
\node [circle,fill=yellow,draw=black,rotate=90,minimum size=5pt,inner sep=0pt, outer sep=0pt]  at (9.3, -6.525) {};
\node [dot,color=blue] (152) at (32, -10) {};
\draw [line width=2pt,color=blue](103.center) to (152.center);
\node [dot,color=blue] (172) at (9, -6) {};
\node [dot,color=blue] (173) at (10.25, -7.5) {};
\node [dot,color=blue] (174) at (13.25, -8.5) {};
\node [dot,color=blue] (151) at (34, -8.5) {};
\draw [line width=2pt,color=blue](150.center) to (151.center);
\draw [line width=2pt,color=blue,bend left] (171.center) to (172.center);
\draw [line width=2pt,color=blue,bend right=15, looseness=0.75] (172.center) to (173.center);
\draw [line width=2pt,color=blue,bend right=15] (173.center) to (174.center);

\node [dot,color=blue] (212) at (13.75, -3.5) {};
\node [dot,color=blue] (213) at (9.75, -2) {};
\node [dot,color=blue] (214) at (11.25, -3) {};
\draw [line width=2pt,color=blue,bend left=15] (211.center) to (213.center);
\draw [line width=2pt,color=blue,bend right=15, looseness=0.25] (213.center) to (214.center);
\draw [line width=2pt,color=blue,bend right=15, looseness=0.50] (214.center) to (212.center);
\node [circle,fill=yellow,draw=black,rotate=90,minimum size=5pt,inner sep=0pt, outer sep=0pt]  at (9.3, -1.75) {};
\node [dot,color=blue] (148) at (31, 1.75) {};
\draw [line width=2pt,color=blue](60.center) to (148.center);
\node [dot,color=blue] (168) at (9, -2.25) {};
\node [dot,color=blue] (169) at (10.25, -0.75) {};
\node [dot,color=blue] (170) at (13.25, 0.25) {};
\node [dot,color=blue] (149) at (34, 0.25) {};
\draw [line width=2pt,color=blue](147.center) to (149.center);
\draw [line width=2pt,color=blue,bend right, looseness=0.75] (167.center) to (168.center);
\draw [line width=2pt,color=blue,bend left=15, looseness=0.75] (168.center) to (169.center);
\draw [line width=2pt,color=blue,bend left=15] (169.center) to (170.center);

\node [dot,color=blue] (179) at (-12.75, -2.25) {};
\node [dot,color=blue] (180) at (-15.25, -4) {};
\node [dot,color=blue] (181) at (-17.5, -4.) {};
\draw [line width=2pt,color=blue,bend right=15, looseness=0.75] (183.center) to (179.center);
\draw [line width=2pt,color=blue,bend left=15, looseness=0.50] (179.center) to (180.center);
\draw [line width=2pt,color=blue](180.center) to (181.center);
\node [dot,color=blue] (197) at (-12.75, -1) {};
\node [dot,color=blue] (198) at (-15.25, -2.75) {};
\node [dot,color=blue] (199) at (-17.5, -3) {};
\draw [line width=2pt,color=blue,bend right=15, looseness=0.75] (200.center) to (197.center);
\draw [line width=2pt,color=blue,bend left=15, looseness=0.50] (197.center) to (198.center);
\draw [line width=2pt,color=blue](198.center) to (199.center);

\node [dot,color=blue] (202) at (-12.75, -7.5) {};
\node [dot,color=blue] (203) at (-15.25, -6.25) {};
\node [dot,color=blue] (204) at (-17.5, -6) {};
\draw [line width=2pt,color=blue,bend left=15, looseness=0.75] (205.center) to (202.center);
\draw [line width=2pt,color=blue,bend right=15, looseness=0.50] (202.center) to (203.center);
\draw [line width=2pt,color=blue](203.center) to (204.center);
\node [dot,color=blue] (207) at (-12.75, -6.25) {};
\node [dot,color=blue] (208) at (-15.25, -5) {};
\node [dot,color=blue] (209) at (-17.5, -5) {};
\draw [line width=2pt,color=blue,bend left=15, looseness=0.75] (210.center) to (207.center);
\draw [line width=2pt,color=blue,bend right=15, looseness=0.50] (207.center) to (208.center);
\draw [line width=2pt,color=blue](208.center) to (209.center);
\node [semicircle,draw=blue,fill=red,rotate=90,minimum size=5pt,inner sep=0pt, outer sep=0pt] (0) at (-3.8, 2) {};
\node [semicircle,draw=blue,fill=red,rotate=90,minimum size=5pt,inner sep=0pt, outer sep=0pt] (0) at (-3.8, -5) {};
\node [circular sector,draw=blue,fill=red,rotate=0,minimum size=8pt,inner sep=0.1pt, outer sep=0pt] (0) at (3., 0.) {};
\node [circular sector,draw=blue,fill=red,rotate=0,minimum size=8pt,inner sep=0.1pt, outer sep=0pt] (0) at (3., -6.8) {};
\node [semicircle,draw=blue,fill=red,rotate=90,minimum size=5pt,inner sep=0pt, outer sep=0pt]   (0) at (21,-1.3) {};
\node [circular sector,draw=blue,fill=red,rotate=0,minimum size=8pt,inner sep=0.1pt, outer sep=0pt] (0) at (28.1, -4.7) {};
\node [semicircle,draw=blue,fill=red,rotate=90,minimum size=5pt,inner sep=0pt, outer sep=0pt]   (0) at (-15,-13.) {};
\node at (-4, -13.){\footnotesize : $\dfrac{\pi}{2}$ achromatic phase Delay};
\node [circular sector,draw=blue,fill=red,rotate=0,minimum size=8pt,inner sep=0.1pt, outer sep=0pt] (0) at (8.4, -13.) {};
\node at (20, -13.){\footnotesize : $\dfrac{\pi}{4}$ achromatic phase Delay};
\node [color=blue] at (13.6,-2){\scriptsize$_1^1C$};
\node [color=blue] at (13.6,-6.2){\scriptsize $_2^1C$};
\node [color=blue,draw=blue,fill=white,minimum size=7pt,inner sep=1pt, outer sep=0pt]  at (31.2, 2.5) {\scriptsize $_1^1A$};
\node [color=blue,draw=blue,fill=white,minimum size=7pt,inner sep=1pt, outer sep=0pt]  at (33.8, 0.2){\scriptsize $_1^1B$};
\node[color=black,draw=black,fill=white,minimum size=7pt,inner sep=1pt, outer sep=0pt] at (31.2, -1.5){\scriptsize $_1^2A$};
\node[color=black,draw=black,fill=white,minimum size=7pt,inner sep=1pt, outer sep=0pt] at (33.8, -4){\scriptsize$_1^2C$};
\node[color=black,draw=black,fill=white,minimum size=7pt,inner sep=1pt, outer sep=0pt] at (31.2, -6.5){\scriptsize $_1^2B$};
\node [color=blue,draw=blue,fill=white,minimum size=7pt,inner sep=1pt, outer sep=0pt]  at (33.8, -8.5){\scriptsize $_2^1A$};
\node [color=blue,draw=blue,fill=white,minimum size=7pt,inner sep=1pt, outer sep=0pt]  at (31.2, -10.7){\scriptsize $_2^1B$};
\end{tikzpicture}

%% file: 4-suite-new.tex
\subsection{Anisopistonic  variance \label{sec:Anis-error}}

The anisopistonic variance $\sigma_{A}^2(\theta)$ is the variance of the difference between the piston observed in two different directions with angular separation $\theta$. \cite{2008A&A...477..337E} developed and tested an analytical expression of  $\sigma_{A}^2(\theta)$ as a function of D and the seeing parameters $r_0$, $L_0$ and the isoplanatic angle $\theta_{0}$. Here, we summarize the key steps and parameters of that analysis for the commodity of the reader.

\figref{fig:GeometeryAnis} shows the general geometry of the problem. The variance $\sigma_{A}^2$ of the differential piston difference is:
\begin{figure}
	\centering
	\input{fig/geom-param.tikz}
\caption{The geometrical parameters to compute the contribution of a turbulent layer at
		altitude $h_m$ to the difference of piston between two directions separated by $\theta$. $ D $ is the
		telescope diameters and $\Delta$ is the baseline.}
	\label{fig:GeometeryAnis}
\end{figure}
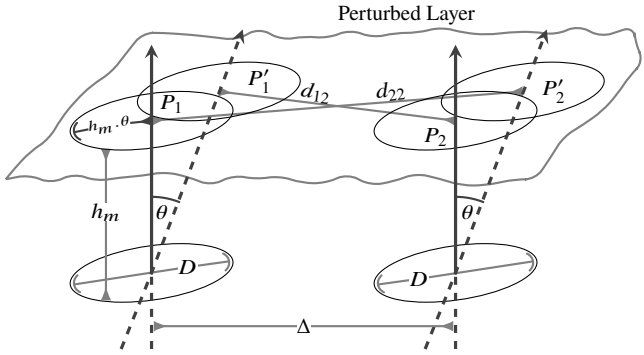
\begin{eqnarray}
\sigma_{A}^{2}&=&\left\langle\left[\left(P_{2}-P_{1}\right)-\left(P_{2}^{\prime}-P_{1}^{\prime}\right)\right]^{2}\right\rangle \\\nonumber
&=&4\left\langle P_{1}^{2}\right\rangle-4\left\langle P_{1} P_{2}\right\rangle+2\left\langle P_{1} P_{1}^{\prime}\right\rangle+2\left\langle P_{2} P_{1}^{\prime}\right\rangle+2\left\langle P_{1} P_{2}^{\prime}\right\rangle,
\end{eqnarray}
where $ P_{i}$ and $ P^\prime_{i}$ are the average OPD of the layer over the pupil $i$ observed in two directions separated by an angle $\theta$.

Assuming $\left\langle P_{1}^{2}\right\rangle=\left\langle P_{2}^{2}\right\rangle, \quad\left\langle P_{1} P_{2}\right\rangle=\left\langle P_{1}^{\prime} P_{2}^{\prime}\right\rangle$, and $\left\langle P_{1} P_{1}^{\prime}\right\rangle=\left\langle P_{2} P_{2}^{\prime}\right\rangle$ for a homogeneous and
isotropic turbulence. Considering that $\Delta \gg h_m \theta$, we can write $d_{12} \simeq d_{21} \simeq \Delta$ and $\left\langle P_{1} P_{2}\right\rangle \simeq\left\langle P_{1} P_{2}^{\prime}\right\rangle \simeq\left\langle P_{2} P_{1}^{\prime}\right\rangle$ and hence:
\begin{equation}
\sigma_{A}^{2}=4\left\langle P_{1}^{2}\right\rangle+2\left\langle P_{1} P_{1}^{\prime}\right\rangle.
\end{equation}

The variance $\left\langle P_{1}^{2}\right\rangle$ and the covariance $ \left\langle P_{1} P_{1}^{\prime}\right\rangle$ are computed from the integration of OPD differences over the pupil of diameter D. Assuming a Von Karman model for the structure function of the phase, \cite{Takato:95} write:
\begin{eqnarray}
\sigma_{A}^{2}(\theta)&=&0.02294 \pi^{2 / 3} \lambda^{2}\left(\frac{D}{r_{0}\left(h_m\right)}\right)^{5 / 3} \label{equ:sigmap}\\\nonumber
&\times&\int_{0}^{\infty} \frac{2 J_{1}^{2}(x)}{x\left[x^{2}+\left(\pi \frac{D}{L_{0}\left(h_m\right)}\right)^{2}\right]^{11 / 6}}\left[1-J_{0}\left(2 x \frac{h_m \theta}{D}\right)\right] \d x,
\end{eqnarray}
where $r_0(h_m)$ and $L_0(h_m)$ are the Fried parameter and the outer scale for the turbulence in the layer at altitude $h_m$. For each layer of thickness $\Delta h$, we have: 
\begin{equation}
\lambda^{2} r_{0}\left(h_m\right)^{-5 / 3}=\frac{C_{N}^{2}\left(h_m\right) \Delta h}{0.06}.
\label{equ:hmax}
\end{equation}

In the standard  near-field approximation,  the overall OPD variance is the sum of the single layer variances and putting equation \eqref{equ:hmax} in \eqref{equ:sigmap} yields:
\begin{eqnarray}
\sigma_{A}^{2}(\theta)&=&3 \pi^{\frac{2}{3}} D^{\frac{5}{3}} \int_{0}^{\infty} \int_{0}^{\infty} \frac{2 J_{1}^{2}(x) C_{N}^{2}(h)}{x\left[x^{2}+\left(\pi \frac{D}{L_{0}\left(h_m\right)}\right)^{2}\right]^{\frac{11}{6}}}\nonumber\\&\times&\left[1-J_{0}\left(2 x \frac{h_m \theta}{D}\right)\right] \d x \d h.
\label{equ:sigmap_final}
\end{eqnarray}

\cite{2008A&A...477..337E} has developed an analytical expression of equation \eqref{equ:sigmap_final} based on a development of the integral in convergent series using a technique based on the convolution theorem of the Mellin transform introduced by \cite{1994JOSAA..11..379S}. This development yields:
\begin{equation}
\sigma_{A}^{2}(\theta)=\pi^{\frac{2}{3}} \lambda^{2}\left(\frac{D}{r_{0}}\right)^{-1 / 3}\left(\frac{\theta}{\theta_{0}}\right)^{2} P_{A}\left(\frac{\pi D}{L_{0}}\right),
\label{equ:varp}
\end{equation}
where we have introduced the isoplanatic angle $\theta_{0}$ from \cite{1982JOSA...72...52F}:
\begin{equation}
\theta_{0}=0.058 \lambda^{6 / 5}\left[\int_{0}^{\infty} h^{5 / 3} C_{N}^{2}(h) \d h\right]^{-3 / 5},
\end{equation}
$P_A$ is a polynomial function with parameters dominated by the ratio between the telescope diameter and the outer scale $\frac{\pi D}{L_0}$. \cite{2008A&A...477..337E} defines a small aperture regime by $ D<\frac{\alpha L_0}{\pi}$ where:
\begin{eqnarray}
P_{A}\left(\frac{\pi D}{L_{0}}\right)&=&8.8410^{-2}\left[0.216-0.225\left(\frac{\pi D}{L_{0}}\right)^{\frac{1}{3}}\right.\nonumber\\
&+&\left.0.122\left(\frac{\pi D}{L_{0}}\right)^{2}-0.096\left(\frac{\pi D}{L_{0}}\right)^{\frac{7}{3}}+0.02\left(\frac{\pi D}{L_{0}}\right)^{4}\right. \nonumber\\
&-&\left.0.014\left(\frac{\pi D}{L_{0}}\right)^{\frac{13}{3}}\right].
\label{equ:paDL}
\end{eqnarray}
\begin{figure}
	\centering
	\includegraphics[width=\linewidth]{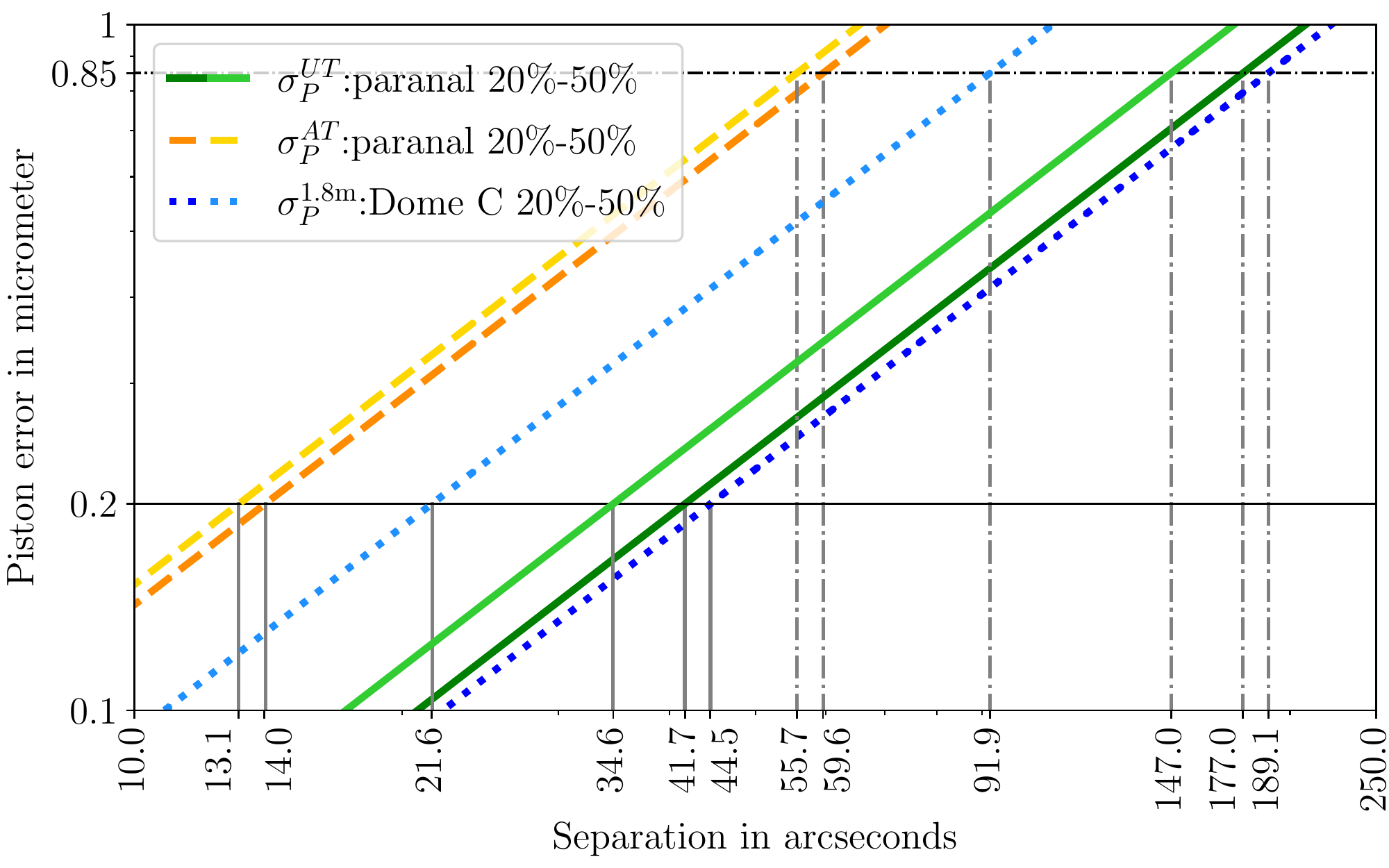}
	\caption{Anisopistonic error as a function of the separation between the observed target and the off-axis star used for fringe tracking in logarithmic scale. In dashed lines, yellowish with \SI{8}{\meter} telescopes (\textit{UTs}) at \textit{Paranal} in the best 20 per cent seeing (upper curve) and in median seeing (lower curve). In full-line greenish with \SI{1.8}{\meter} telescopes (\textit{ATs}) at \textit{Paranal}, for best 20 per cent seeing (upper curve) and median seeing (lower curve). In dotted line blueish, at Dome C for \SI{1.8}{\meter} telescopes placed above the \SI{30}{\meter} ground layer. The list of numbers below the abscissa axis is the Isopistonic angle.}
	\label{fig:theta-max}
\end{figure}

For $\alpha<1.3$, the difference between the results of equation \eqref{equ:paDL} and equation \eqref{equ:sigmap_final} are smaller than 5 per cent and they reach 10 per cent for $\alpha\sim2.6$, this allows using equation \eqref{equ:paDL} for all the cases considered in this paper. In equation \eqref{equ:paDL}, almost all the sensitivity of $\sigma_{A}^2$ to the exact vertical distribution of turbulence described by the profile $C_n^2(h)$ is included in the terms $r_0$, $L_0$, and $\theta_{0}$ evaluated on the ground. However, very special turbulent profiles such as a very strongly dominant layer (near the ground or at the troposphere) might introduce deviations.

\figref{fig:theta-max} gives the anisopistonic error computed from equation \eqref{equ:paDL} for the three cases considered here. \textit{UTs} and \textit{ATs} at \textit{Paranal}, in the median seeing conditions ($L_0=\SI{16}{\meter}$, $r_0=\SI{0.76}{\meter}$ ) and best 20 per cent seeing condition ($L_0=\SI{22}{\meter}$, $r_0=\SI{0.53}{\meter}$ ) and \SI{1.8}{\meter} telescopes at the Dome C.  It allows reading the values of the isopistonic angle $\theta_{P}(\lambda=2 \mu m)$ defined by $\sigma_{A}\left[\theta_{P}(\lambda)\right]=\frac{\lambda}{10}$ that are given in  \tabref{tab:2}.

Note the very strong dependence of the isopistonic angle on the telescope diameter. The isopistonic angle is very large and the maximum separation between the observed target and the guide star are more likely to be limited by the hardware at the telescope or by AO isoplanetism than by the isopistonic angle in the $ K $ band.
Equation \eqref{equ:varp} and	\figref{fig:GeometeryAnis} gives the anisopistonic error for a target at zenith. To compute its value at a zenith distance $z$, we use the dependence of $r_0$­ and $\theta_{0}$­ with that distance:
\begin{equation}
\left\lbrace \begin{array}{ll}
r_{0}(z)=r_{0}(0)(\cos z)^{3 / 5} ; \\
\theta_{0}(z)=\theta_{0}(0) \cos(z).
\end{array}\right. 
\end{equation}

The outer scale $L_0$­ is a parameter describing the turbulence \textquoteleft where it is\textquoteright\ that does not change
with the zenith distance.
\begin{table}
	\centering
	\caption{Isopistonic angle, in different sites and seeing conditions, for a $\lambda/10$ error at $\lambda=2\mu m$. The isopistonic angle is proportional to the wavelength.}
	
	{\footnotesize 	\begin{tabular}{L{1.83cm}L{1.6cm}L{1.6cm}L{1.6cm}}
			\hline
			& D=\SI{8}{\meter}, \textit{Paranal} & D=\SI{1.8}{\meter}, \textit{Paranal} & D=\SI{1.8}{\meter} , Dome C \\ \hline
			\multicolumn{1}{L{1.83cm}}{Best 20 per cent} &  \SI{41.7}{\arcsecond}         &  \SI{14.}{\arcsecond}           &  \SI{44.5}{\arcsecond}       \\
			seeing& & &\\
			\multicolumn{1}{L{1.73cm}}{Median seeing}    &  \SI{34.6}{\arcsecond}          &  \SI{13.1}{\arcsecond}            &  \SI{21.6}{\arcsecond}          \\ \hline
	\end{tabular}}
	\label{tab:2}
\end{table}

\subsection{Servo loop variance \label{sec:serv-loop}}
\label{section-loop-error}
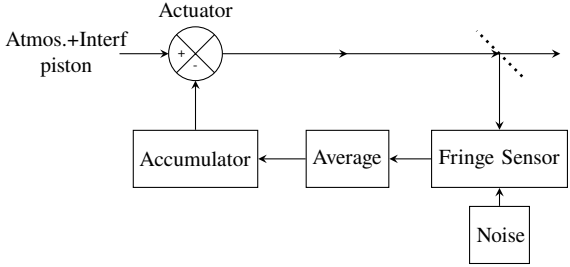
\begin{figure}
	\centering
	\begin{tikzpicture}[minimum size=0.75cm]
			\draw [line cap=round] (-0.25,-0.25) -- (0.25,0.25) (-0.25,0.25) -- (0.25,-0.25);
			\draw (0,0) circle (0.353553);
			\node[anchor=north] at (0, 0.2) {\tiny -};
			\node[anchor=east] at (0.2, 0) {\tiny +};
			\draw [->, > = stealth] (-1, 0) -- (-0.353553, 0) ;
			\node at (-1.7,0.2){\small Atmos.+Interf };
			\node[anchor=north] at (-1.7,0.20){\small piston};
			\node[anchor=south] at (0.0,0.20){\small Actuator};
			\draw [->, > = stealth] (0.353553, 0) -- (2, 0);
			\draw [->, > = stealth] (1.9, 0) -- (4, 0);
			\draw [dotted,line width=1] (4.0+0.3, -0.3) -- (4.0-0.3, 0.3);
			\draw [->, > = stealth] (4.0, 0) -- (4.8, 0);
			\draw [->, > = stealth] (4.0, 0) -- (4.0, -1);
			\node[draw,anchor=north, text width=1.6cm] (a) at (4.0, -1) {\small Fringe Sensor};
			\node[draw,anchor=north, text width=0.9cm] (b) at (2, -1) {\small  Average};
			\node[draw,anchor=north, text width=1.45cm] (c) at (0, -1) {\small Accumulator};
			\node[draw,anchor=north, text width=0.6cm] (d) at (4.0, -2) {\small  Noise};
			\draw[->, > = stealth] (a)to[right](b);
			\draw[->, > = stealth] (b)to[left](c);
			\draw[->, > = stealth] (d)to[left](a);
			\draw [->, > = stealth] (c) -- ((0,-0.353553);
\end{tikzpicture}
		\caption{The simple servo loop of FTs used in this paper.
		}\label{fig:Servo}  
\end{figure}

In this paper, we will consider only the simplest servo loop that is sketched in \figref{fig:Servo}. The fringe sensor measures the phase delay introduced by the atmosphere and the interferometer after the last piston correction applied by the piezo actuator. The measure is affected by the detection noise discussed in the previous section. The resulting error signal is averaged over a dwell time $\mathcal{T}_F$ and subsequently passed to an accumulator that adds the measured error to the previous piston actuator position.

Actual FTs, such as the GFT, use more sophisticated phase delay controllers based on a state-space auto-regressive representation of the atmospheric and vibration perturbation with a Kalman filter optimization. Here we consider that such a device only allows approaching the performances of the simple integrator affected only by the atmospheric piston that we analyze in the following. This might be a fair or even quite optimistic representation of the current status of the GFT on \textit{UTs} that fails to reach the longest frame times allowed by the atmosphere. It is probably a slightly pessimistic description of a \textit{GRAVITY+} optimum situation where vibrations and injection instabilities have been damped below the effects of the atmospheric piston by systems independent from the FT. The simple loop used here has the great advantage to allow an analytical expression of the effects of the atmospheric piston.

Frequency-domain models for each of the components of the loop shown in \figref{fig:Servo} can be found in many papers, e.g. \cite{2000SPIE.4006.1116F}. If we note $ \underline{Y}_{in}(f) $ and $ \underline{Y}_{out}(f) $, the input and output signal of the servo loop system in the frequency domain, the succession of elements in  \figref{fig:Servo} translates into the following equation :
\begin{eqnarray}
	\underline{Y}_{out}(f)= \underline{Y}_{in}(f)-
	\underline{H}_{pz}(f)\underline{H}_{s}(f)\underline{H}_{a}(f)\times\underline{Y}_{out}(f).
\end{eqnarray}

This gives the square modulus of the
servo loop error-input transfer function
\begin{eqnarray}
	H_{loop}(f)=
	\left|\frac{1}{1+\underline{H}_{pz}(f)\underline{H}_{s}(f)\underline{H}_{a}(f)}\right|^2.
\end{eqnarray}

The fringe sensor (sampling and averaging) system includes an integration period $\mathcal{T}_F$  and an average delay time of $\mathcal{T}_L$ between the input and the output signals. If the computation time is negligible, we have $\mathcal{T}_L\simeq\mathcal{T}_F/2$. Then the transfer function for the fringe sensor is given by:
\begin{eqnarray}
	\label{Sampling}
	\underline{H}_{s}(f)= \frac{\sin(\pi f\mathcal{T}_F)}{j\pi f\mathcal{T}_F}\exp(-j\pi f
	\mathcal{T}_F)= \frac{1-\exp(-j2\pi f \mathcal{T}_F)}{2j\pi f \mathcal{T}_F}.
\end{eqnarray}

The accumulator adds the incremental error signal to the current piezo-mirror position. This is represented by the function
\begin{eqnarray}\label{accum}
	\underline{H}_{a}(f)= \frac{1}{2j\pi f\mathcal{T}_F}.
	\label{equ:H-laa}
\end{eqnarray}

The piezo-actuator system is  generally very fast and has a nearly instantaneous response, yielding $ \underline{H}_{pz}=1$, and hence:
\begin{eqnarray}
	H_{loop}(f)&=&
	\left|\frac{1}{1+\underline{H}_{s}(f)\underline{H}_{a}(f)}\right|^2\\
	&=& \frac{(2\pi f \mathcal{T}_F)^4}{(2\pi f \mathcal{T}_F)^4-4\sin^2(\pi f\mathcal{T}_F)[(2\pi f \mathcal{T}_F)^2-1]}\nonumber.
\end{eqnarray}

This last expression is approximated by  \cite{1994JOSAA..11..288P} as:
\begin{eqnarray}
	H_{loop}(f)\simeq \frac{(2\pi f \mathcal{T}_F)^2+\frac{4}{89}(2\pi f \mathcal{T}_F)^4}{1+ \frac{2}{45}(2\pi f \mathcal{T}_F)^4},
	\label{equ:H-loop}
\end{eqnarray}
and the residual piston variance introduced in the servo loop system by the change of the atmospheric piston through the loop cycle is then given by:
\begin{eqnarray}\label{Hloop}
	\sigma_L^2=\left(\frac{\lambda}{2\pi}\right)^2\,2\int_0^{\infty}{H_{loop}(f)W_{\phi}(f)}{\rm{d}}f,
\end{eqnarray}
where $ W_{\phi}(f) $ is the power spectrum of the piston phase $\phi=2\pi p_{ij}/\lambda$. That power spectrum and that integral have been computed by \cite{2000SPIE.4006.1116F} assuming a Von Karman spectrum with outer scale $L_0$ for the random atmospheric refractive index fluctuations, which were supposed to be isotropic within each turbulent layer. The temporal variations of the piston are computed using a multilayer Taylor model, with turbulent layers drifting with the wind through the telescope aperture faster than they change internally. It writes:
\begin{eqnarray}
	\sigma_L^2= A_0sec(z) D^{-1/3}\mathcal{T}_F^2\upsilon^2\left[1+A_1(D,L_0)+A_2(D,L_0, \Delta)\right],
\end{eqnarray}
where $z$ is the zenith distance and $ \upsilon^2=\int_0^{\infty}{C_N^2(h)v^2(h)}\d h $ is the $2^{\text{nd}}$ order moment of the velocity components in the direction of the interferometric baseline that provides an upper limit for $ W_{\phi}(f) $.

The parameters $A_1$, $A_2$, and $A_3$ have different expressions in the different regimes set by the relative size of the outer scale $L_0$ and of the telescope diameter $D$. In all cases considered here, the telescope diameter is much smaller than turbulence outer scale and then \cite{2000SPIE.4006.1116F} derives the following values of  $ A_0 $, $ A_1 $, and $A_2$: 
\begin{eqnarray}\left\{
	{\begin{array}{*{20}l}
			{A_0 = 0.643}  \\
			{A_1 = 3.61\times10^{-2}\left(L_0/D\right)^{8/3}\times} \\
			{~~~~~~~~~~ _2F_1(4/3,3/2;5/2;-0.09\left(L_0/D\right)^2)}  \\
			{A_2 = -1.07\times10^{-2}\left(L_0/D\right)^{8/3}\left(1-0.474(D/\Delta)^{1/3}\right)\times}\\
			{~~~~~~~~~~_2F_1(4/3,3/2;5/2;-0.09\left(L_0/D\right)^2)}  \\
	\end{array}}\right.,
\end{eqnarray}
where $ \Gamma(x) $ is the gamma function and $ _2F_1 $ is the Gauss hypergeometric function given by \cite{1994JOSAA..11..379S}:
\begin{eqnarray*}
	_2F_1[\alpha,\beta;\gamma;z] =
	\sum_{k=0}^{\infty}{\frac{(\alpha)_k~(\beta)_k}{(\gamma)_k}\frac{z^k}{k!}}~~~~;
	\mathrm{where}~~~(a)_k = \frac{\Gamma(a + k)}{\Gamma(a)}
\end{eqnarray*}

\begin{figure*} 
	\centering
	\subfigure[HFT-\textit{UT} \textit{Paranal}]{%
		\label{fig:HFT-UT-OT-aranal}%
		\includegraphics[width=0.48\linewidth]{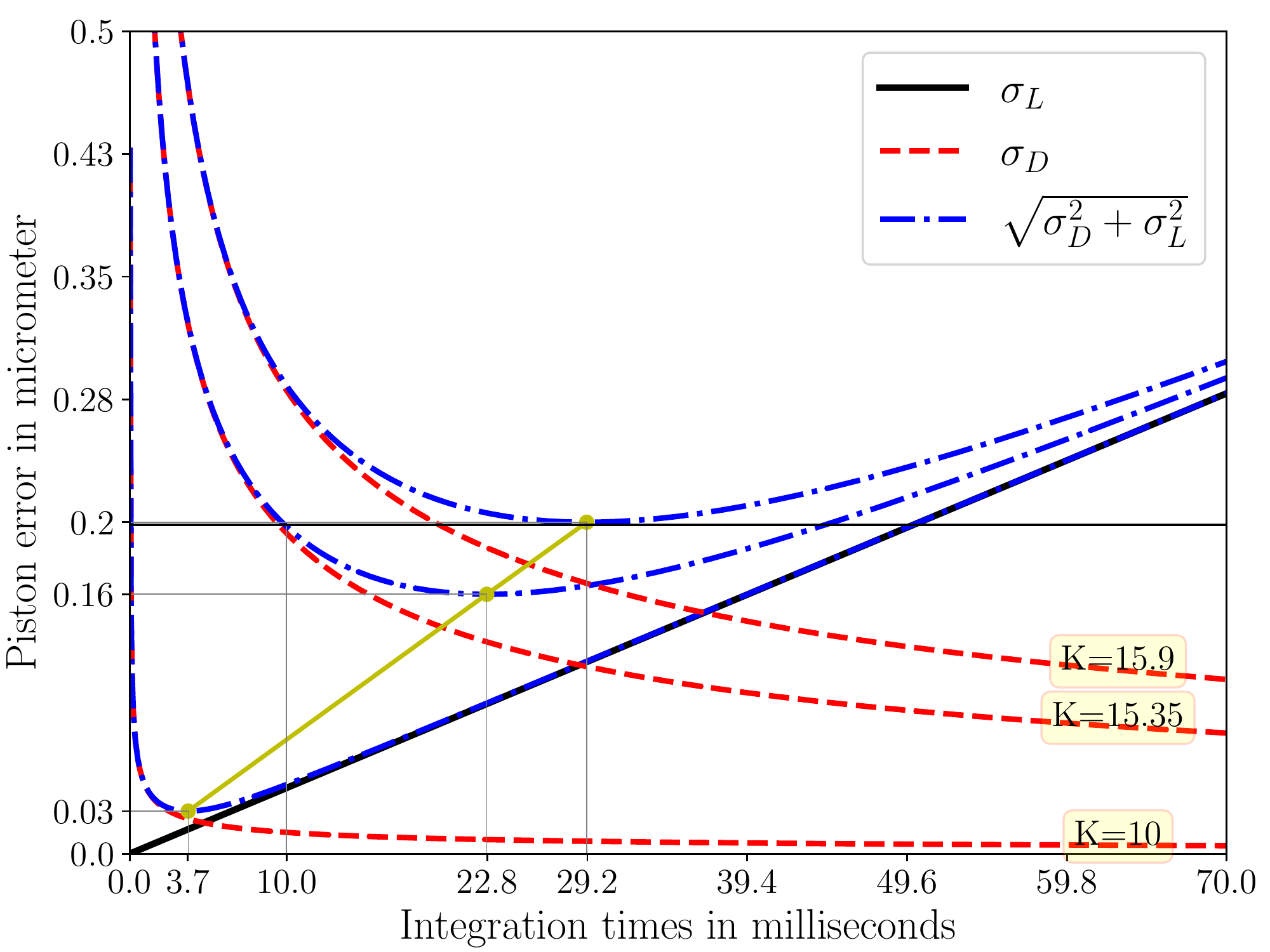}}
	\subfigure[GFT-\textit{UT} \textit{Paranal}]{%
		\label{fig:GFT-UT-OT-aranal}%
		\includegraphics[width=0.48\linewidth]{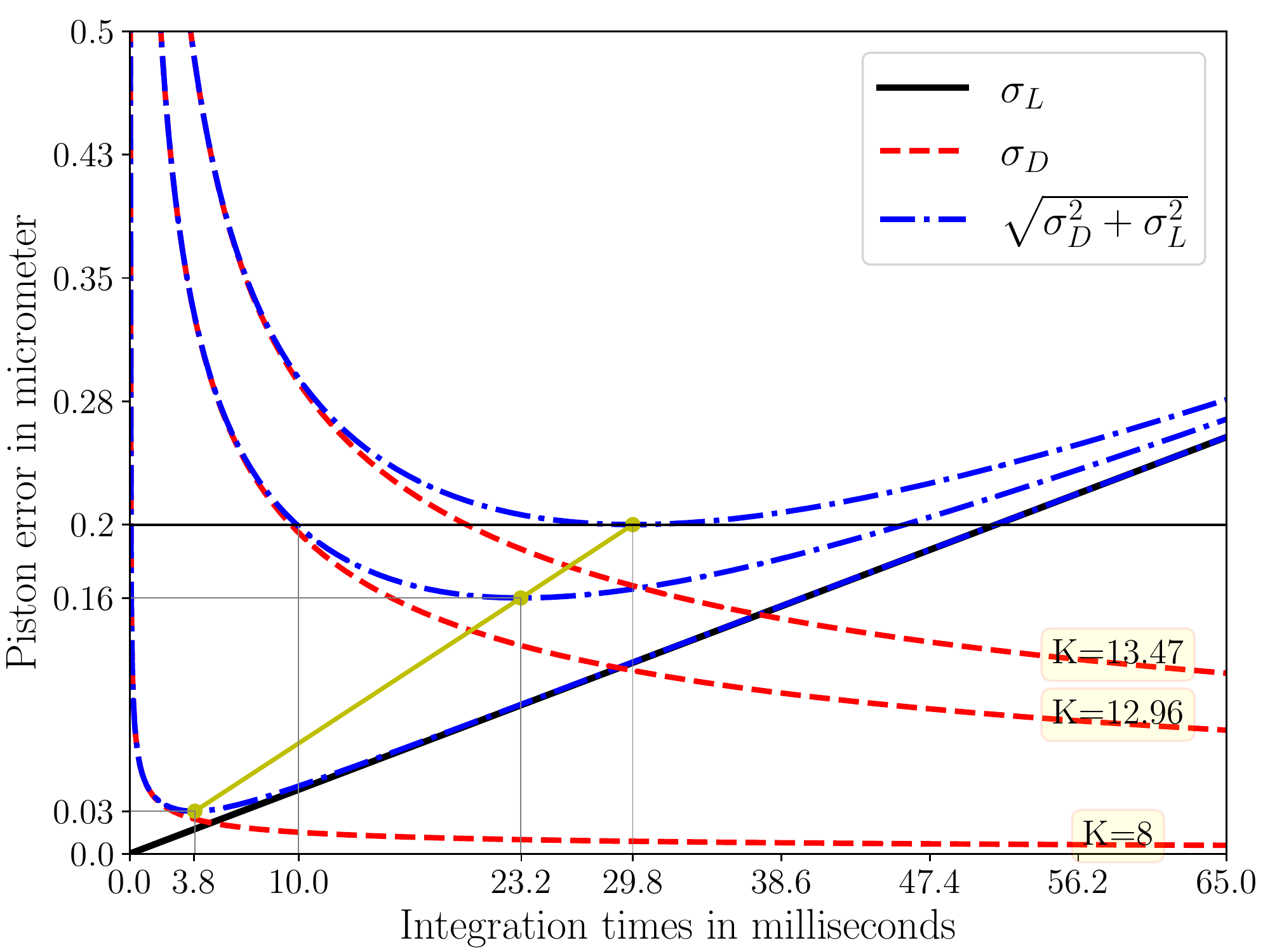}}
	\caption{Tracking error as a function of the frame time, for various magnitudes $K$ (blue dot\textendash dashed curves). The black line is the loop error and the red dashed lines are detection errors. The optimum frame times, which minimize the tracking error for a given magnitude, are indicated in green.}
	\label{fig:optimum-time}
\end{figure*}

Introducing the average velocity
$$ \bar{v}_2 = \left[\frac{\int_0^{\infty}{C_N^2(h)v^2(h)}{\rm{d}}h}{\int_0^{\infty}{C_N^2(h)}{\rm{d}}h}\right]^{1/2}$$
and the Fried parameter $ r_0$ defined by $ \int_0^{\infty}{C_N^2(h)}{\rm{d}}h = 0.06\,\lambda^2\, r_0^{-5/3}
$, we can write
\begin{eqnarray}
	\sigma_L^2 \simeq \,0.06~A_0sec(z)
	\lambda^2\,r_0^{-5/3}D^{-1/3}\mathcal{T}_F^2(\bar{v}_2)^2\left[1 + A_1 + A_2\right],
\end{eqnarray}
which can be expressed in terms of the atmospheric coherence time $ \mathcal{T}_0 $ as the following:
\begin{eqnarray}\label{varloop}
	\sigma_L^2 = \,5.76~10^{-3}~A_0sec(z)
	\lambda^2\,\left(\frac{r_0}{D}\right)^{1/3}\frac{\mathcal{T}_F^2}{\mathcal{T}_0^2}\left[1 + A_1 + A_2\right],
\end{eqnarray}

As all parameters in equation \ref{varloop} are constant for given seeing conditions, at least when $\mathcal{T}_F \ll L_0/\bar{v}_2$ i.e. the time needed for an outer scale to be drifted away. Then the loop error introduced by the atmospheric piston is proportional to the fringe sensor dwell time $\mathcal{T}_F$.

\subsection{Tracking error on the guide star and optimum frame time\label{Optimum-time}}
For the source used for fringe sensing and tracking, we will have a residual piston error with
variance $\sigma_{F}^2=\sigma_{D}^2+\sigma_{L}^2$ if we assume that the fringe sensing error and the loop error are
independent random variables. We have seen that $\sigma_{D}^2$ is a function of $\sigma_{L}^2$ but in a stationary
tracking regime the tracking noises introduced by the detection and the servo loop are independent in each frame. For a given star, $\sigma_{D}^2(\mathcal{T}_{{F}})$ decreases with the frame time $\mathcal{T}_{{F}}$ while we have seen in the previous section that $\sigma_{L}^2(\mathcal{T}_{{F}})$ is proportional to $\mathcal{T}_{{F}}$. \figref{fig:optimum-time} shows the combination of the detection and loop errors as a function of the frame time for various source magnitudes for the GFT and HFT fringe tracking concepts. We see that for each magnitude we have an optimum frame time that minimizes the tracking error. For a given tracking quality specification (for example $\frac{\lambda}{10}=\SI{0.2}{\micro\meter}$ in the plot), we have the optimum frame time and the corresponding limiting tracking coherent magnitude.
From \figref{fig:optimum-time} for the \textit{UTs} at \textit{Paranal}, and from similar figures for \textit{ATs} at \textit{Paranal} and \SI{1.8}{\meter} telescopes at Dome C, we derive the optimum exposure times and limiting magnitudes given in \tabref{tab:YT1}. 

Note that in this fringe tracking model where we would be limited only by the optical turbulence, for a $\frac{\lambda}{10}=\SI{0.2}{\micro\meter}$ tracking precision in the $ K $ band, the optimum exposure time  is quite large with \textit{UTs},  of the order of \SI{31}{\milli\second}, almost independently from the fringe tracking design and number of collected photons.

With the GFT and HFT values given in Section \ref{sec:HFT}, this yields limiting magnitudes of $\mathrm{K}=13.5$ with the GFT and $\mathrm{K}=15.9$ with the HFT. This optimum exposure time sets a goal for the damping of \textit{VLTI} vibrations and for the stability of AO that should be compatible with such frame times. As we cannot be sure that such long exposure times will
indeed be permitted by \textit{VLTI} and AO perturbations, we have also considered a maybe more realistic limit at $\mathcal{T}_{{F}}=\SI{10}{\milli\second}$ that yields $\mathrm{K}\sim13$ with the GFT and $\mathrm{K}\sim15.3$ with the HFT. We see that the limitation to a shorter frame time as a moderate impact as around the optimum time $\mathcal{T}_{{F}}$, the variance $\sigma_{D}^2(\mathcal{T}_{{F}})$ does not vary much in a relatively large range.

\begin{table*}
	\centering
		\caption{Seeing parameters, optimum frame time $\mathcal{T}_{opt}$ (optimum $ \mathcal{T}_{F} $), limiting magnitude for $\mathcal{T}_{opt}$ and limiting magnitude for 10 ms to achieve $\SI{0.2}{\micro\meter}$ for different telescope diameters, sites, and fringe tracking performances.}
	\label{tab:YT1}
	\begin{tabular}{p{2.2cm}C{2.1cm}C{1.9cm}C{0.84cm}C{1cm}C{1cm}C{1cm}C{2cm}}
		\hline
		\rule{0pt}{2.5ex} 
	Tel. D, site, and FT&\rotatebox{0}{ Seeing} \rotatebox{0}{conditions} &\rotatebox{0}{Seeing(\SI{}{\arcsecond}) } & \rotatebox{0}{$\mathcal{T}_0$(\SI{}{\milli\second}) }  & \rotatebox{0}{$\mathcal{T}_{opt}$(\SI{}{\milli\second}) } &  \rotatebox{0}{$K_{lim}^{\mathcal{T}_{opt}(\SI{}{\milli\second})}$ } & \rotatebox{0}{$K_{lim}^{\SI{10}{\milli\second}}$ } & \rotatebox{0}{Outer scale $L_0(\SI{}{\meter})$}\\ \midrule
		\multirow{2}{*}{\SI{8}{\meter} \textit{Paranal} GFT} & 
		Median & 0.7  & 3.2
		 & 20.4   & 13.22 & 12.92 & 22 \\ 
		&Best (20 per cent)  &0.6 & 4.4
		 &  29.8 &  13.47 & 12.96  & 16\\ 
		\multirow{2}{*}{\SI{8}{\meter} \textit{Paranal} HFT} & 
		Median &  0.7 &3.2
		 &20.2		  & 15.62 & 15.31  &22\\  
		& Best (20 per cent) & 0.6 & 4.4
		 & 29.2 &15.9  & 15.35 & 16\\
		\multirow{2}{*}{\SI{1.8}{\meter} \textit{Paranal} GFT} & 
		Median & 0.7 & 3.2
		 &10.2  & 9.48 & 9.46 & 22\\ 
		& Best (20 per cent) & 0.6 & 4.4
		 & 15 & 9.75 & 9.65 & 16\\ 
		\multirow{2}{*}{\SI{1.8}{\meter} \textit{Paranal} HFT} & 
		Median & 0.7  & 3.2
		 & 10.2 & 11.82 &11.80  &22\\ 
		&Best (20 per cent)  & 0.6 & 4.4
		 & 14.7 &12.15 & 12 & 16\\
		\multirow{2}{*}{\SI{1.8}{\meter}  Dome C HFT}
		& Median     & 0.84 & 7.9
		&  35.9  & 14.4 & 13.4  &7.34\\
		& Best (25 per cent) & 0.43 &12
		 & 53.8 & 14.8 &13.4 & 5 \\ \hline
	\end{tabular}
\end{table*}

\begin{figure}
	\includegraphics[width=\linewidth]{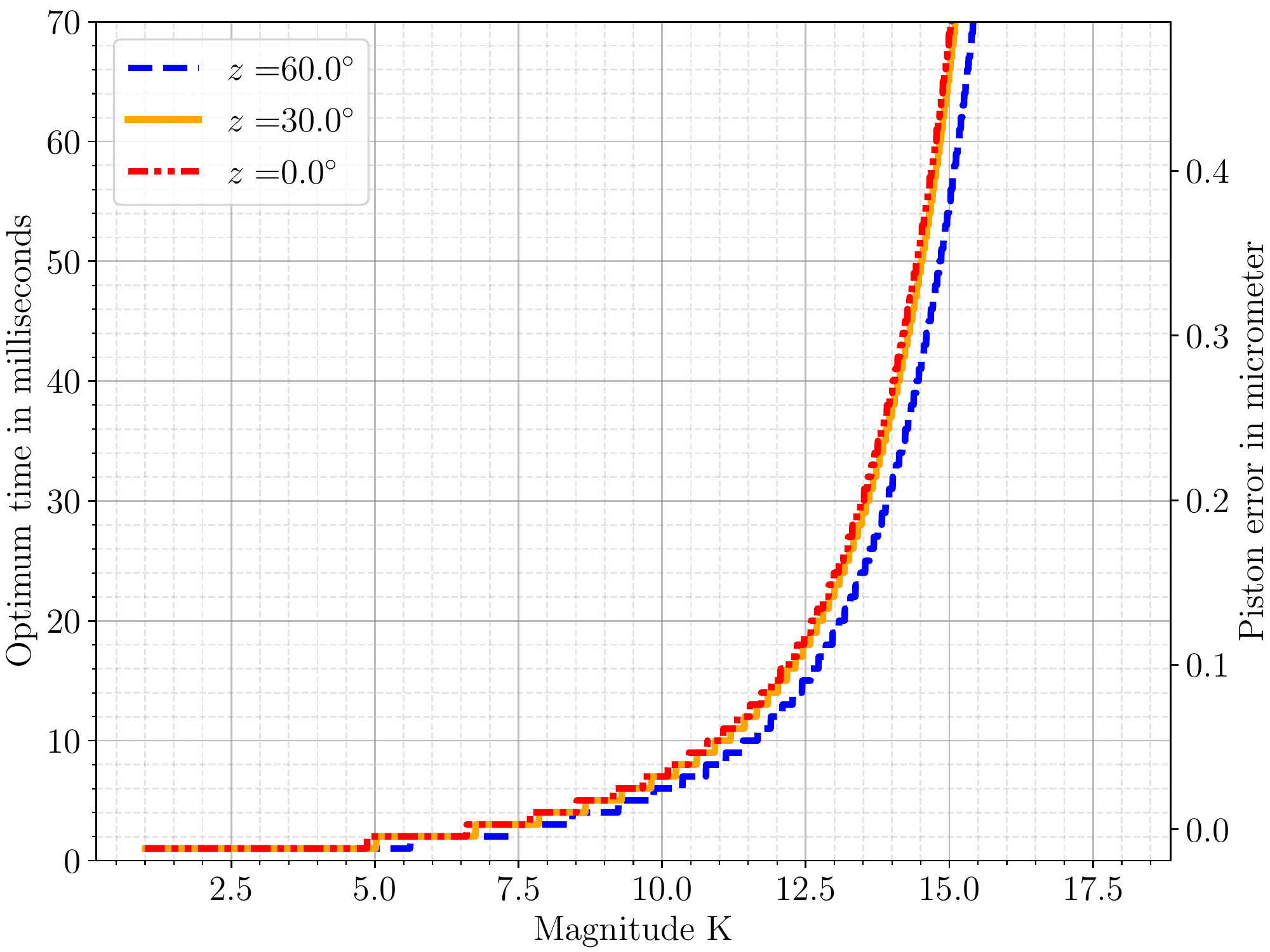}
	\caption{Optimum frame time and piston error as a function of the magnitude $ K $ for GFT-\textit{UT}, for various
		zenith distances $z=$0,\SI{30}{\degree} and \SI{60}{\degree}.}
	\label{fig:optimum-time-fit}
\end{figure}

For a given FT, site, and seeing conditions, each $ K $ magnitude sets an optimum frame time $\mathcal{T}_{{FO}}$,
which minimizes $\sigma_F^2({\mathcal{T}_{{FO}}}) =\sigma_D^2({\mathcal{T}_{{FO}}})+\sigma_L^2({\mathcal{T}_{{FO}}}) $. For computational efficiency, we define a function $\mathcal{T}_{{FO}}=f(K,z)$ that gives the optimum frame time as a function of the magnitude K and the zenith distance, as illustrated by \figref{fig:optimum-time-fit}, and we fit it with a polynomial function. Then we use the proportionality between $\sigma_{F}=a\mathcal{T}_{{FO}}$ indicated in \figref{fig:optimum-time}.

\subsection{Chromatic variation of the phase delay \label{Chrom-OPD}}

For fringe tracking in one spectral band around wavelength $\lambda_{t}$ for observations in another
spectral band around wavelength $\lambda_{o}$ we have to consider the chromatic dispersion of the OPD
between the different interferometric beams. An immediate application would be the use on the
\textit{VLTI} of fringe tracking in the $ K $ band (GFT) or in a combination of $ J $,$ H $, and $ K $ bands (HFT) for
observations with \textit{MATISSE} in the $ L $, $ M $, and $ N  $ bands. Another application will be a $ J $ band
instrument on the \textit{VLTI} with the same FTs. And a new PFI interferometer would
have to track in the near-IR for observations in the $ N $ and $ Q $ bands.

The inter-band piston differences are due to the chromatic differences in the air refraction index. For interferometers like the \textit{VLTI} with delay lines in the air, the largest chromatic effect will result from the large difference in air path between the telescopes and the focal laboratory. For all interferometers, including those with delay lines in evacuated pipes, the turbulent atmosphere introduces homogeneity that has a chromatic component. In the near-IR, the available models \citep{2007JOptA...9..470M} for the variation of the refractive index $n(\lambda,T,P,H)$ with wavelengths as a consequence of variation of temperature, pressure, and humidity allowed the correct computation and removal of the OPD between the $J$, $H$, and $K$ bands in the \textit{VLTI} instrument \textit{AMBER} \citep{2007A&A...464....1P} introduced by the difference in air path in the
Delay Line Tunnels. They also allowed \citep{2004SPIE.5491..577V} to evaluate the phase delay chromatic noise introduced by the dry air turbulence at $\sim \SI{2e-4}{\radian}$ and by the wet air turbulence at $\sim \SI{2e-2}{\radian}$. These noise levels have no impact on the fringe tracking quality, as it was confirmed by successful fringe tracking in the $ H $ band with FINITO \citep{2004SPIE.5491..528G} for \textit{AMBER} $ J $, $ H $, and $ K $ bands. Therefore, the chromatic OPD effects between the near-IR bands should be considered only for high dynamic observations such as extrasolar planets characterization that are beyond the scope of this paper. A $ J+H+K $ FT should therefore only correct for slowly varying chromatic OPD terms to avoid SNR losses due to fringe contrast reduction in broadband channels.
Fringe tracking in the $ K $ band for observations in the $ N $ band has been first implemented on the Keck interferometer \citep{2013PASP..125.1226C}. The use of the \textit{GRAVITY} FT in the $ K $ band to stabilize the fringes of \textit{MATISSE} in the $ L $, $ M $, and $ N $ bands has also been recently commissioned at the \textit{VLTI}. With the Keck interferometer \citep{2006SPIE.6268E..17K} reported group delay fluctuations of the order of \SI{10}{\micro\meter} PTV and \SI{3}{\micro\meter} RMS in $ N $ while fringes were tracked in $ N $. In the same paper, he describes and tests a method to predict that chromatic OPD term from the difference between phase delay and group delay in the $ K $ band. That difference
is dominated by variations of the water vapor column height that also produce the differences between the $ K $ band an $ N $ band phase delay difference. This method allowed the prediction and correction of the chromatic phase delay difference below \SI{1}{\micro\meter}. On the \textit{VLTI}, while we were accurately tracking in the $ K $ band with a residual piston smaller than \SI{0.2}{\micro\meter}, we observed in $ N $ band at \SI{8.5}{\micro\meter} fast phase delay fluctuations of about \SI{0.5}{\micro\meter} on top of slower group delay fluctuations of about \SI{2}{\micro\meter}. At \SI{12.5}{\micro\meter}, these numbers are typically multiplied by 3. Applying Koresko’s method to compute and correct these chromatic effects from the difference observed between the GFT phase and group delays in $K$, we reduce the chromatic OPD fluctuations to
$\lambda/100$ at \SI{8.5}{\micro\meter}, $\lambda/57$ at \SI{10.5}{\micro\meter}, and $\lambda/33$ at \SI{12.5}{\micro\meter}. These are rms values estimated on a sample of eight calibrators observed in a range of fair to good seeing conditions. We do not yet have the database nor the model to evaluate these chromatic residuals in all conditions, so we will use the conservative value of
\begin{equation}
\sigma_{H}=\SI{0.2}{\micro\meter},
\label{equ:sigma-chroma}
\end{equation}
for the chromatic OPD error added by tracking in $ K $ band for observations at \SI{8.5}{\micro\meter}.

\section{Sky coverage maps}
\subsection{Definition and utility}
To find an off-axis guide star for a specific target for a given tracking specification, for example $\sigma_{T}(\lambda_o)<\lambda_{o}/N_{spec}$, we have to consider all possible guide stars in a radius  $\theta_{p}(\lambda_{o})$ around the target and compute the tracking residual for each candidate using equation \eqref{equ:ST:obs:tra} combined with equations (\ref{equ:read-out}, \ref{equ:sigmap_final}, \ref{equ:H-laa} and \ref{equ:sigma-chroma}). The target can be tracked off-axis if we find at least one guide star such as:
\begin{eqnarray}
\begin{array}{l}
\sigma_{D}^{2}\left(\lambda_{t}, J, H, K, \mathcal{T}_{F}\right)+\sigma_{L}^{2}\left(\lambda_{t}, \mathcal{T}_{F}\right)<\left(\frac{\lambda_{t}}{N_{t r a}}\right)^{2} \text { and } \\
\sigma_{D}^{2}\left(\lambda_{t}, J, H, K, \mathcal{T}_{F}\right)+\sigma_{L}^{2}\left(\lambda_{t}, \mathcal{T}_{F}\right)+\sigma_{H}^{2}\left(\lambda_{t},\lambda_{o}\right)\dots\\+\sigma_{A}^{2}\left(\lambda_{o}\right)<\left(\frac{\lambda_{o}}{N_{s p e c}}\right)^{2}
\end{array},
\end{eqnarray} 
where J, H and K are the guide star magnitudes in the corresponding spectral bands and $\mathcal{T}_{{F}}$ is the frame
time that can be the optimum frame time set by the guide star magnitude, as explained in Section \ref{Optimum-time} or a fixed frame time set by the FT limits. This approach is used in Section \ref{AGN-Examp} where we discuss two examples of target lists for large AGN programs on the \textit{VLTI}.

To be able to evaluate the sky coverage for any kind of observation, with target lists yet unknown, it is practical to compute sky coverage maps that give the probability to find at least one usable guide star for any target in any region of the sky. These maps can be used to guide the preselection of science target lists and also to specify the required performances of the FT and the wide field selector, which has an impact on their design or development plan.

\subsection{Methodology\label{sec:Methodology}}

The first step is to define a database of possible guide stars. We have used the 2$^{\text{nd}}$ data release
of \textit{Gaia} survey \citep{2018A&A...616A...1G}\footnote{\url{https://gea.esac.esa.int/archive/documentation/GDR2/}} that covers astrometry, photometry, radial velocities, and astrophysical parameters of 1.692.919.135 extrasolar objects. Although, this survey does not give directly the K magnitudes, we expected it to be more complete for faint stars than the 2MASS catalogue which was initially specified to reach only magnitude $\mathrm{K}\simeq14.3$ \citep[]{2006AJ....131.1163S}, while we planned to explore precisely the domain between $\mathrm{K}=14$ and $\mathrm{K}=16.5$. Histograms of the number of stars as a function of the K magnitude showed that both the \textit{Gaia} DR2 and 2MASS list seemed complete up to K=17 for \textit{Gaia} and $\mathrm{K}\simeq16.5$ for 2MASS. Although the magnitudes of many of the targets fainter than $\mathrm{K}\simeq15$ are flagged as very uncertain in the 2MASS catalogue, it should give quite similar sky coverage maps. Finally, the justification to stay with the \textit{Gaia} data base is its astrometric quality. With the maximum \textit{VLTI} baseline of \SI{200}{\meter}, an error of \SI{1}{\arcsecond} on the position of the guide star results in an error of \SI{1}{\milli\meter} on the zero OPD on the science fringes. This is much larger than the typical low resolution coherence lengths that range from \SI{60}{\micro\meter} to \SI{300}{\micro\meter}  from $K$ to $N$. Scanning to find the science fringes is problematic on faint targets. The fringe tracking specification of \SI{10}{\micro\meter}, to be sure to be within the coherence lenght, corresponds to an astrometric precision of \SI{10}{milli\arcsecond}, which is achieved by \textit{Gaia} even on the faintest targets \citep{2018A&A...616A...2L} but well below the nearly \SI{1}{\arcsecond} 2MASS astrometric precision. Moreover, the \SI{1}{milli\arcsecond} \textit{Gaia} astrometric precision would allow a science instrument astrometry of \SI{1}{milliarcsecond} that would allow the registration of the images obtained in the different science spectral bands with the resolution of the $J$ band. \textit{Gaia} observed 1.381.964.755 sources in the blue colour  index ($G_{BP}$-band: \SIrange{330}{680}{\nano\meter}) and 1.383.551.713 objects in the red colour  indexes ($ G_{RP} $ - band: \SIrange{630}{1050}{\nano\meter}). To switch from \textit{Gaia} bands ($G_{RP}$ and $G_{BP}$) to $K$,  we used the 2MASS quadratic
photometry relationship from \cite{2018A&A...616A...4E}, as follows:
\begin{eqnarray}
\mathrm{G-K}&=&-0.1885+2.092 \cdot\left(\mathrm{G}_{\mathrm{BP}}-\mathrm{G}_{\mathrm{RP}}\right)\nonumber\\&&-0.1345 \cdot\left(\mathrm{G}_{\mathrm{BP}}-\mathrm{G}_{\mathrm{RP}}\right)^{2}.
\end{eqnarray}

This formula is valid only if the difference between the blue and red colour  index is between $0.25$ and $5.5$ ($0.25<\mathrm{G_{BP}-G_{RP}}<5.5$). This validity condition also sets a high probability to be observing a source without too strong IR excesses that would compromise the K magnitude estimate. We selected sources that have a magnitude $\mathrm{K}\leq 17$, which is one magnitude above the best limit of our FTs. 

This leaves us with a database of 814.728.004 targets as an input for our sky coverage maps (including 498.346.813 targets brighter than $\mathrm{K}\leq16$). Note that we do not have a guarantee that all these sources are good calibrators for interferometric observations as we cannot be sure that they are single unresolved stars. However, an overwhelming majority should be good enough for fringe tracking. In the range of magnitudes $\mathrm{K}$ from 10 to 16 that are relevant to our study, it is extremely unlikely that sources without strong IR excess are resolved enough to lower the contrast of fringes
on \SI{100}{\meter} baselines. Even a supergiant like Betelgeuse would have an angular size smaller than 0.1 milliarcsecond (mas) if it is far enough to have an observed magnitude $\mathrm{K}>10$. Stars brighter than $\mathrm{K}\sim 5$ could start to be resolved, but their number is negligible (and many of them are already out of the \textit{Gaia} observation list to avoid saturating its detectors). A small number of binary stars would not be suitable for fringe tracking, but they should be eliminated only if the separation is smaller than the single aperture Airy radius in $K$ (\SI{50}{mas} for D=\SI{8}{\meter}) and the magnitude difference is smaller than 1.
This would eliminate a very small fraction of the candidates, without significantly changing the sky coverage percentages.

Our database contains only the $K$ band magnitude. As our HFT options uses also some flux in $ H $ and in $ J $, we have used fixed magnitude differences that are the mean values for our reduced database:
\begin{equation}
H-K=0.1\quad \text{and } \quad J-K=0.5.
\end{equation}

In future versions of our sky coverage map calculator, we will implement a database containing magnitude estimates in the other bands. They will be necessary to validate specific guide stars but should have a very marginal impact on sky coverage maps.

To estimate the probability to find a guide star in any position of the sky, independently from any specific target list, we divide the sky in sampling pixels of $\left(\SI{10}{\arcsecond}\right)^2$ of fixed surface whatever the declination (i.e. the sampling step in right ascension is $(\frac{\SI{10}{\arcsecond}}{\cos{(\delta)}})^2$. This sampling step is set by the limits of the RAM available to compute the maps. As the stellar density does not evolve on much smaller angular scales except in very special areas of the sky, we estimate that it is more than sufficient for the general sky coverage maps that we are considering here, and indeed a sampling of $\left(\SI{30}{\arcsecond}\right)^2$ gives very similar results. Then we examine each one of the sources in our database. From their $K$ magnitude, we define an optimum frame time $\mathcal{T}_{{FO}}$ as in Section \ref{Optimum-time}. If there is a set frame time $\mathcal{T}_{{FS}}$, we use the optimum frame time only if it is smaller than the set time:
\begin{equation*}
\mathcal{T}_{F}=\min \left(\mathcal{T}_{F O}, \mathcal{T}_{F S}\right).
\end{equation*}

Then we have $\sigma_{F}^{2}\left(\lambda_{t}, J, H, K, \mathcal{T}_{F}\right)=\sigma_{D}^{2}\left(\lambda_{t}, J, H, K, \mathcal{T}_{F}\right)+\sigma_{L}^{2}\left(\lambda_{t}, \mathcal{T}_{F}\right)$. If
$\sigma_{F}^{2}\left(\lambda_{t}, J, H, K, \mathcal{T}_{F}\right)>\left(\frac{\lambda_{t}}{N_{t r a}}\right)^{2}$, the tracking noise is too large and this star does not contribute to the sky coverage. If it could contribute, we examine all the pixels in its vicinity within a radius $\theta_{P}(\lambda_{o})$. From the angular distance $\theta$ between the source and the centre of the sky map pixel, we compute $\sigma_{A}(\lambda_{o})$. Then, if $\lambda_{t}\neq\lambda_{o}$, we add the fixed $\sigma_{A}(\lambda_o)^2$ defined in Section \ref{Chrom-OPD}. If the resulting total variance from equation \eqref{equ:ST:obs:tra} $\sigma_{T}^{2}<\left(\frac{\lambda_{o}}{N_{s p e c}}\right)^{2}$ we set the examined pixel to 1, which means that there is at least one guide star for a target in that pixel. Finally, the sky coverage map is given in larger pixels of $(\SI{0.5}{\degree}^2$ by dividing the number of $(\SI{10}{\arcsecond})^2$ pixels set to 1 by the total number of  $(\SI{10}{\arcsecond})^2$ pixels in that $(\SI{0.5}{\degree}^2$ pixel, which gives the probability for a target randomly placed in that sky map pixel to find at least one usable off-axis guide star.

We have investigated the following sites and telescopes:
\begin{itemize}
	\setlength{\itemindent}{0.8cm}
\item[(i)]  8 m telescopes (\textit{UTs}) at \textit{Paranal} Observatory.
\begin{itemize}
	\setlength{\itemindent}{0.8cm}
\item[(a)] GFT and HFT in a \textit{GRAVITY}+ context, for observations in the $ K $ band.
\item[(b)] GFT and HFT in a \textit{GRAVITY}+ context, for observations in the $ N $ band with
\textit{MATISSE}.
\end{itemize}
\item[(ii)]   1.8 m telescopes (\textit{ATs}) at \textit{Paranal} Observatory.
\begin{itemize}
	\setlength{\itemindent}{0.8cm}
\item[(a)] GFT and HFT for observations in the $ K $ band.
\item[(b)] GFT and HFT for observations with \textit{MATISSE} in the $ N $ band.
\end{itemize}
\item[(iii)]  PFI with 1.8 m telescopes at a site like \textit{Paranal}.
\begin{itemize}
	\setlength{\itemindent}{0.8cm}
\item[(a)] HFT for observations in the $ N $ band.
\end{itemize}
\item[(iv)]  PFI with 1.8 m telescopes at Dome C in Antarctica, assuming that the telescopes are
above the 30 m ground layer.
\begin{itemize}\setlength{\itemindent}{0.8cm}
\item[(a)] HFT for observations in the $ K $ and $ N $ bands.
\end{itemize}
\end{itemize}

We have considered median seeing conditions as well as the seeing conditions that occur in the best 20 per cent of the time. The corresponding parameters of these sites are given in \tabref{tab:YT1}.

\begin{figure*}
	\centering
	\begin{tabular}{m{0.88\textwidth} m{0.05\textwidth} m{0.07\textwidth}}
		\subfigure[Sky coverage for the GFT on \textit{UTs} for an optimum exposure time and the 20 per cent best seeing conditions.\label{fig:SC1a}]{	\includegraphics[width=0.88\textwidth]{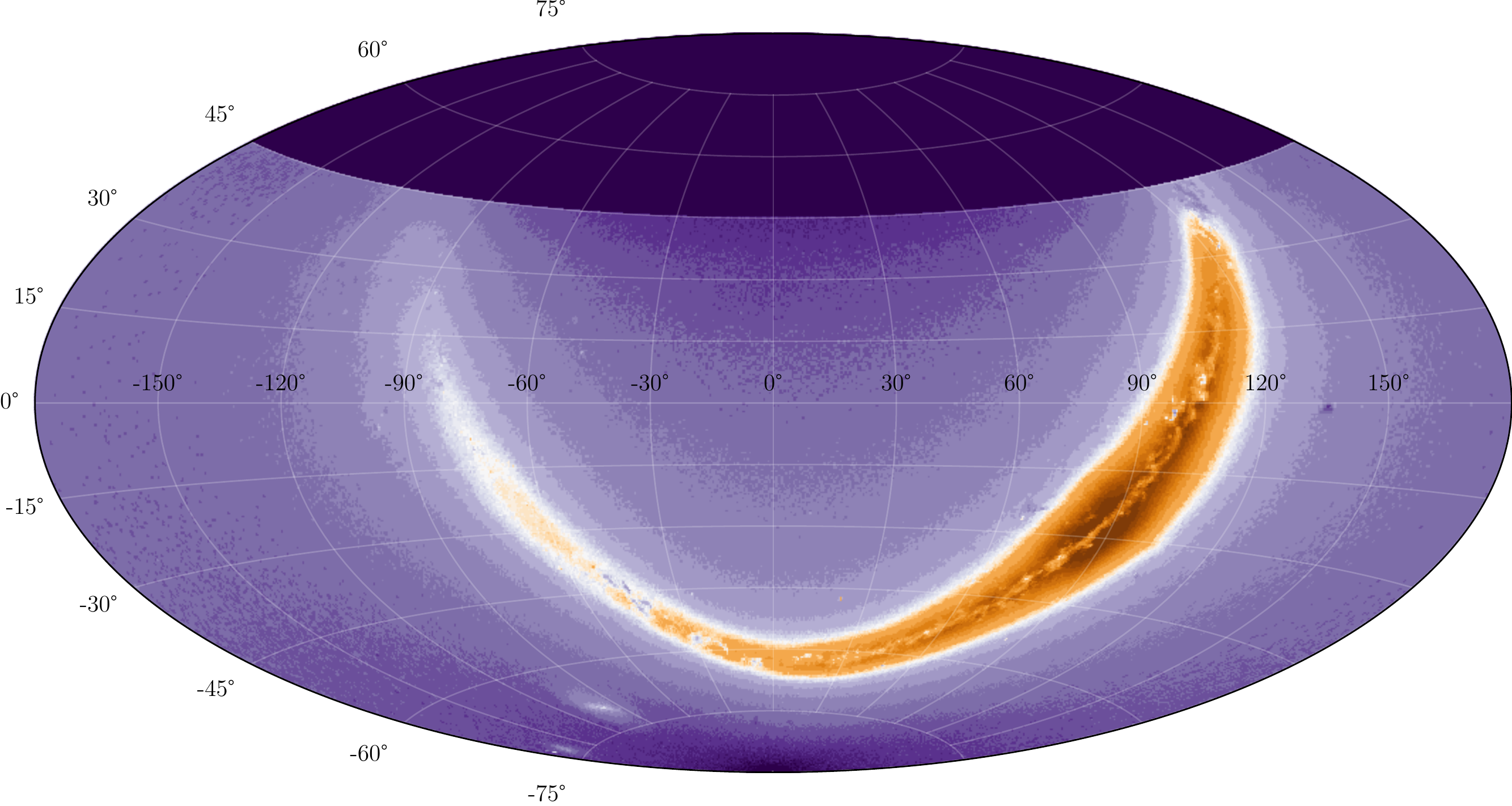}}
		
		\subfigure[Sky coverage difference between optimum and 10 ms exposure time (best 20 per cent seeing).\label{fig:SC1b}]{
			\includegraphics[width=0.44\textwidth]{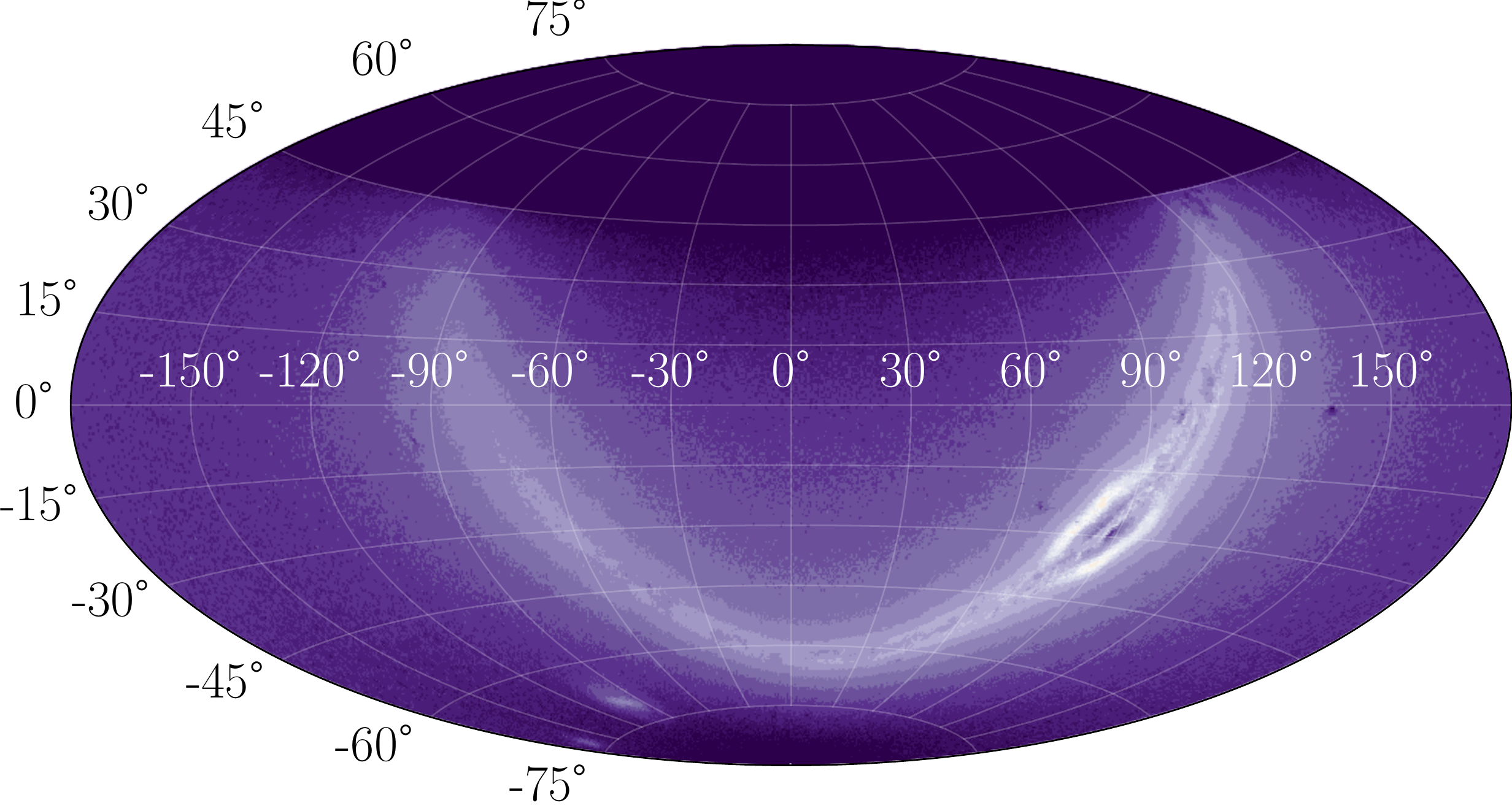}}
		\subfigure[Sky coverage difference between 20 per cent best and median seeing (optimum exposure time).\label{fig:SC1c}]{
			\includegraphics[width=0.44\textwidth]{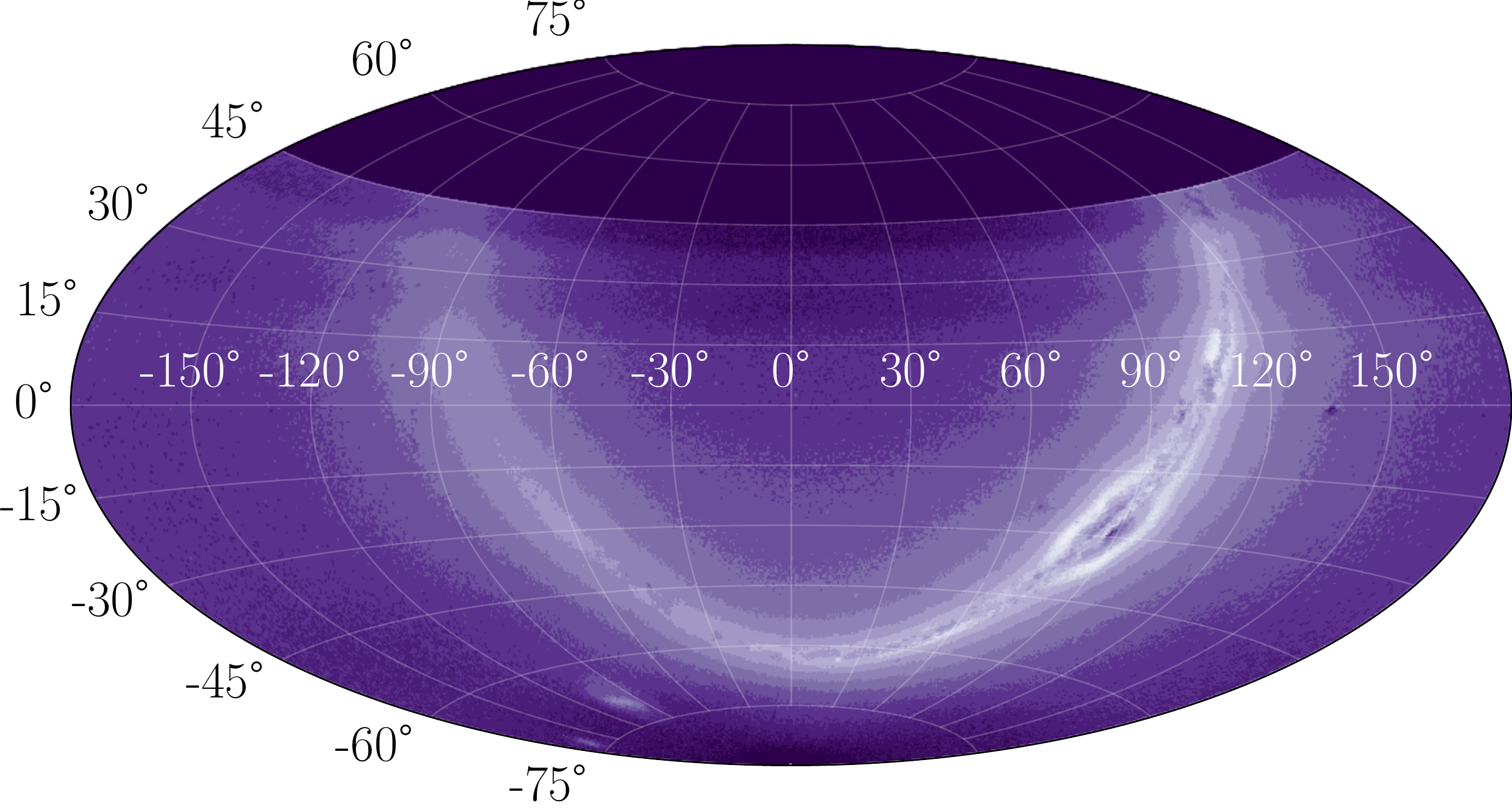}}
		&\includegraphics[width=0.07\textwidth]{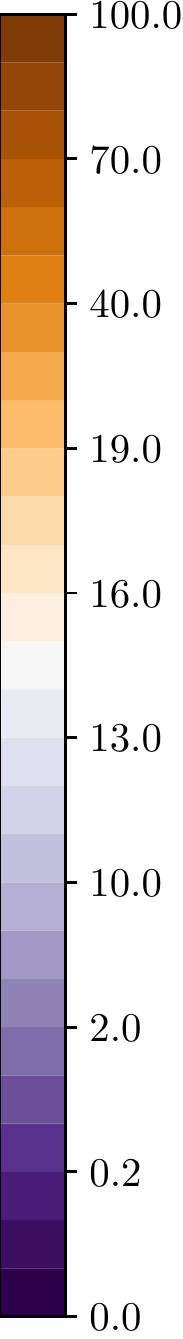}&\rotatebox{-90}{\hspace*{-1.5cm}Sky Coverage}\\
	\end{tabular}
	\caption{Sky Coverage expected with the \textit{GRAVITY+} upgrade and the current \textit{GRAVITY} FT tracker. The top image shows the sky coverage with an optimum exposure time (that can reach 30 ms) in the best 20 per cent seeing conditions. The sky coverage remains limited far from the Galactic Plane as it drops below 10 per cent at $\pm\SI{30}{\degree}$ of the Galactic Plane in the best case. However, this is sufficient to undertake a large AGN program as we will see  in Section \ref{sec:QSO-list}. The two	lower images show the impact of a frame time limited to \SI{10}{\milli\second} (left) and a seeing in median conditions (right). The impact is significant without radically changing the sky coverage.}
	\label{fig:SC1}
\end{figure*}

\subsection{{VLTI} with {UTs}\label{sec:VLTI-UTS}}
\figref{fig:SC1a} shows the sky coverage that should be obtained with the GFT for observations in the $ K $ band, with the seeing expected in the best 20 per cent conditions. The sky coverage remains limited far from the Galactic Plane as it drops below 10 per cent at $\pm\sim\SI{30}{\degree}$ of the Galactic Plane. However, this is sufficient for a program based on a large number of sources evenly spread over the sky, as we will see it for a large AGN program in Section
\ref{sec:QSO-list} below.
	
\figref{fig:SC1b} shows the difference between observations with the optimum frame time, which can reach \SI{30}{\milli\second} (see \figref{fig:optimum-time}), and observations with a maximum frame time set at \SI{10}{\milli\second}. The gain offered by the optimum frame time is significant as it extends the 10 per cent limit by about \SI{10}{\degree} away from the Galactic Plane. \figref{fig:SC1c} shows the difference between the sky coverage in the best 20 per cent seeing conditions and in median seeing conditions. The seeing conditions do not change the general shape of the sky coverage, but should be considered carefully for targets with a guide star at the limit of the performances

\begin{figure*}
	\centering
	\begin{tabular}{m{0.88\textwidth} m{0.05\textwidth} m{0.07\textwidth}}
	\includegraphics[width=0.88\textwidth]{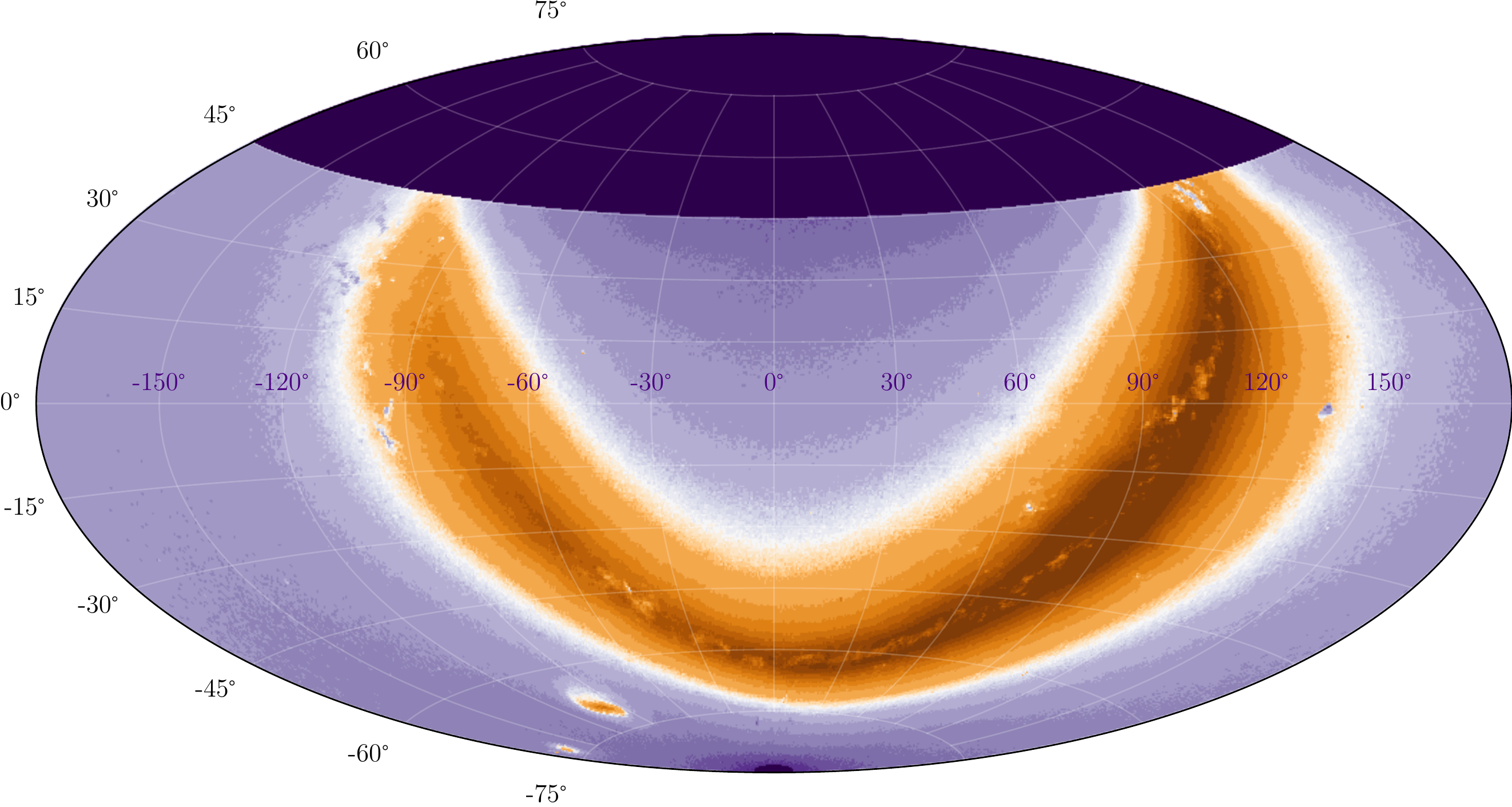}
		&\includegraphics[width=0.06\textwidth]{fig/bar}&\rotatebox{-90}{\hspace*{-1.4cm}Sky Coverage}\\
	\end{tabular}
	\caption{Sky coverage in the best 20 per cent seeing conditions and optimum exposure time with the new generation HFT
		 evaluated in Section \ref{sec:HFT}. This figure illustrates the strong gain in sky coverage allowed by a gain of two magnitudes on the fringe tracking sensitivity limits.}
	\label{fig:SC2}
\end{figure*}

\figref{fig:SC2} shows the sky coverage with the new generation FT evaluated in Section \ref{sec:HFT}. The sky coverage improvement offered by the gain of two magnitudes on the tracking limit is quite substantial, with a sky coverage larger than 10 per cent of the sky observable from \textit{Paranal}.

\begin{figure*}
	\centering
	\begin{tabular}{m{0.88\textwidth} m{0.05\textwidth} m{0.07\textwidth}}
		\subfigure[GFT for \textit{MATISSE} - max separation = 1 arcminute. \label{fig:SC3a}]{\includegraphics[width=0.44\textwidth]{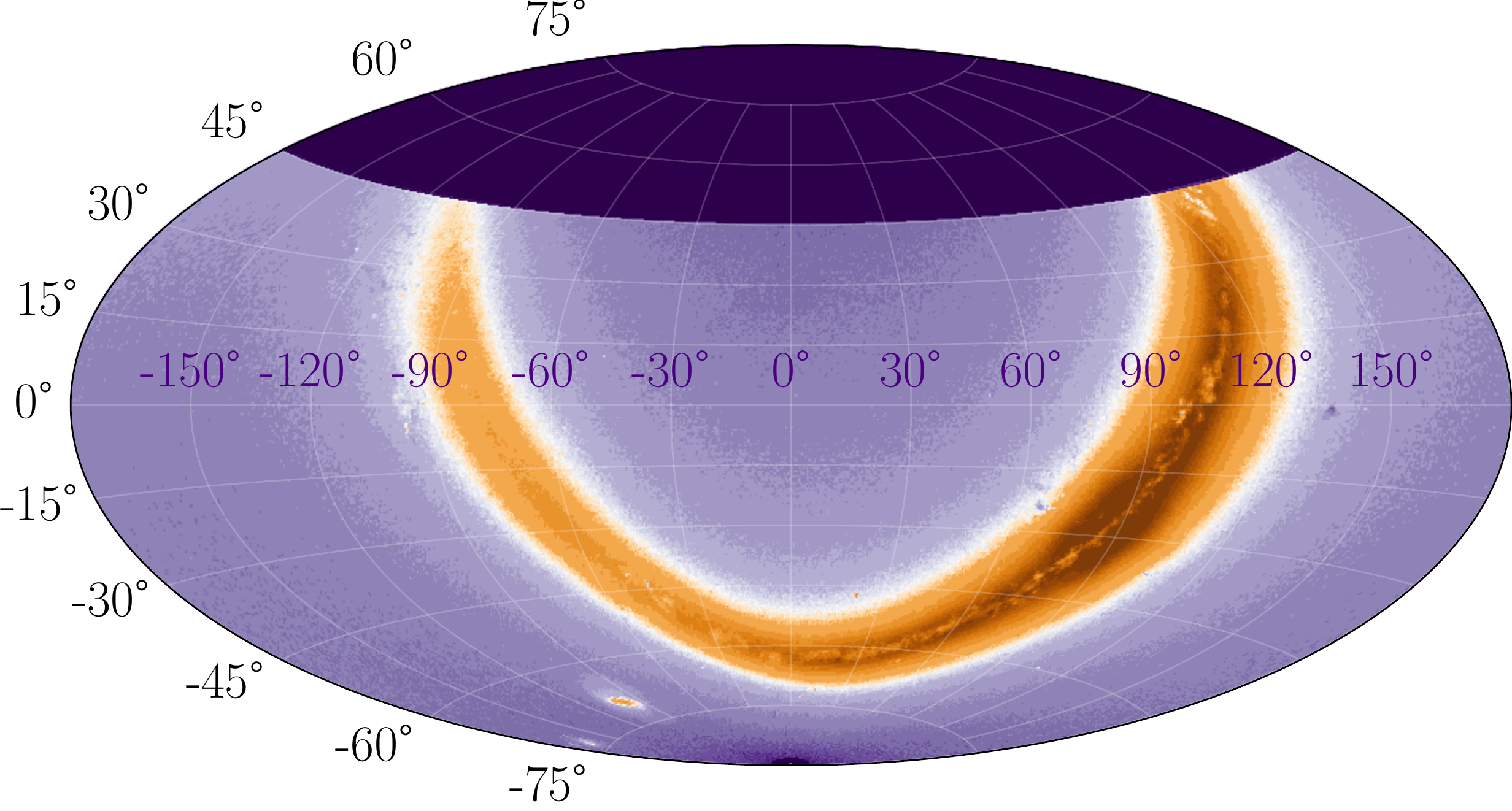}}
		\subfigure[HFT for \textit{MATISSE} - max separation = 1 arcminute. \label{fig:SC3b}]{\includegraphics[width=0.44\textwidth]{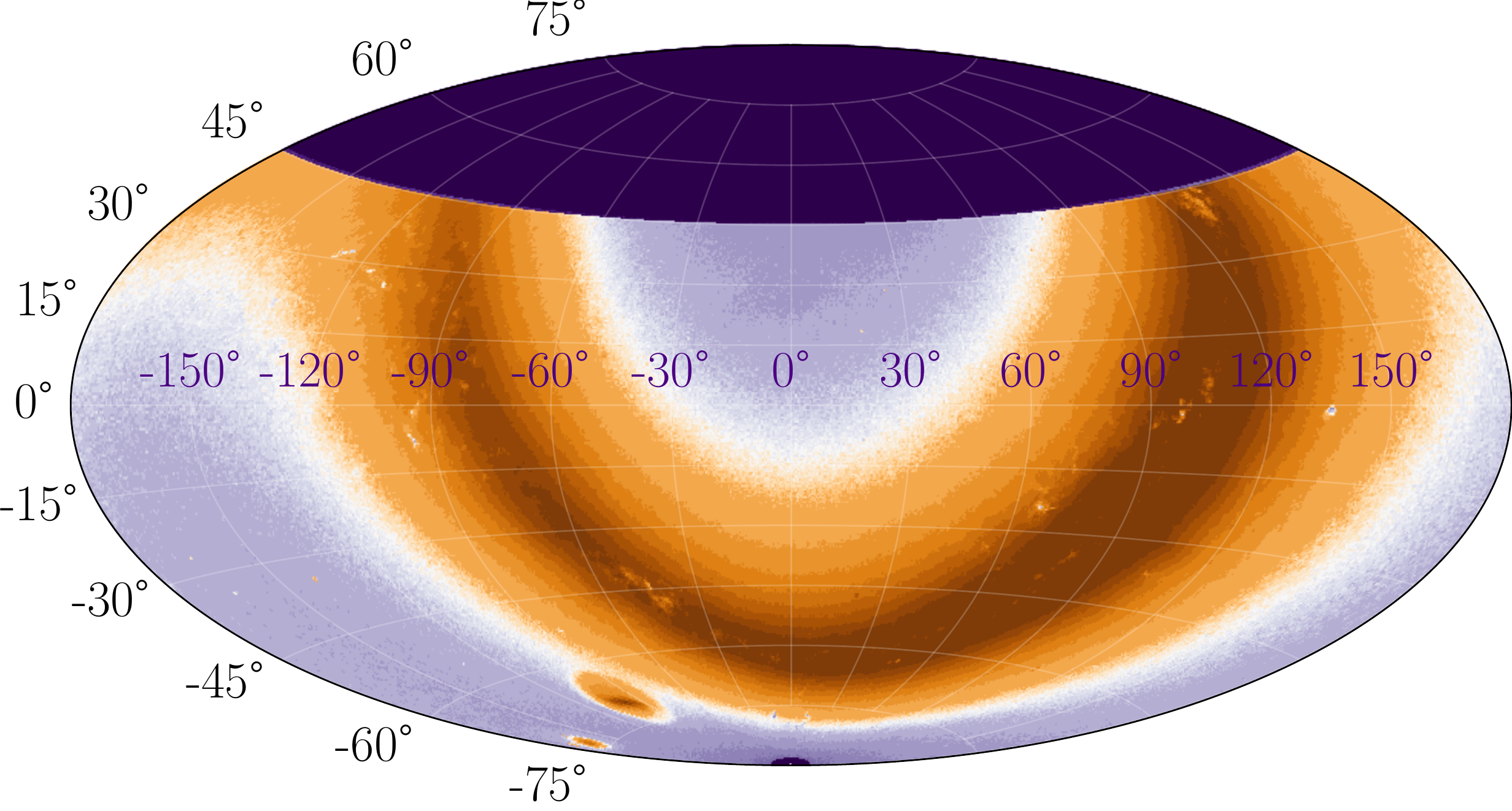}}
		\subfigure[GFT for \textit{MATISSE} - max separation = 2 arcminutes. \label{fig:SC3c}]{\includegraphics[width=0.44\textwidth]{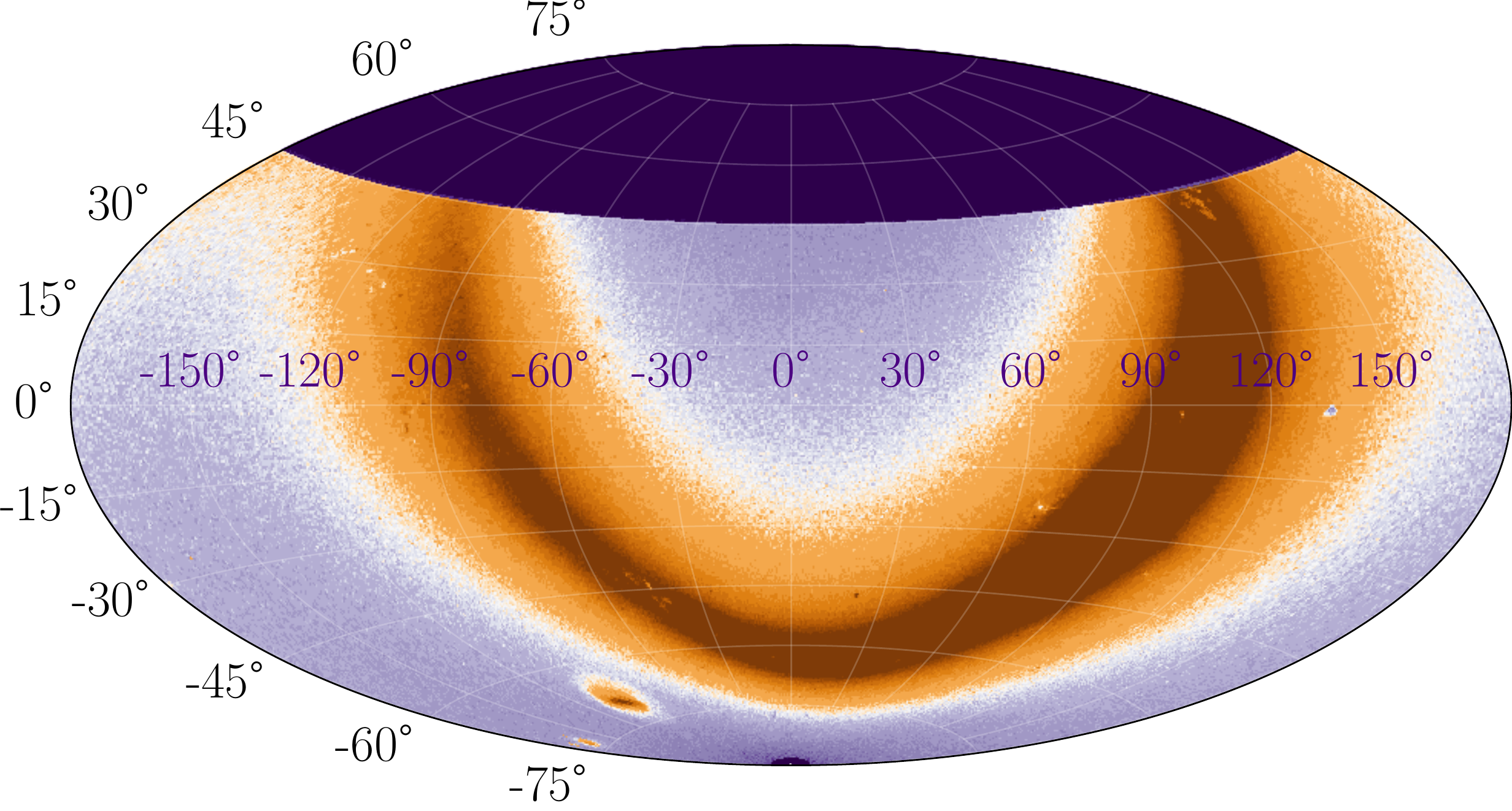}}
		\subfigure[HFT for \textit{MATISSE} - max separation = 2 arcminutes. \label{fig:SC3d}]{\includegraphics[width=0.44\textwidth]{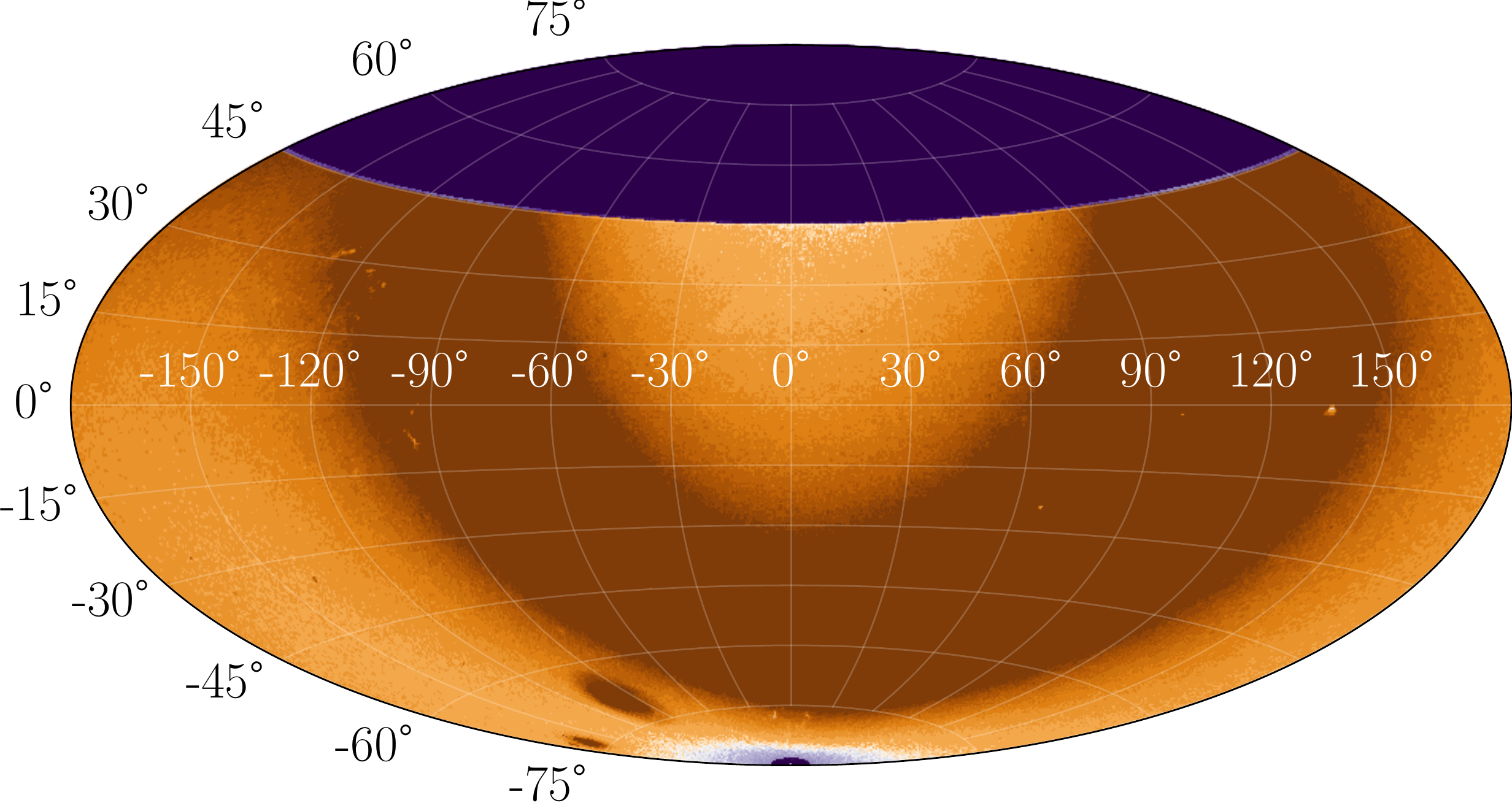}}
		&\includegraphics[width=0.06\textwidth]{fig/bar}&\rotatebox{-90}{\hspace*{-1.4cm}Sky Coverage}\\
	\end{tabular}
	\caption{Sky coverage for tracking in the $ K $ band with the GFT (top-left) and the HFT (top-right) for observations in the $ N $ band with \textit{MATISSE}. Optimum exposure time and best	20 per cent seeing. Top row: target separation limited to 1 arcminute. Bottom row: target separation limited to 2 arcminutes that remain smaller than the isopistonic angle with \textit{UTs}, which is strongly relaxed for $ N $ band observations with \textit{UTs}.}
	\label{fig:SC3}
\end{figure*}

\figref{fig:SC3} shows the sky coverage with the GFT and the HFT for observations in the $N$ band with \textit{MATISSE}, in the same conditions, with a separation between the target and the guide star limited to \SI{1}{\arcminute}. That sky coverage is only slightly better than this for $K$ band observations. However, the isopistonic angle for observations in the $N$ band with \textit{UTs} is \SI{174}{\arcsecond} at \SI{8.5}{\milli\second} in median seeing conditions and can reach \SI{230}{\arcsecond} in the
best 20 per cent conditions. Extending the off-axis range to these values yields a spectacular increase in the sky coverage as shown by Figs \ref{fig:SC3c} and \ref{fig:SC3d} where the separation is still limited to \SI{2}{\arcminute} for hardware realism.

\subsection{VLTI with {ATs}\label{sec:VLTI-ATS}}
\begin{figure}
	\centering
	\includegraphics[width=1\linewidth]{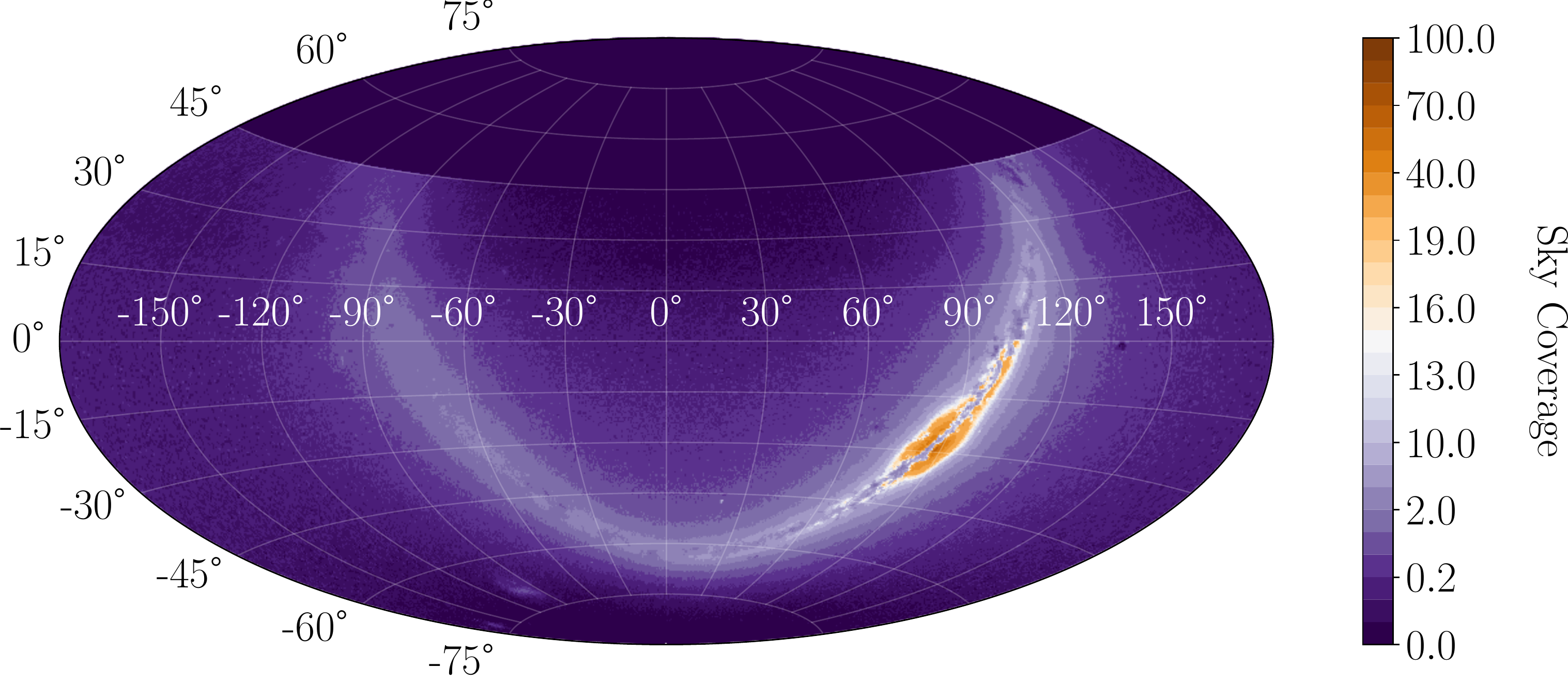}
	\caption{Best Sky coverage with an HFT on \textit{ATs} (or with a \SI{1.8}{\meter} class telescopes at a PFI interferometer in a site like \textit{Paranal}) for observations in the $K$ band. The sky coverage exceeds 10 per cent only within $\pm\SI{10}{\degree}$ around the Galactic Plane. With the GFT, the sky coverage in the same conditions is negligible.}
	\label{fig:SC4}
\end{figure}

\begin{figure}
	\centering
	\includegraphics[width=1\linewidth]{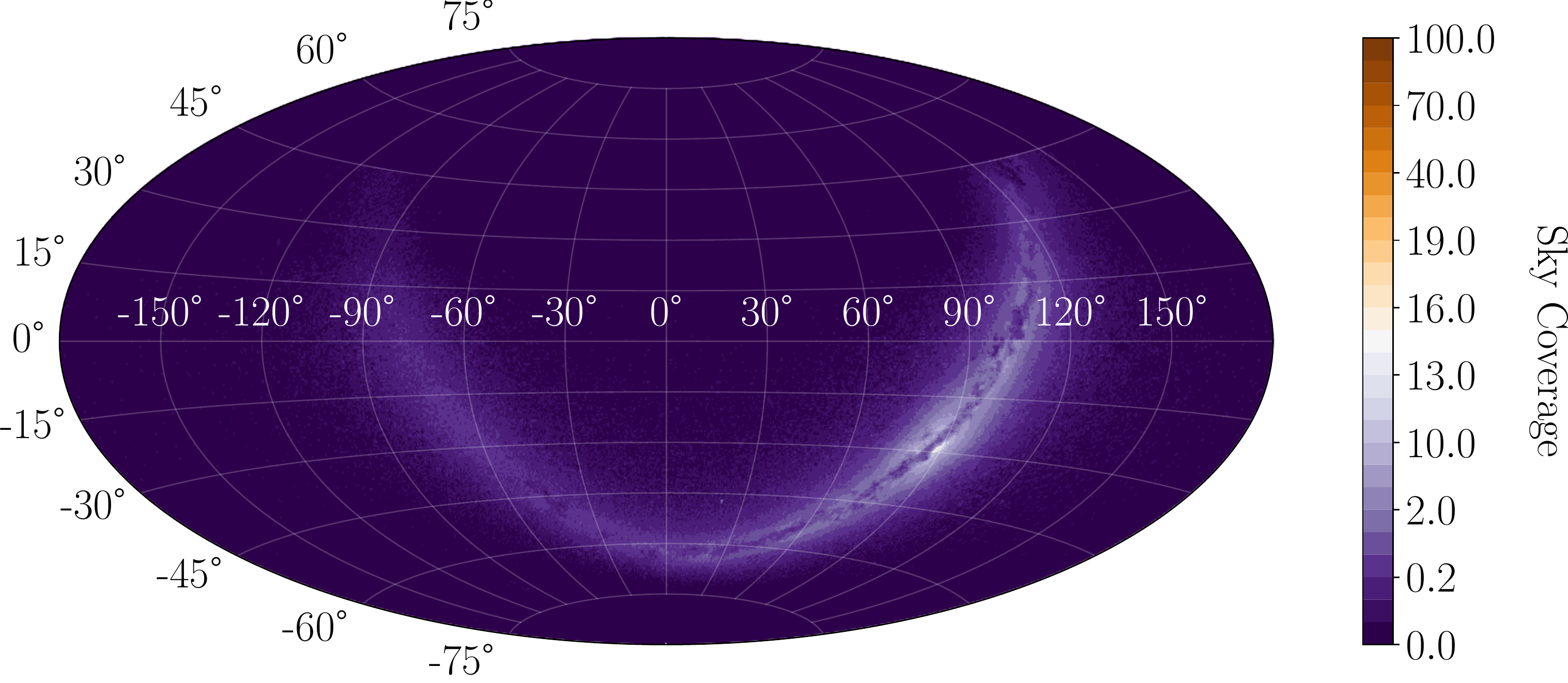}
	\caption{Best sky coverage with a GFT on \textit{ATs} for observations with \textit{MATISSE}. That sky coverage would be achievable with the GFT \gmleft as it is\gmright with a periscopic system allowing  feeding in \textit{GRAVITY} and \textit{MATISSE} two sources with separations up to \SI{1}{\arcminute}. The sky coverage might seem modest but it immediately opens 	very interesting science programs as it would allow imaging and high spectral resolution observations for already a few nearby AGNs, as shown in \figref{fig:AGN6}.}
	\label{fig:SC5}
\end{figure}

We have also investigated the sky coverage that can be expected with the \textit{ATs}. It is reduced first because the fringe tracking limiting magnitude is very substantially reduced, by 3.2 magnitudes if nothing changes except the telescope diameter. The smaller apertures also result in  faster turbulence and hence increased loop error $\sigma_{L}$ and in a much smaller isopistonic angle and increased anisopistonic error $\sigma_{A}$. As a consequence, the sky coverage with the GFT on the \textit{ATs} for observations in the $ K $ band is less than a marginal, with less than 1 per cent even in the highest guide star density. With the HFT we have values that can exceed 20 per cent near the Galactic Centre and 10 per cent almost everywhere very close to the Galactic Plane, as shown in \figref{fig:SC4}. This might be worth it for classes of programs with targets near the Galactic Plane. For observations in the $ N $ band, we have a strong increase in sky coverage, as shown by \figref{fig:SC5}. With the GFT we will have a limited but sometimes usable sky coverage near the Galactic Plane and we will see on the AGN example below that this already provides some high-interest targets to boost the image reconstruction capability of \textit{MATISSE} with the \textit{ATs} in the $N$ band. This will include YSOs that are too resolved for \gmleft on-axis\gmrightd

\begin{figure}
	\centering
	\includegraphics[width=1\linewidth]{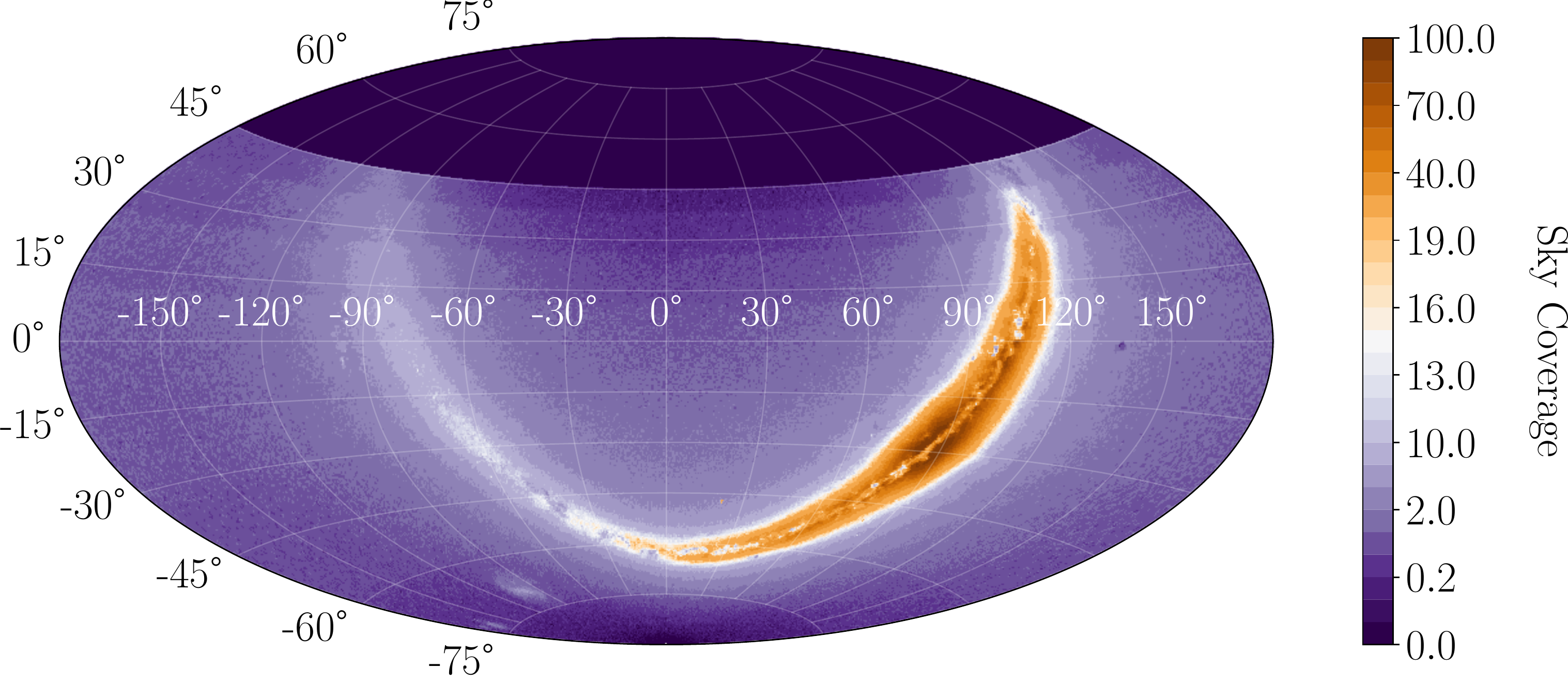}
	\caption{Best sky coverage with an HFT for observations in the $N$ band with \textit{ATs} in \textit{Paranal} or with PFI with 1.8 m apertures in a \textit{Paranal} like site. The overall transmission was assumed to be the same as with the \textit{VLTI}. In this figure as well as in Figs \ref{fig:SC4} and \ref{fig:SC5}, we assumed an optimum exposure time and the 20 per cent best seeing conditions.}
	\label{fig:SC6}
\end{figure}

\subsection{PFI with 1.8 m telescopes in a {Paranal} like site\label{sec:PFI}}

The sky coverage with an HFT for \textit{ATs} at \textit{Paranal} shown in  \figref{fig:SC4} for the $K$ band, and \figref{fig:SC6} for the $N$ band, should be extremely similar to this with a new generation interferometer, like the PFI, with any number of telescopes in a site with the same seeing conditions as \textit{Paranal}.
Indeed with an HFT, the fringe tracking performances are independent of the number of apertures, at least for unresolved targets such as the off-axis guide stars. This sky coverage for $ N $ band observations is more than sufficient for the main science goal of PFI which is the detailed imaging of protoplanetary discs. Showing that this can be achieved with \SI{1.8}{\meter} class telescopes has indeed a remarkable impact on the cost and hence the feasibility of such an array.
The preliminary studies of the PFI also showed that its overall transmission could be substantially improved with regard to the \textit{VLTI}, because PFI would be an optimized interferometer unlike the \textit{VLTI} that is an interferometric mode added to a general-purpose multitelescope observatory.

\section{Example of application: an AGN large program}
\label{AGN-Examp}

To give a more concrete example of the potential of off-axis fringe tracking we have considered two target lists for a science program on AGNs. Such a science program would include several components that are briefly described in the following with their requirements.
\subsection{Phase differential interferometry of Broad Line Regions \label{Interf-BLR}}

The spectro-astrometric study of AGN BLRs by phase differential interferometry (PDI) in emission lines, proposed by \citep{2001CRASP...2...67P} and successfully achieved by \textit{GRAVITY} \citep{2020A&A...635A..92G}, can constrain the kinematics, size, and geometry of a spatially unresolved BLR using the displacement of the AGN photocentre with wavelengths through an emission line (that is proportional to the differential phase on unresolved sources). It can be used to evaluate the mass of the SMBH in the centre of the AGN \citep{10.1093/mnras/stu2613}. Combined with Reverberation Mapping that gives an equivalent linear size of the BLR, the angular size given by differential interferometry yields a direct distance measurement of the BLR \citep{10.1093/mnras/staa2988}. A study on a large number of BLRs up to $ z=2 $ is one of the main motivations for the \textit{GRAVITY}+ program. The key condition to realize it is the ability to observe a large number of Seyfert 1 AGNs and QSOs up to magnitude $\mathrm{K}\leqslant 17$.

\textit{GRAVITY}+ intends to use the \textit{GRAVITY} science instrument and is focused on off-axis fringe tracking in the $K$ band with the GFT and science observation at medium ($ R=1500 $) or high ($ R=4500 $) spectral resolution in the $ K $ band. Observing strong emission lines in the $ K $ band requires targets with high enough redshifts to have Paschen $\alpha$, $\beta$, $\gamma$, and $\delta$, Helium I or even $H\alpha$ lines shifted in the $K$ band. High redshift means distant and that implies that bright enough targets are large and hence difficult to perform Reverberation Mapping on. A complement to the \textit{GRAVITY}+ $K$
band program would be the $J$ band instrument proposed by \cite{2019vltt.confE..37P} that allows observing the same lines with a much smaller redshift and hence on closer and smaller targets more likely to allow RM measurements. The condition for these observations are:
\begin{itemize}
	\setlength{\itemindent}{0.8cm}
	\item[(i)] To have an off-axis target for fringe tracking in the near-IR.
	\item[(ii)] A fringe tracking precision $\sigma_{T}\simeq\SI{0.2}{\micro\meter}$ for observations in the $K$ band.
	\item[(iii)] A fringe tracking precision $\sigma_{T}\simeq\SI{0.1}{\micro\meter}$ for observations in the $J$ band.
\end{itemize}

In addition to these fringe tracking requirements, we need the PDI SNR to be sufficient.
On unresolved targets (i.e. with size $\Omega$ much smaller than the resolution limit $\lambda/B$) the differential phase \textit{signal} is given by \citep{1986JOSAA...3..634P}  
\begin{equation}
\phi(\lambda)=2\pi~\epsilon(\lambda)~(B/\lambda),
\end{equation}
where $\epsilon(\lambda)$ is the photocentre variation of the source at the wavelength $\lambda$. In an emission line, the photocentre variation with regard to the nearby continuum (where the photocentre is set to zero) is $\epsilon(\lambda)=k(\lambda)\Omega$ where $k(\lambda)$ depends from the geometry of the target and the strength of the line. For emission lines much stronger than the continuum, $k_{max}\simeq0.5$. The differential phase \textit{noise} is $\sigma_{\phi}(\lambda)=\frac{1}{SNR_{C}(\lambda)\sqrt{2}}$ \citep{2020SPIE11446E..0LP} and the Phase Differential Interferometry $SNR_{PDI}$ for unresolved targets can thus be written as:
\begin{equation}
S N R_{P D I}(\lambda) \simeq \pi\sqrt{2}\dfrac{\Omega_{B L R}}{\lambda / B} S N R_{C}(\lambda).
\label{equ:AGN1}
\end{equation}

In the following, we should consider that a condition for a good spectro-astrometric information is given by:

\begin{equation}
 \min \left[\frac{\Omega_{B L R}}{\lambda / B}, 1\right]S N R_{C}(\lambda)>10,
\label{equ:AGN1b}
\end{equation}

The term $ \min \left[\frac{\Omega_{B L R}}{\lambda / B}, 1\right] $   expresses the fact that the differential phases stop increasing with the size $\Omega$ on resolved objects. The criteria given here correspond to an error on the BLR angular size of $\simeq1\mathrm{\ per\ cent}$ in the case of thin flat Keplerian BLR observed in a strong emission line \citep{10.1093/mnras/stu2613}.

\subsection{Amplitude differential interferometry of BLRs}

The interpretation of spectro-interferometric measures alone as well as combined with Reverberation
Mapping alone is strongly model dependent and different BLR geometries can give very different SMBH masses and distance estimates from the same measurements as shown by \cite{10.1093/mnras/stu2613}. The same paper shows that amplitude differential interferometry (ADI), i.e. the variations of the amplitude of the coherent flux, or of the visibility modulus through the line very strongly constrains and improves the mass and distance estimates. On unresolved targets the differential visibility drop $1-V(\lambda)$ is proportional to $\left(\frac{\Omega}{\lambda/B}\right)^2$ and the $ SNR_{AD1} $ can be estimated by:
\begin{equation}
\operatorname{SNR}_{A D I}(\lambda) \simeq \min \left[\left(\frac{\Omega_{B L R}}{\lambda / B}\right)^{2}, 1\right] \operatorname{SNR}_{C(\lambda)}>10,
\label{equ:AGN2}
\end{equation}

The condition $SNR_{A D I}(\lambda) \simeq10$ corresponds for example to a precision of $\simeq\SI{6}{\degree}$ on the inclination of a thin flat BLR.

\subsection{Imaging of broad line regions \label{Img-BLR}}

To definitively validate mass and distance estimates by PDI, even completed by ADI, we need to reconstruct images of BLRs on
a sample of sources to solve all possible geometric ambiguities. The largest BLRs are expected
to have angular sizes of the order of \SI{1}{milliarcsecond} or smaller and this is confirmed by the first successful measurements by \textit{GRAVITY}. Imaging these structures would require baselines
of \SI{300}{\meter} in the visible, \SI{600}{\meter}  in the $J$ band, and more than \SI{1}{\kilo\meter} in the $K$ band  with tracking quality of respectively \SI{0.06}{\micro\meter}, \SI{0.01}{\micro\meter}, and \SI{0.02}{\micro\meter}. This is out of reach for any existing optical interferometer. The \textit{VLTI's} current and final maximum baselines are \SI{135}{} and \SI{200}{\meter} and the CHARA interferometer, which would have a sufficient resolution in the visible with its baselines up to 330 m, cannot be sensitive enough with its \SI{1}{\meter} apertures. 
In Section \ref{sec:five} below we will discuss with some more
details the sensitivity conditions for an optical interferometer to make
images of BLRs.


In optical interferometry, the key condition to make images is to obtain closure phase measurements. The closure phase decreases as the third power of the ratio $\frac{\Omega}{\lambda/B}$ and the condition to have a closure phase usable for image reconstruction becomes a condition on the closure phase $SNR_{CP}$:
\begin{equation}
S N R_{C P}(\lambda) \simeq \min \left[\left(\frac{\Omega_{B L R}}{\lambda / B}\right)^{3}, 1\right] \operatorname{SNR}_{C(\lambda)}>10.
\label{equ:AGN3}
\end{equation}

The possibility to retrieve images from unresolved targets is very limited.

\subsection{Constraining the structure of the dust torus with \textit{MATISSE}}

A key component of the unified model of AGNs \citep{1993ARA&A..31..473A} is the dust torus that obscures the central emission source in type 2 AGNs, where it is seen \gmleft edge-on\gmrightc\ and does not in type 1 AGNs, where it is sufficiently \gmleft face-on\gmrightd\   Optical Interferometry in the mid-IR with \textit{MIDI} \citep{2013A&A...558A.149B} complicated the long held image of a simple equatorial \gmleft torus\gmrightc\ which is now thought to be clumpy and with polar components probably blown away by radiation pressure (the dusty wind) as well as other associated inflows and outflows. The complexity of that structure is confirmed by the first observations of AGN dust tori by \textit{GRAVITY} \citep{2017Msngr.170...10G} and \textit{MATISSE}. Understanding the structure of the dust in a large sample of AGNs with a large range of luminosities would help constrain and complement the BLR studies described in the previous section by explaining how the accretion disc around the central SMBH exchanges material with the galaxy. IR Reverberation
Mapping, which gives linear dust torus sizes, combined with the angular measures of optical interferometry yields direct distance measurements as demonstrated by \cite{2014Natur.515..528H}. But like for BLRs, the precision of these distance measurements strongly depends on the geometry of the torus. The two topics are strongly correlated, as the geometry of the gas in the BLR might be a prolongation of this of the surrounding dust.

One of the primary science goals of \textit{MATISSE} \citep{2014Msngr.157....5L} is the study of that innermost dust \gmleft torus\gmright in AGNs. In spite of the complexity revealed by optical interferometry, the inner part of the dust structure is likely to be dominated by an inclined (and more or less skewed) ring, complemented by a dusty biconical outflow, or \gmleft wind \gmrightc with a strength related to the central source luminosity and hence SMBH mass. In the $K$ and $L$ band, we would be dominated by the bright inner rim near or shortly behind the dust sublimation radius; at longer wavelengths we see colder dust.
\textit{MATISSE}, alone or assisted by the GFT, will be able to observe a few dozen AGN tori and will obtain images of 2 or 3 nearby AGNs. However, that number of targets, as well as the range of available physical conditions in the AGN, is very strongly limited by the current sensitivity limits of the \textit{VLTI} AO and GFT. So the \textit{MATISSE} scientific potential will be very strongly enhanced by the \textit{GRAVITY}+ upgrade.

Similarly to the BLRs, the dust torus observations of \textit{MATISSE} can be classed in the three categories \gmleft imaging\gmrightc \gmleft ADI\gmrightc\ and \gmleft PDI\gmright according to the ratio size over resolution $\frac{\Omega}{\lambda/B}$ and equations (\ref{equ:AGN1b}, \ref{equ:AGN2}, and \ref{equ:AGN3}) apply by changing the wavelength domain and replacing the size of the BLR $\Omega_{B L R}$ by this of the torus
$\Omega_T(\lambda)$. A key difference is that the torus is much larger than the BLR and imaging, i.e. $\Omega_T(\lambda)>\frac{\lambda}{B}$ applies on much large number of targets. In the $ K $ band, we will be dominated by the inner rim of the dust. \cite{2006ApJ...639...46S} finds that the infrared reverberation mapping size is very close to the dust sublimation radius which is proportional to the square root of the bolometric luminosity. \cite{2011A&A...527A.121K} find an interferometric size in the $K$ band just slightly larger, with an excess that seems loosely correlated to the luminosity. The apparent size of the torus changes with the wavelength of observation and can reach a $\lambda^2$ dependence for optically thin dust directly illuminated by the central source. The overview of the observation of 23 AGNs
by \textit{MIDI} by \cite{2013A&A...558A.149B} reveals a $\frac{\Omega_T(\lambda_{N})}{\Omega_T(\lambda_{K})}$ size ratio that ranges
from $\frac{\lambda_{N}}{\lambda_{K}}$ to $\left(\frac{\lambda_{N}}{\lambda_{K}}\right)^2$ with a median value of $\frac{\Omega_T(\lambda_{N})}{\Omega_T(\lambda_{K})}=9$. We will use these $\Omega_T(\lambda_{K})\propto L^{1/2}$ and $\frac{\Omega_T(\lambda_{N})}{\Omega_T(\lambda_{K})}=9$ relations to check what kind of observations \textit{MATISSE} can obtain on some of the AGNs that are considered in the next sections.

\subsection{Large list of Quasars\label{sec:QSO-list}}

We have considered two lists of AGNs to investigate various aspects of the potential of
\textit{GRAVITY} and \textit{MATISSE} observations in the \textit{GRAVITY}+ context with off-axis tracking
either by the GFT or by the HFT.

\begin{figure}
	\centering
	\includegraphics[width=1\linewidth]{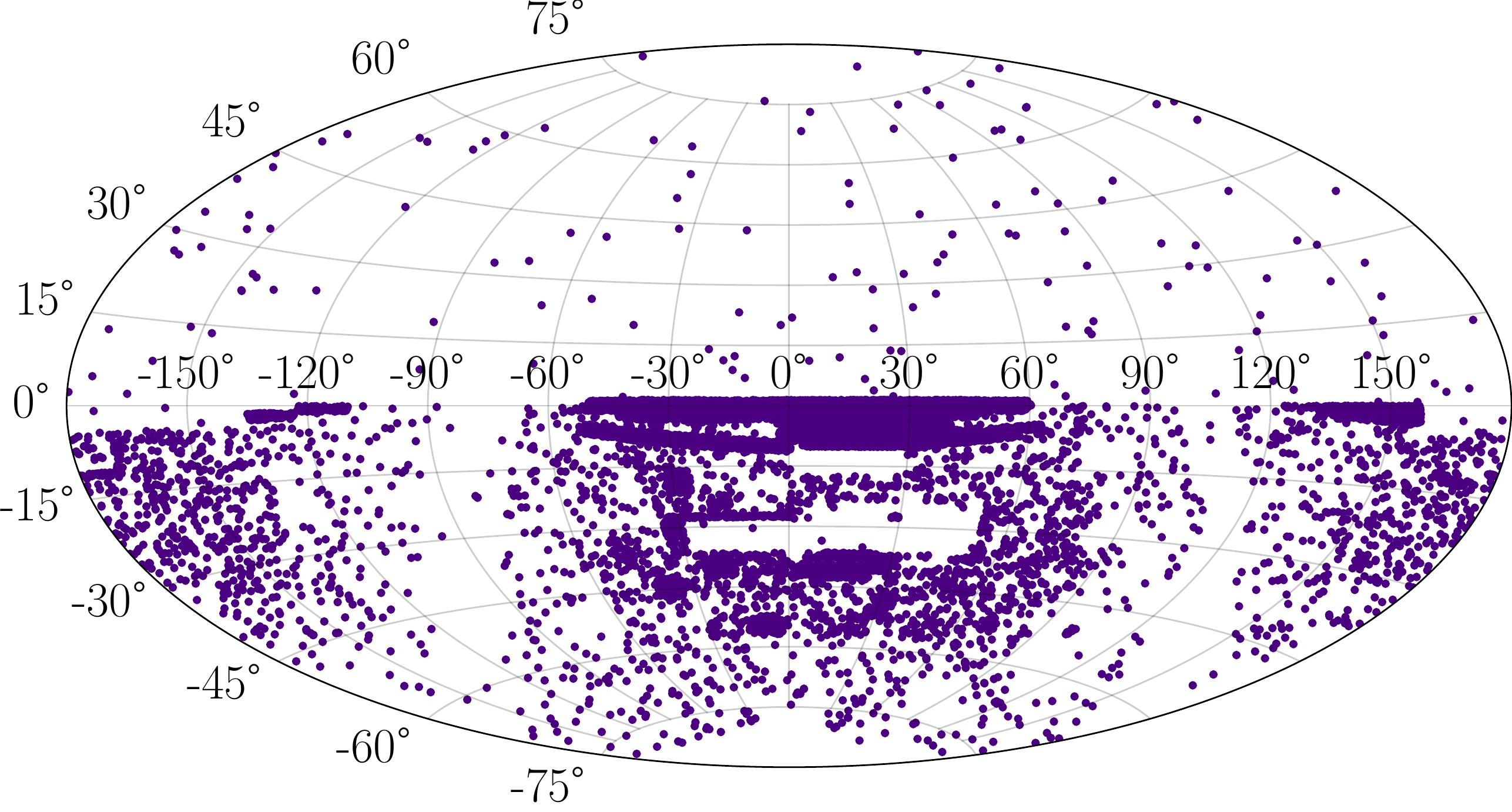}
	\caption{The list of 15799 QSOs used to
		evaluate the potential of the overall Quasar program.
		Note the very strong survey	biases\label{fig:AGN1}}
\end{figure}

The first list contains 15799 Quasars observable with the \textit{VLTI}. We utilized the \textsc{python}
packages \textsc{astroquery} \cite[]{2019AJ....157...98G}, \textsc{astropy} \cite[]{2018AJ....156..123A}, and \textsc{numpy} \cite[]{harris2020array}. We queried Simbad astronomical database \citep[]{2000A&AS..143....9W} for
every object classed as a QSO. We selected the QSO classification because this reduces the relative contamination from the host galaxy, minimizing the difference between the catalogue $K$ magnitude and this of the actual central source that is seen by the \textit{VLTI}. Our list is then a subset of the AGNs observable from the \textit{VLTI} but a fraction of targets in this list will be eventually fainter than K=17.  
We crossmatched these QSO coordinates with the VISTA Hemisphere Survey \citep[VHS;][]{2013Msngr.154...35M}. VHS has a Ks limiting magnitude of 18.1 making it appropriate for this study that we restricted it to $K\leqslant17$. From VHS, we found the $ Ks $ band magnitude using an aperture of \SI{1}{\arcsecond} and removing any objects with a Ks magnitude greater than 17. The VHS is almost strictly limited to targets with negative declination and misses all targets in the Northern hemisphere that would be observable from \textit{Paranal}. It also seems to contain some other very strong survey biases, as
shown by \figref{fig:AGN1}, but it provides a large number of potential candidates distributed all over the southern hemisphere and we considered it as a usable input to test the number of potential candidates for various types of observations. We have to note that there is a relative deficit of targets near the Galactic Plane and a massive over-representation of targets near declination zero.

The targets for which we find a guide star are labelled as a function of the quality of tracking
offered by the best guide star that minimizes the error budget of equations (\ref{eq:6} and \ref{equ:ST:obs:tra}). Tracking
qualities of  \SI{0.2}{\micro\meter},  \SI{0.1}{\micro\meter} and  \SI{0.06}{\micro\meter} allow observations in the $ K $, $ J $ and $ R $ bands respectively, while tracking qualities of  \SI{0.85}{\micro\meter},  \SI{0.42}{\micro\meter} and  \SI{0.28}{\micro\meter} are good for observations in $N$, $M$ and $L$ respectively.

\begin{figure*}
	\centering
	\subfigure[\label{fig:AGN2a}]{\includegraphics[width=0.48\textwidth]{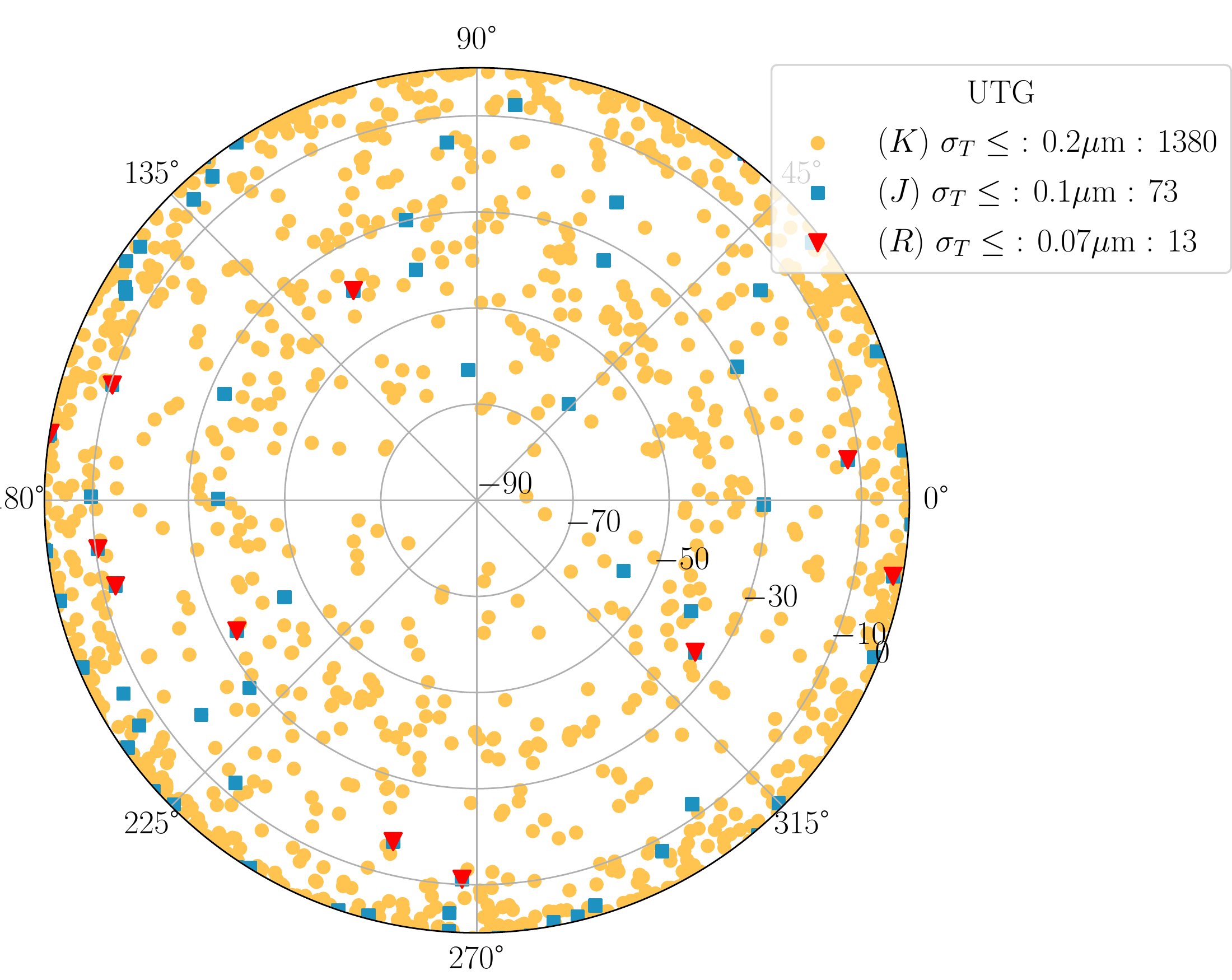}}
	\subfigure[\label{fig:AGN2b}]{\includegraphics[width=0.48\textwidth]{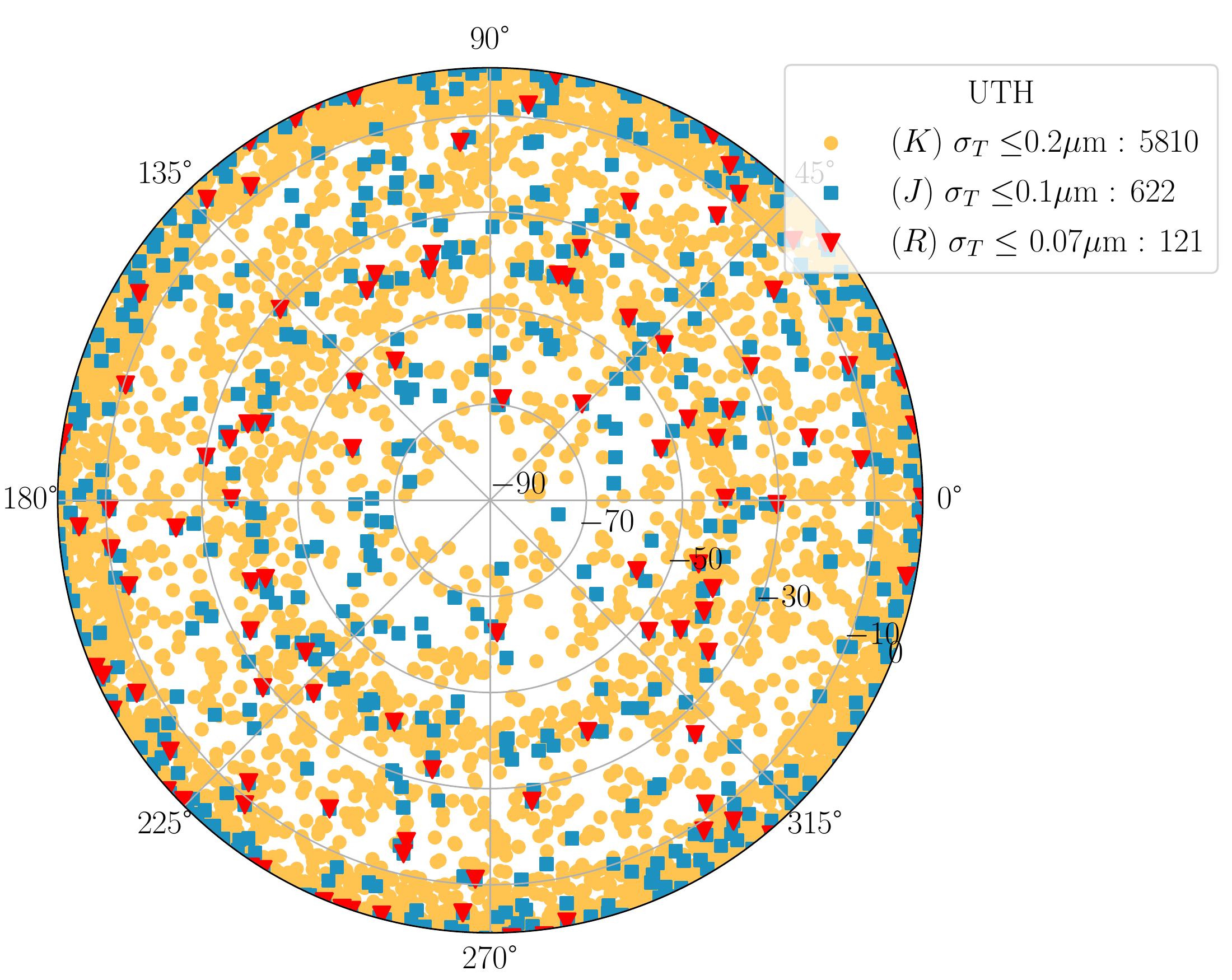}}
	\caption{Targets from our list of $ 15\,799 $ QSOs observable in the near-IR with the \textit{UTs} with the GFT (a) and the HFT (b). The orange dots are for targets observable only in the $ K  $ band, the blue squares  for targets that would  also be observable in the $ J $ band, and the red triangles for targets potentially observable in the visible.}
	\label{fig:AGN2}
\end{figure*}

 Figs \ref{fig:AGN2a} and \ref{fig:AGN2b} show the targets that would have guide stars with the GFT and the HFT on \textit{UTs} in the best 20 per cent seeing conditions. The GFT would allow observations in the $ K $ band of 1380 of the initial $ 15\, 799 $ QSOs, i.e. 8.3 per cent. This is more than enough to support a large program for the observation of BLRs in $K$. These targets are selected strictly on the availability of a guide star and should therefore represent a statistically correct distribution of QSOs characteristics. We should also note that the initial list of QSOs used here is quite biased against a sky coverage concentrated near the Galactic Plane. The HFT would allow  observing 6073 targets in $ K $, i.e. 38 per cent of the initial list. Among these, 658 targets would have sufficient tracking quality for a good observation in the $ J $ band. The HFT would be really needed for an AGN BLR program in the $ J $ band with \textit{UTs}. For observations in median seeing conditions, the numbers of accessible targets drop to 832 targets with the GFT and 4267 with the HFT.

\begin{figure*}
	\centering
	\subfigure[\label{fig:AGN3a}]{\includegraphics[width=0.48\textwidth]{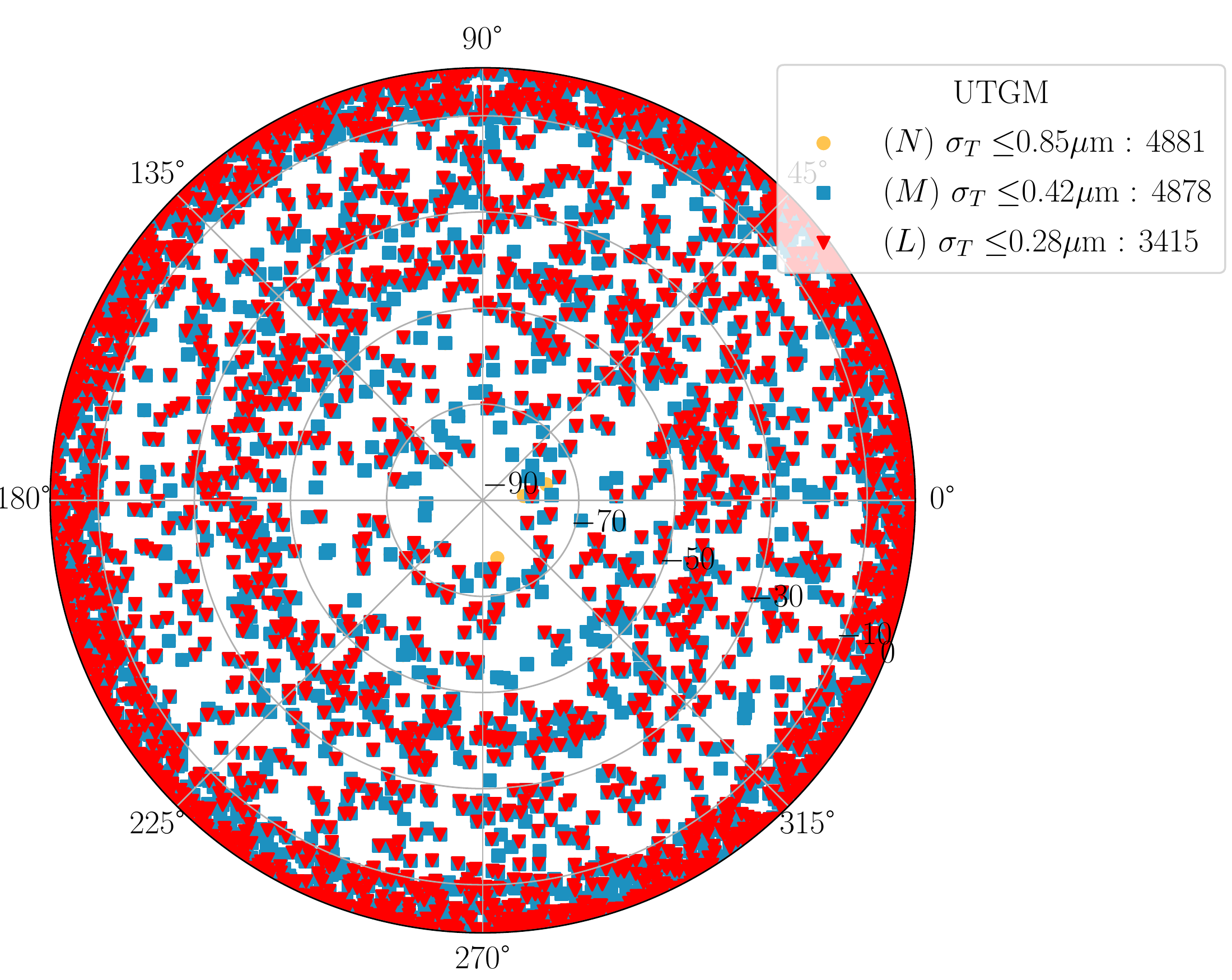}}
	\subfigure[\label{fig:AGN3b}]{\includegraphics[width=0.48\textwidth]{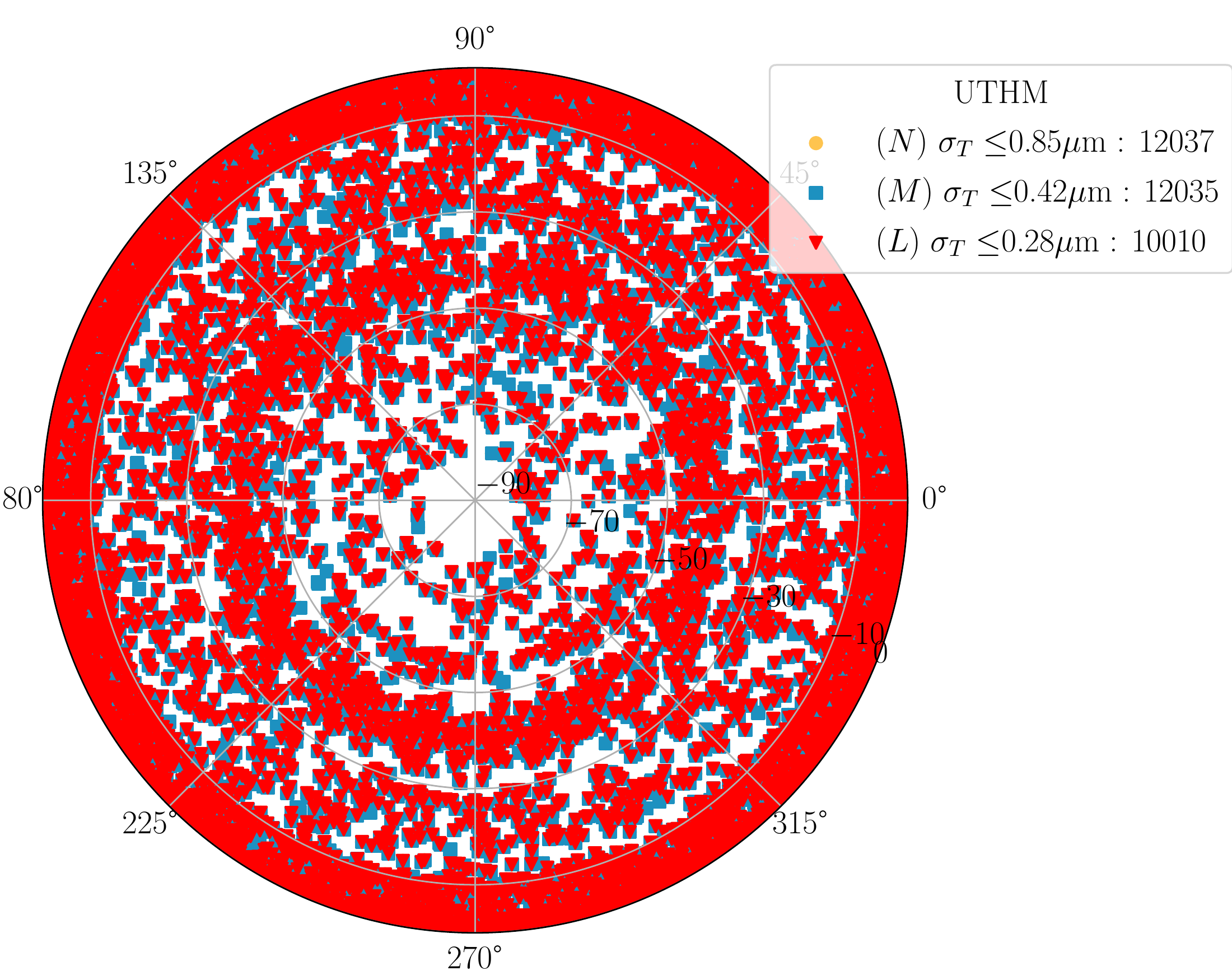}}
 \caption{Targets from our list of $ 15\,799 $ QSOs observable in the mid-IR with \textit{MATISSE} on the \textit{UTs} with the GFT (a) and the HFT (b). The red triangles are for 	targets observable in the $ L $, $M$, and $N $ band, the blue squares for targets observable in the $ M $, and $ N $ band, and the orange dots for targets observable only in the $ N $ band.\label{fig:AGN3}}
\end{figure*}

Figs \ref{fig:AGN3a} and \ref{fig:AGN3b} show the number of targets that could be observed using \textit{MATISSE} with GFT tracking (left) and an HFT tracker (right). With the GFT, 4881 targets would be accessible to \textit{MATISSE}, including 3415 well tracked enough for good observations in $ L $ and 4878 suitable for good observations in $ M $. Here, the number of targets is dominated by the necessity to have a $ \lambda/10 $ tracking on the guide star in $ K $ to avoid fringe jumps that would be
very destructive for $ L $ and M observations. The HFT would allow observing $12\,037$ targets, i.e. 76 per cent of the initial list and $10\,010$ targets (63 per cent) would be tracked well enough for observations in all \textit{MATISSE} bands.

\begin{figure*}
\centering
\subfigure[\label{fig:AGN4a}]{\includegraphics[width=0.48\textwidth]{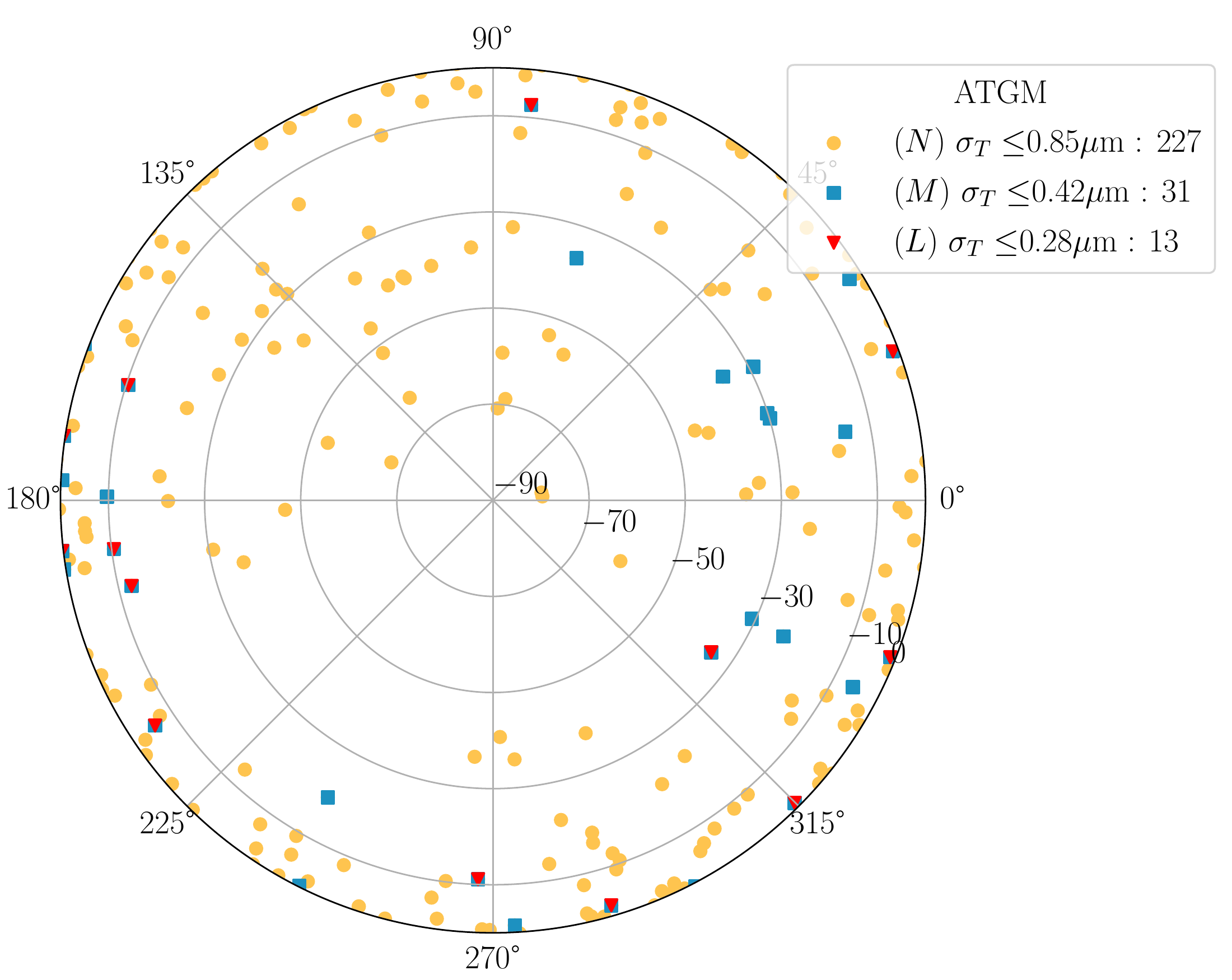}}
\subfigure[\label{fig:AGN4b}]{\includegraphics[width=0.48\textwidth]{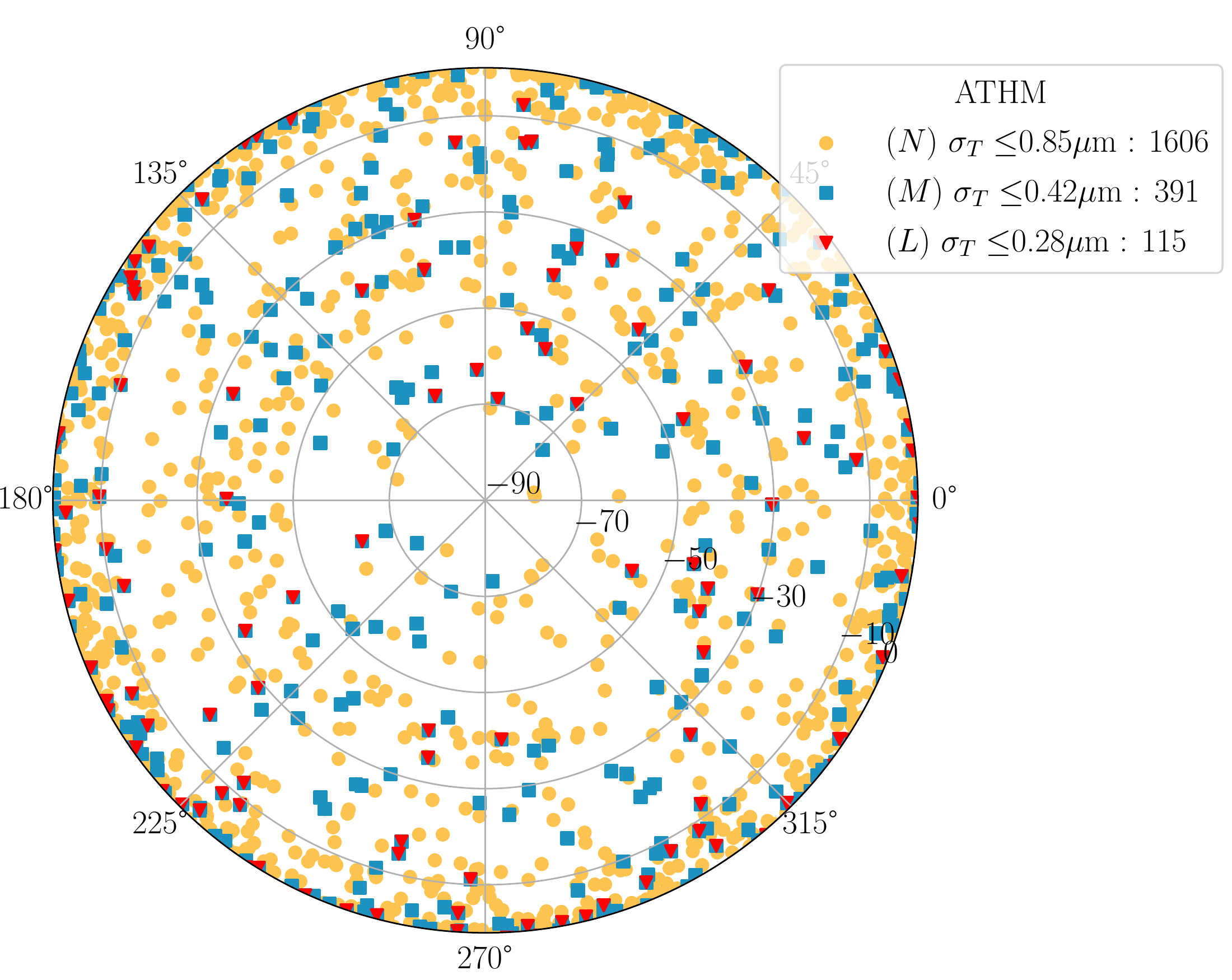}}
\caption{Targets from our list of $ 15\,799 $ QSOs observable in the mid-IR with the GFT (left) and the HFT (right) on \textit{ATs}. The orange dots are for targets observable only in the $ N $  band, the blue squares for targets observable in the $N$ and $ M $ band. The red triangles are for targets observable in the $ L $, $M$, and $N$ band.}\label{fig:AGN4}												 
\end{figure*}

With \textit{ATs}, the GFT would allow observing 227 QSOs with \textit{MATISSE} in $ N $ (\figref{fig:AGN4}). This is an extremely small fraction (1.4 per cent) of the initial list but would be quite sufficient for a very significant program. Note that on \textit{ATs}, we only need  to install a simple upgrade allowing it to feed \textit{MATISSE} with a target up to 1 arcminute away from the GFT guide star. The fact that the \textit{AT} AO would almost certainly fail to track on these QSOs (we have not checked the range of off-axis tracking for the AO of the \textit{ATs}) is not a showstopper in the $N$ band as \SI{1.8}{\meter} telescopes are diffraction limited at \SI{10}{\micro\meter} in good seeing conditions. One might just need a fast tip-tilt tracker at the entrance of \textit{MATISSE}. With the HFT, a quite large numbers of QSOs would be accessible to \textit{MATISSE}.

\begin{figure*}
	\centering
	\subfigure[\label{fig:AGN5a}]{\includegraphics[width=0.48\textwidth]{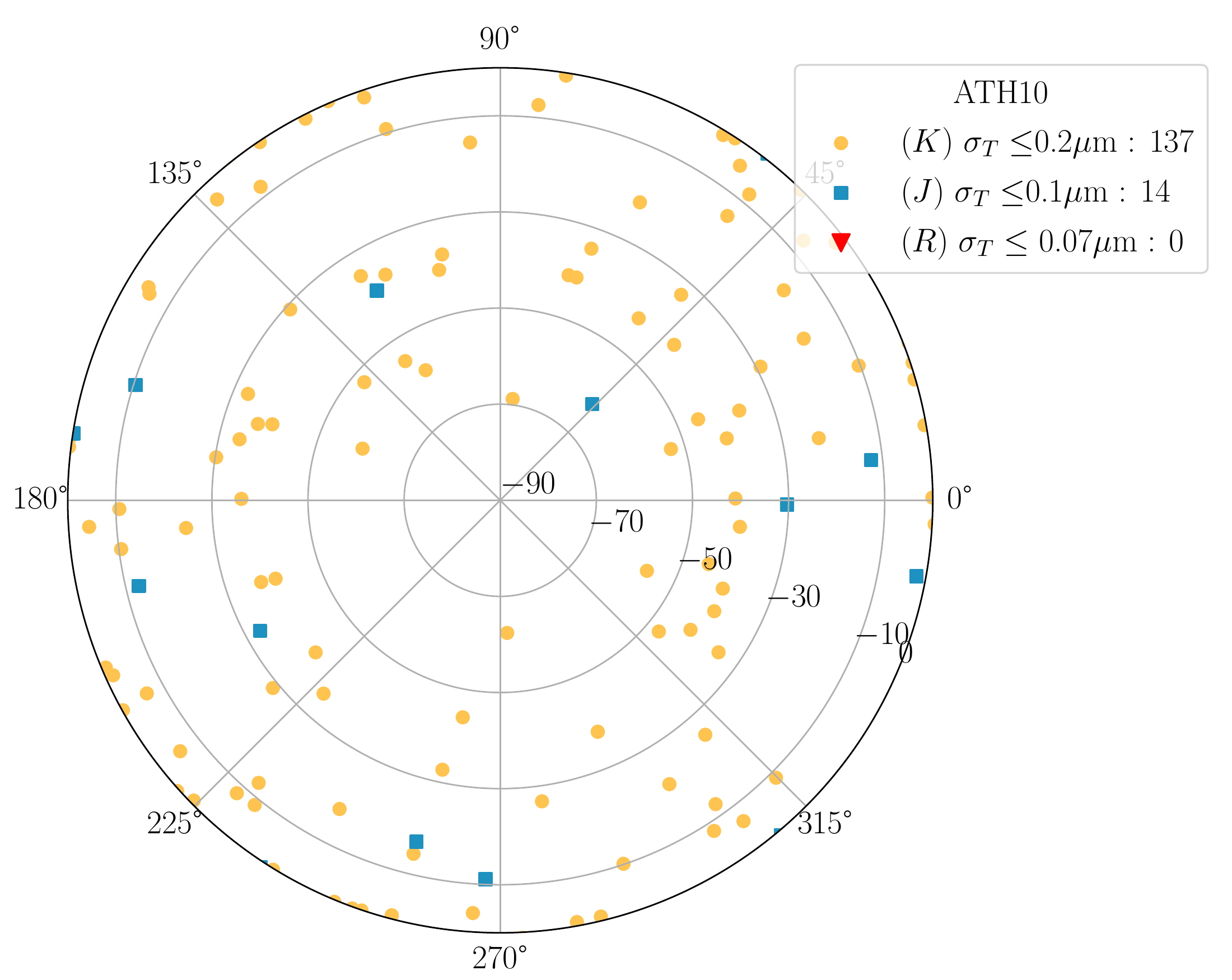}}
	\subfigure[\label{fig:AGN5b}]{\includegraphics[width=0.48\textwidth]{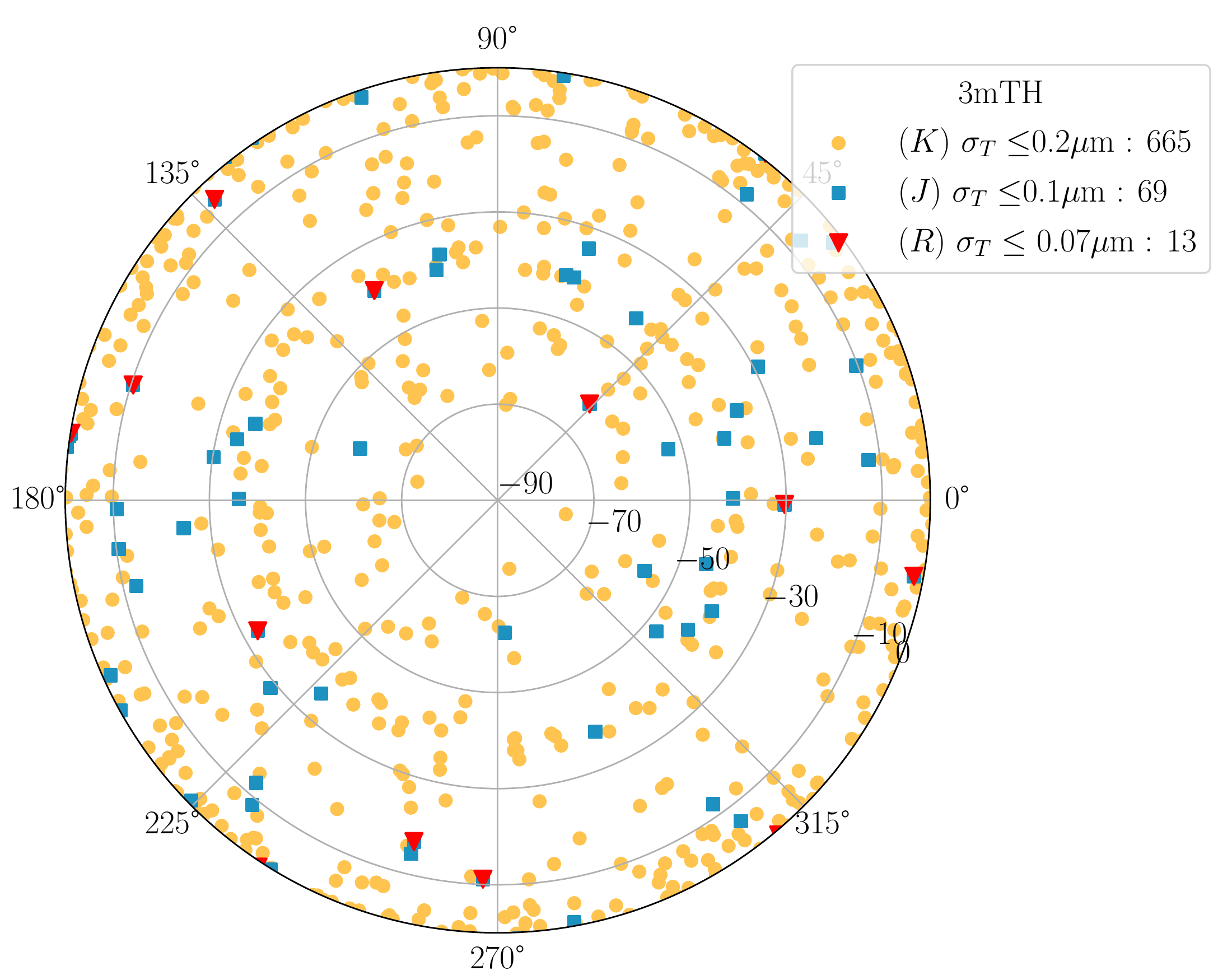}}
\caption{Targets that would be observable with an HFT on a new generation
interferometer with overall transmission 10 per cent and \SI{1.8}{\meter} telescopes (left) and \SI{3}{\meter}
telescopes (right) in a site with the same seeing as \textit{Paranal}. The orange dots are for targets observable only in the $K$ band, the blue squares  for targets that would  also be observable in the $ J $ band, and the red triangles for targets potentially observable also in the visible.}\label{fig:AGN5}
\end{figure*}

As explained at the end of Section \ref{sec:VLTI-ATS}, the figures for the HFT on the \textit{ATs} also apply to a new generation interferometer with \SI{1.8}{\meter} apertures on a site similar to \textit{Paranal}. We see that such an interferometer would have access to a substantial program on AGN torus in the mid-IR with quite enhanced imaging capability with regard to the \textit{VLTI} but that the near-IR and the visible would be out of reach.
If an optimum design of PFI allows an overall transmission of 10 per cent instead of the 2 per cent maximum considered here for the \textit{VLTI}, this situation could be improved, as indicated by \figref{fig:AGN5} for an interferometer with 1.8 m telescopes and \SI{3}{\meter} telescopes. We see that the visible would be accessible only for interferometers with quite large apertures, which quite certainly makes it out of any reasonable expectation. However, an array of \SI{1.8}{\meter} telescopes with optimized transmission would have a potential for the imaging of some BLRs in the $ K $ and $ H $ bands assuming that there are candidates with sufficient red shifts in the list.

\subsection{A shortlist of nearby AGNs}

We have used the shortlist of 331 nearby (less than 100 Mpc) AGNs with well-defined characteristics selected by \cite{2020MNRAS.494.1784A} to further investigate the feasibility of observations with \textit{MATISSE}. We will first select the targets that have guide stars suitable for off-axis tracking and observations with \textit{MATISSE}. As this list was not flux cut in magnitude like the QSO list, some targets may not be bright enough to observe even with the long coherent integration offered by the off-axis tracking.

All sources are brighter than magnitude 17 in the Wise W1 filter (\SI{3.4}{\micro\meter}), from the wise all-sky survey, except for 18 AGN which had no reported counterpart within 3 arcseconds. From the list of known AGN and R90 AGN candidates, we calculated the predicted angular size in $ K $ band. First, we determine the bolometric luminosity for each object using the \SI{12}{\micro\meter} luminosity. 
We convert that luminosity, found in \cite{2020MNRAS.494.1784A}, into \SIrange{2}{10}{KeV} luminosity using the MIR-Xray correlation in \cite{2020MNRAS.494.1784A} and we then convert the \SIrange{2}{10}{KeV} luminosity to bolometric using a conversion factor of 10 from \cite{2010MNRAS.402.1081V}. 
We use the $\Omega_T(\lambda_{K})\propto L^{1/2}$ relation confirmed by \cite{2006ApJ...639...46S} where $\Omega_T(\lambda_{K})$ is the physical radius of the dust as modelled by a thin torus and $ L $ is the bolometric luminosity, and R$_{44}$ which is the radius at a luminosity of \SI{10e44}{erg\per\second}. Using that relation, we calculate the physical radius of the dust in $ K $ band and convert the associated diameter to an angular size using the distances provided in \cite{2020MNRAS.494.1784A}. For angular size in the $N$ band, we use the median size ratio $\frac{\Omega_T(\lambda_{N})}{\Omega_T(\lambda_{K})}=9$ established by \cite{2013A&A...558A.149B} from \textit{MIDI} observations. Then we will also use the N band magnitude of the AGN core found in \cite{2020MNRAS.494.1784A} to evaluate the coherent flux SNR that can be achieved on these targets with \textit{MATISSE}.

The commissioning of \textit{MATISSE} \citep{2020SPIE11446E..0LP} showed that in the N band with \textit{UTs}, we obtain a coherent
flux $SNR_{C1}=1$ per coherence time ($\mathcal{T}_C= \SI{250}{\milli\second}$) for each \SI{0.3}{\micro\meter} spectral channel, for a 170 mJy source. In $ N $, we are always dominated by the background photon noise and the $SNR_{C1}$ per frame is proportional to the source flux $ F(jy) $ in Jansky and to the square root of the integration time:  $SNR_{C1}\propto F(jy) \sqrt{\mathcal{T}_{int}} $. If $\mathcal{T}_{int}$ is the integration time allowed by the fringe tracking without fringe jumps and $\mathcal{T}_{tot}$ is the total open shutter time on the source, the coherent flux $ SNR $ will be given by: 
\begin{equation}
S N R_{C}=\frac{\operatorname{SNR}_{C 1}^{2}}{\sqrt{1+2 \operatorname{SNR}_{C 1}^{2}}} \sqrt{\frac{\mathcal{T}_{\text {tot}}}{\mathcal{T}_{\text {int}}}},
\end{equation}
with
\begin{equation}
\operatorname{SNR}_{C 1}=\sqrt{\frac{\mathcal{T}_{\text {int}}}{\mathcal{T}_{c}}} \frac{F(J y)}{0.17}=10^{3.24-0.4 N_{\operatorname{mag}}},
\end{equation}
with $\mathcal{T}_{int}=\SI{10}{\second}$, $\mathcal{T}_{C}=\SI{250}{\milli\second}$, and $\mathcal{T}_{tot}=360\cdot\mathcal{T}_{int}$ we have : 
\begin{equation}
S N R_{C}=\frac{S N R_{C 1}^{2}}{\sqrt{1+2 S N R_{C 1}^{2}}} \cdot \frac{\tau_{\text {tot}}}{\tau_{\text {int}}}=\frac{10^{7.76-0.8 N_{\text {mag}}}}{\sqrt{1+10^{6.78-0.8 N_{\text {mag}}}}},
\label{equ:AGN4}
\end{equation}
where we have used $F(N_{mag}=0) = \SI{47.33}{jy}$ at \SI{9}{\micro\meter}.

With the \textit{ATs}, we lose three magnitudes even if we integrate slightly longer and the same computation gives:
\begin{equation}
S N R_{C}=\frac{S N R_{C 1}^{2}}{\sqrt{1+2 S N R_{C 1}^{2}}} \sqrt{\frac{\tau_{\text {tot}}}{\tau_{\text {int}}}}=\frac{10^{5.36-0.8 N_{\operatorname{mag}}}}{\sqrt{1+10^{4.38-0.8 N_{\operatorname{mag}}}}}.
\label{equ:AGN5}
\end{equation}

Then we use equations (\ref{equ:AGN1b}, \ref{equ:AGN2}, and \ref{equ:AGN3}) to find if the object can be used for imaging, for ADI, or only for PDI. If none of these conditions can be satisfied, the source can be fringe tracked but is too faint to obtain usable data in 1 h with \textit{UTs} and 2 h with \textit{ATs} within our $ SNR=10 $ criteria.

\begin{figure}
	\centering
	\includegraphics[width=\linewidth]{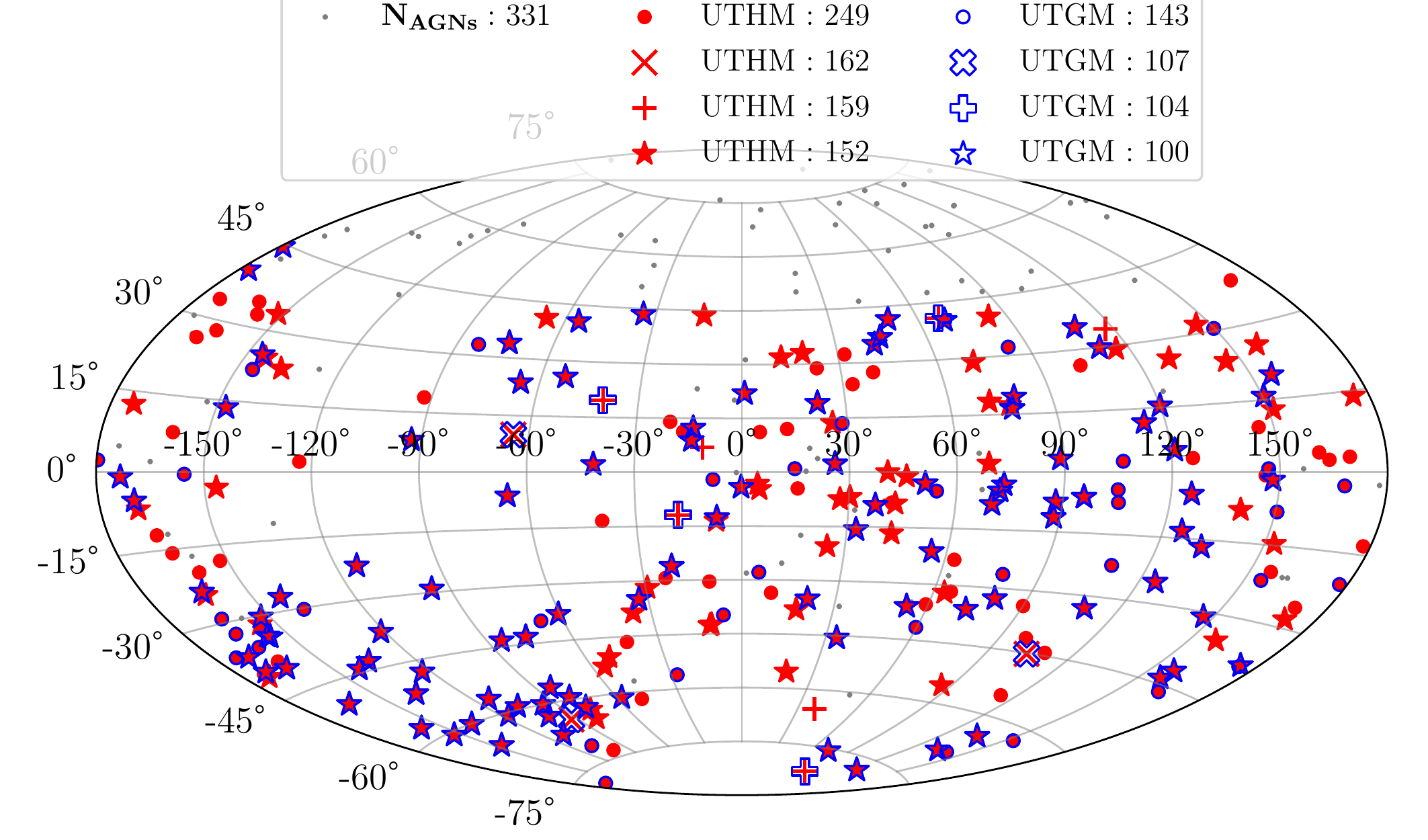}
	\caption{Nearby AGNs that would be observable with \textit{MATISSE} on the \textit{UTs}. The red symbols are for targets made accessible by the HFT. The red dots are for AGNs that have an HFT guide star but are too faint to provide useful information in 1 h  of open shutter time. The red x are for targets restricted to PDI and the red + for targets accessible also by ADI. The red stars are for targets large enough for image reconstruction. The blue symbols have the same meanings for targets that can be observed with the GFT.}
	\label{fig:AGN6}
\end{figure}

\begin{figure}
	\includegraphics[width=\linewidth]{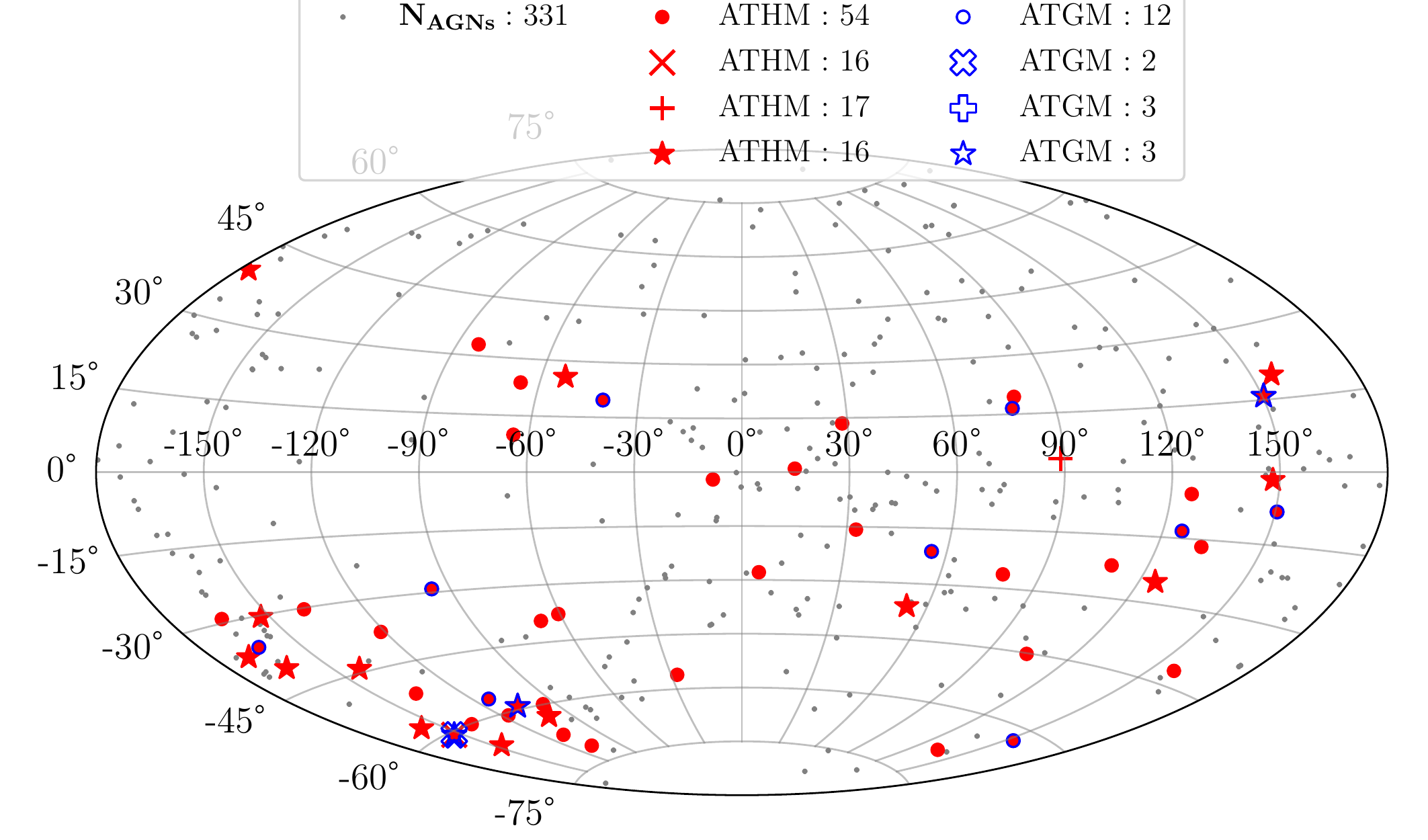}
	\caption{Nearby AGNs that would be observable with \textit{MATISSE} on the \textit{ATs}. The symbols are the same as in \figref{fig:AGN6}. On \textit{ATs} the HFT makes a huge difference.}
	\label{fig:AGN7}
\end{figure}

\figref{fig:AGN6} shows the nearby AGNs from \cite{2020MNRAS.494.1784A} list observable with \textit{MATISSE} on the \textit{UTs}. The HFT would allow off-axis tracking with sufficient precision for $N$ band observations on 249 nearby AGNs, i.e. more than 75 per cent of the targets in the list. 
We would obtain usable information on 162 targets and 152 of them should be large enough for image reconstruction. With the GFT, we would be able to track on 143 targets and to make images in 100 of these. \figref{fig:AGN7} shows the nearby AGNs observable with \textit{ATs}. Quite remarkably among the 54 candidates (i.e. 16 per cent) that could be off-axis tracked, 16 will be bright enough for image reconstruction with the very good u-v coverage permitted by the \textit{ATs}. This illustrates again the strong justification of the HFT for \textit{MATISSE} imaging with the \textit{ATs}. With the GFT, the number of potential image reconstructions from this list drops to 2 but includes very famous and important targets.

\section{Interferometry in Antarctica}
\label{sec:five}

\begin{figure*}
	\centering
	\begin{tabular}{m{0.88\textwidth} m{0.05\textwidth} m{0.07\textwidth}}
		\subfigure[FT-\textit{AT} Dome C \label{fig:AC1a}]{\includegraphics[width=0.44\textwidth]{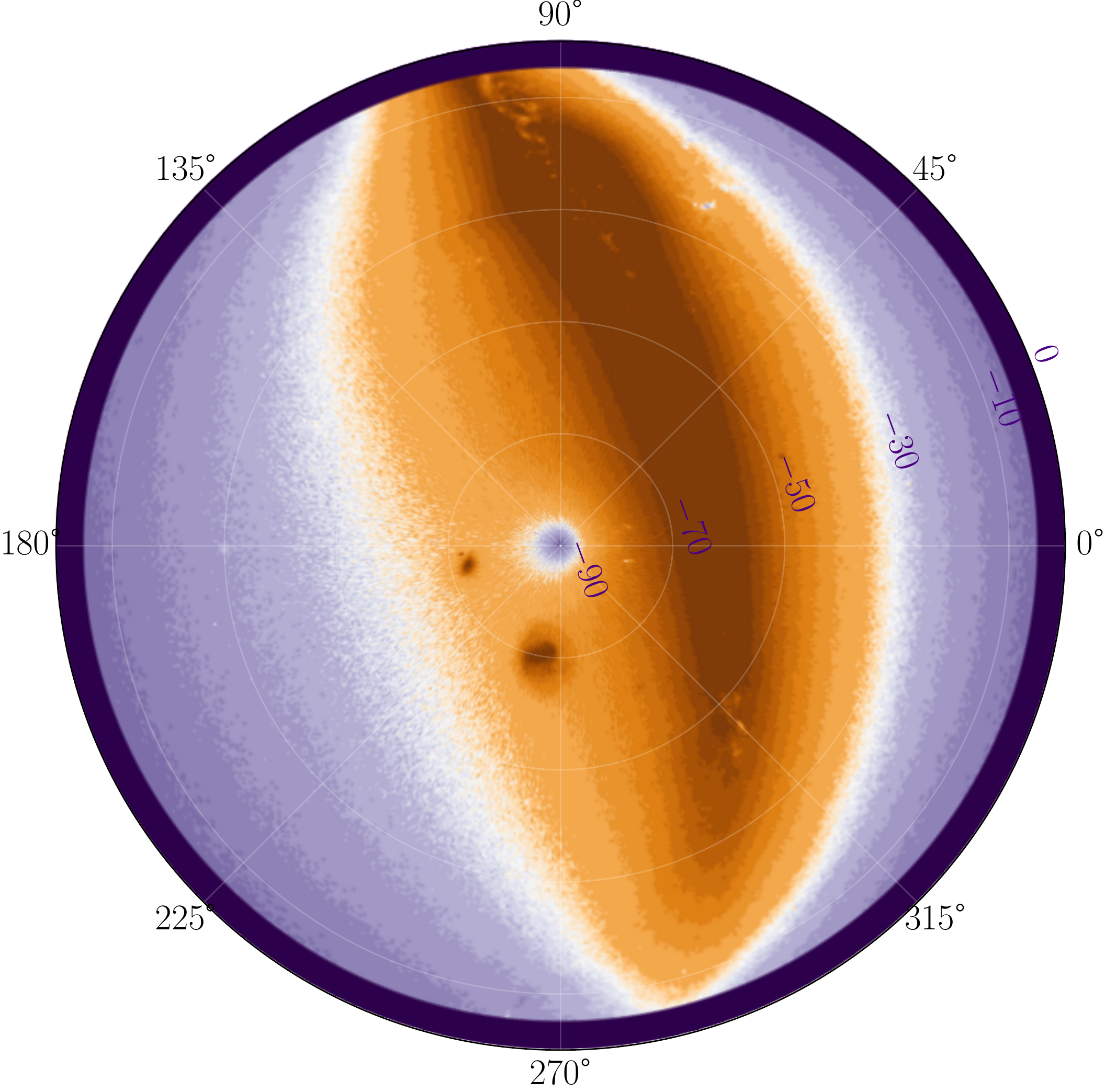}}
		\subfigure[HFT-\textit{AT} \textit{MATISSE} Dome C.\label{fig:AC1b}]{\includegraphics[width=0.44\textwidth]{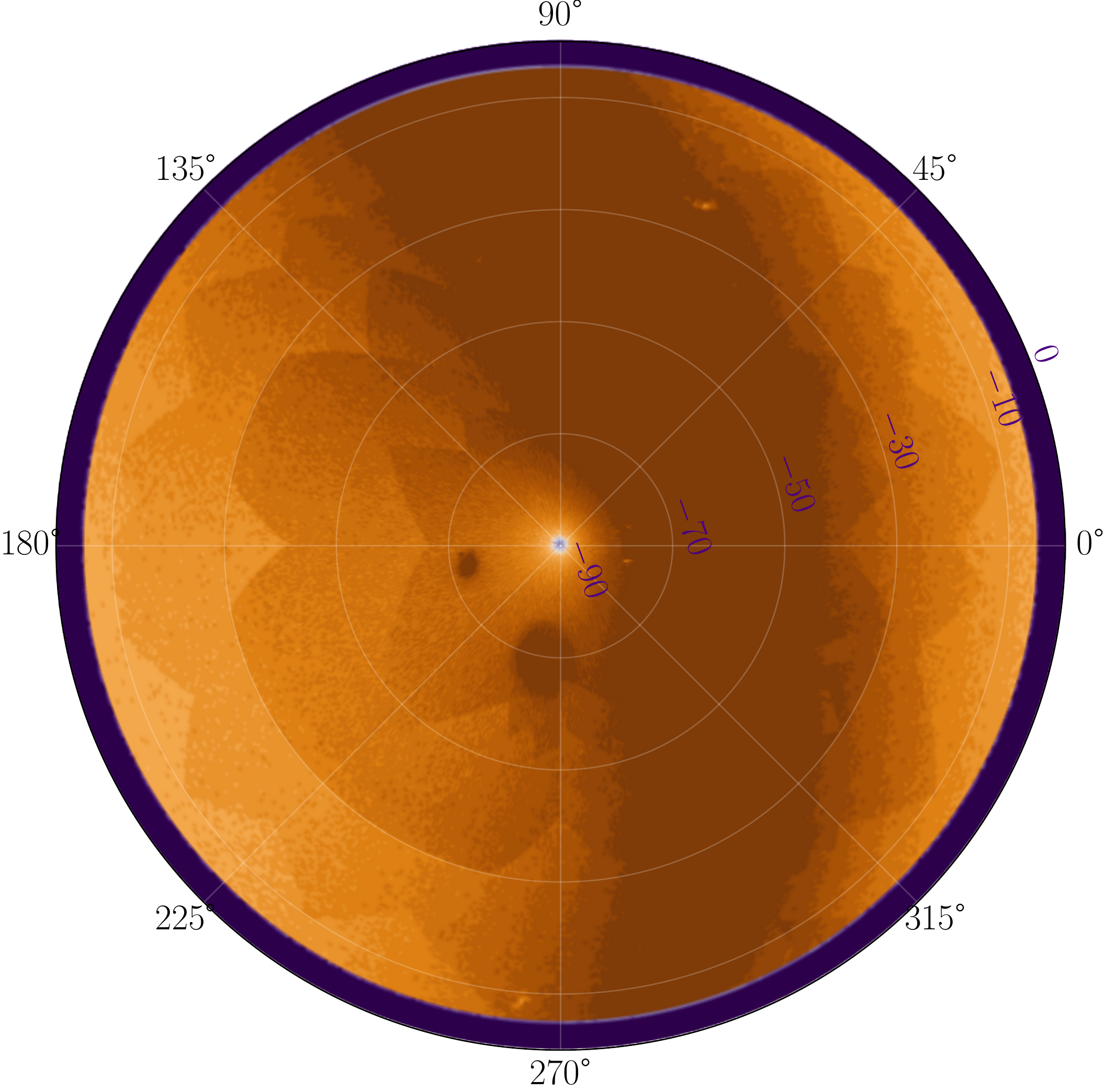}}
		&\includegraphics[width=0.06\textwidth]{fig/bar}&\rotatebox{-90}{\hspace*{-1.4cm}Sky Coverage}\\
	\end{tabular}
	
	\caption{Sky coverage with an HFT on a Dome C interferometer with 1.8 m telescopes
	for observations in the near-IR (right) and in the mid-IR (left)\label{fig:AC1}}.
	\label{fig:hft-at-domecmlmdit-ot}
\end{figure*}

\begin{figure*}
\centering
\includegraphics[width=0.48\linewidth]{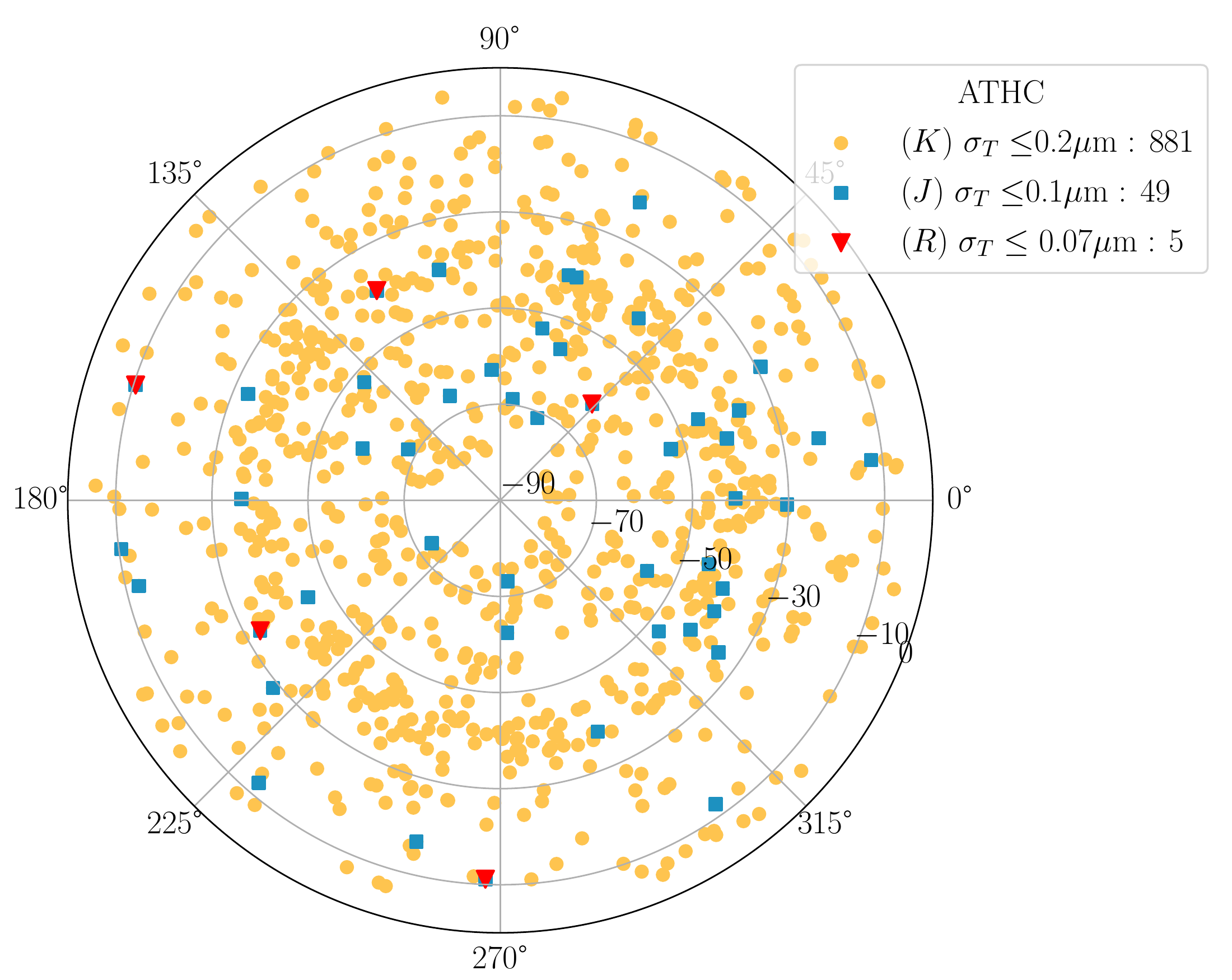}
\includegraphics[width=0.48\linewidth]{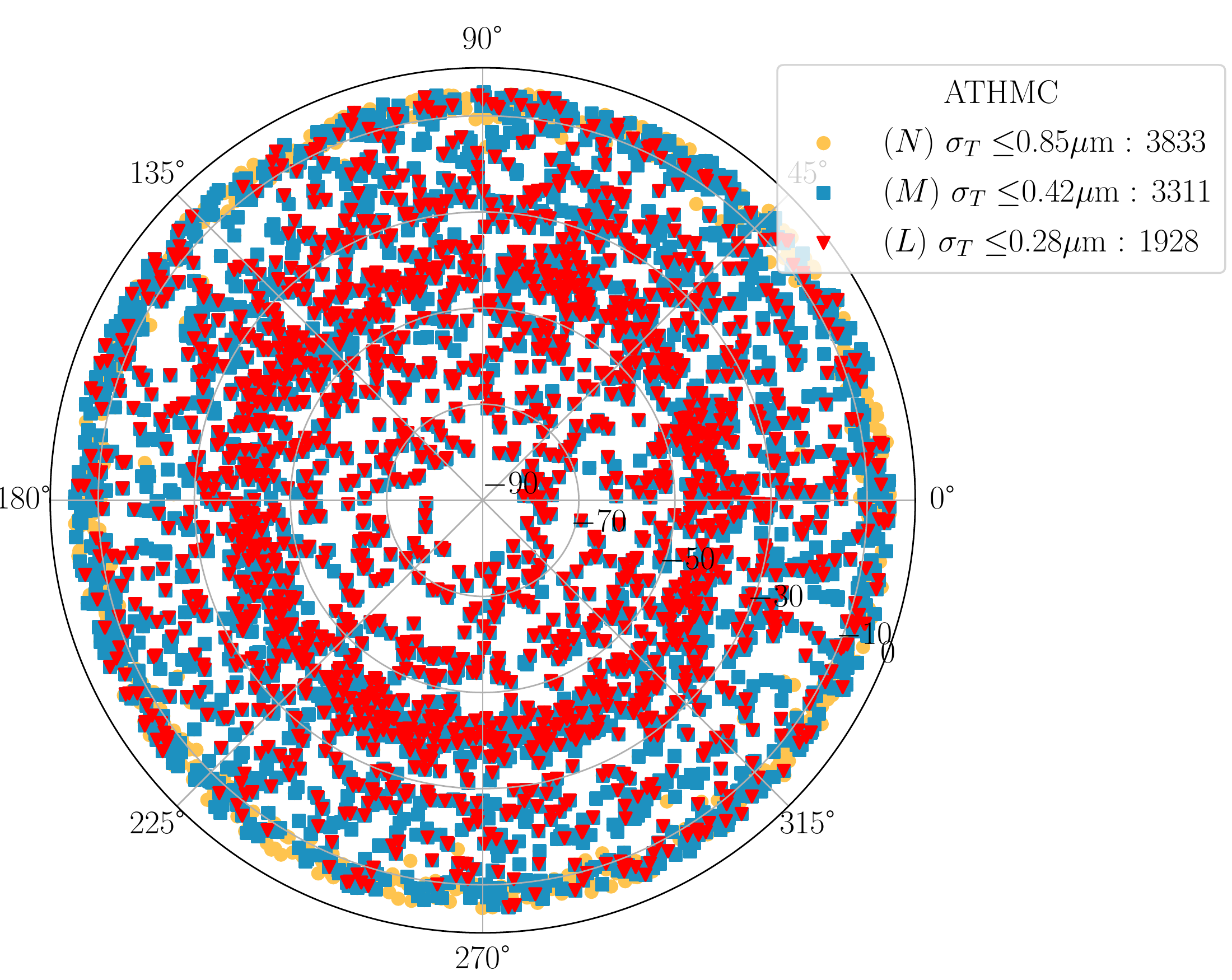}
\caption{{QSOs observable with an HFT on a Dome C interferometer with \SI{1.8}{\meter} telescopes. Left: observations in the near-IR in K (orange dots), H\&K (blue squares) and J\&H\&K (red triangles). Right: observations in the mid-IR in N (orange dots), M\&N (blue squares), and L\&M\&N (red triangles).}\label{fig:AC2}}
\end{figure*}

About 15 yr ago, quite a lot of thinking and site testing has been invested investigating the possibility of a large astronomical facility in Antarctica. The seeing at Dome C is variable and most often quite poor below the thin turbulent ground layer but it becomes often very good above 20 m and almost always excellent above a 30 m. We used the seeing parameters given in \tabref{tab:YT1} obtained from \cite{2009A&A...499..955A,2020MNRAS.496.4822A}. The very small outer scale corresponds to large isopistonic angles even with modest apertures as shown by \figref{fig:theta-max} where we see that the isopistonic angle at Dome C with 1.8~m telescopes would be comparable to this at \textit{Paranal} with 8~m telescopes given that the primary mirrors are above the ground layer. At Dome A, the situation could be even better, with a probably finer ground layer  \cite{2006SPIE.6267E..1KS}. The potentially much larger off-axis potential at Dome C was one of the initial triggers of this work. The \figref{fig:hft-at-domecmlmdit-ot} show the sky coverage that would be obtained with 1.8 m telescopes for $ K $ band and $ N $ band observations with an HFT with still 2 per cent overall transmission. The sky coverage is fair in $ K $ and looks better than the sky coverage with the 8~m telescopes in \textit{Paranal}. For $ N $ band observations, we have a very good sky coverage, larger than 20 per cent in all the observable sky. However, \figref{fig:AC2} shows the fraction of observable stars in the large QSO list and we see that we still are quite below the possibility of a 8 m telescope at \textit{Paranal}.  In particular, we note the very small number of targets that would be observable in the visible. Then, even at Dome C, it seems that a network of 1.8 m telescope would not allow imaging of BLRs in the visible. However, these result might be biased by our QSO list with its the over representation of targets near the equator that would not be observable from Dome C. For YSOs and for observations in the $ N $ band in general, we would be limited by the availability of interesting targets a declination below $\sim-30^\circ$. It seems indeed that this limitation strongly restricts the number of interesting YSO candidates.

With an interferometer of overall transmission 2 per cent and \SI{1.8}{\meter} apertures, the \figref{fig:AC2} shows a high potential of BLR imaging in the $ K $ band. Imaging of AGN BLRs in the visible would require larger apertures or a much improved overall transmission even at Dome C.

%% file: fig/geom-param.tikz
	\begin{tikzpicture}[decoration={random y steps,segment length=3mm,amplitude=1mm}]
	\tikzset{zxplane/.style={canvas is zx plane at y=#1,very thin}}
	\tikzset{yxplane/.style={canvas is yx plane at z=#1,very thin}}
	\begin{scope}[canvas is zx plane at y=0]
	\draw (0,2) circle (1cm);
	\draw (0,-2) circle (1cm);
	\draw [color=gray,thick,(-) ] (0.4,-2.88)--(-0.2,-1.6)node[minimum size=2pt,inner sep=0pt, outer sep=0pt,fill=white,text=black]{$D$}--(-0.4,-1.12);
	\draw [color=gray,thick,(-) ] (-0.4,2.88)--(0.2,1.6)node[minimum size=2pt,inner sep=0pt, outer sep=0pt,fill=white,text=black]{$D$}--(0.4,1.12);
	\end{scope}
	\draw [minimum size=2pt,inner sep=0pt, outer sep=0pt,thick,color=gray,thick,latex reversed - latex reversed] (-2.6,-0.36)--(-2.6,0.8)node[text=black,fill=white] {$h_{m}$}--(-2.6,1.63);
	\begin{scope}[canvas is zx plane at y=2]
	\draw [thick,color=gray,latex reversed - latex reversed] (0,2)--(-0.6,-0.1)node[yshift=-0.7ex,rotate=-8,text=black,above]{$d_{12}$}--(-1,-1.5);
	\draw [thick,color=gray,latex reversed - latex reversed] (0,-2)--(-0.65,0.925)node[yshift=-0.7ex,rotate=7,text=black,above]{$d_{22}$}--(-1,2.5);
	\draw (0,2) node [anchor=north east] {$P_2$} circle (1cm);
	\draw (0,-2)node [anchor=south west] {$P_1$}  circle (1cm);
	\draw (-1,2.5)node [anchor=west] {~~$P_2^\prime$} circle (1cm);
	\draw (-1,-1.5)node [xshift=1ex,yshift=-1ex,anchor=south west] {~~$P_1^\prime$}  circle (1cm);
	\draw [gray,thick,decorate,rounded corners](2,-3.15)--(2,3.7)--(-1.8,3.7)--(-3,3.2)--(-3.,0.2) node[text=black,above] {Perturbed Layer}--(-3,-3.15)--(2,-3.15);
	\draw [color=darkgray,thick,(-latex reversed ] (0.4,-2.88)--(0.15,-2.5)node[minimum size=2pt,inner sep=0pt, outer sep=0pt,fill=white,rotate=8,text=black]{\tiny$h_{_{m}}.\theta$}--(-0.,-2.);
	\end{scope}
	
	\draw [color=darkgray,very thick,->,>=stealth,] (-2,0)--(-2,3);
	\draw [color=darkgray,thick,domain=68:90] plot ({-2+cos(\x)}, {sin(\x)});
	\node at (-1.87,0.8) {$\theta$};
	\draw [color=darkgray,very thick,dashed,->,>=stealth,] (-2.4,-1)-- (-2,0)--(-0.8,3.2);
	\draw [color=darkgray,very thick,->,>=stealth,] (2,0)--(2,3);
	\draw [color=darkgray,very thick,dashed,->,>=stealth,] (1.6,-1)-- (2,0)--(3.2,3.2);
	\draw [color=darkgray,thick,domain=68:90] plot ({2+cos(\x)}, {sin(\x)});
	\node at (2.15,0.8) {$\theta$};
	\draw [color=darkgray,very thick,dashed,] (-2,-1)--(-2,0);
	\draw [color=darkgray,very thick,dashed,] (2,-1)--(2,0);
	\draw [thick,color=gray,latex reversed - latex reversed] (-1.97,-0.7)--(0,-0.7)node[minimum size=2pt,inner sep=0pt, outer sep=0pt,text=black,fill=white] {$\Delta$}--(1.97,-0.7);
	\end{tikzpicture}
	

%% file: 5-conclusion.tex
\section{Discussion and perspectives}

This paper provides tools to estimate the performances of off-axis fringe tracking that include all the atmospheric effects that can impact it. The parametric modelling of the atmospheric effects such as the loop error, related to the evolution of piston with time, and the anisopistonic error, related to the evolution of piston with angular distance, have the great
advantage of allowing the estimation of tracking performance for a given off-axis source from standard seeing parameters measured in \textit{Paranal}, and other observatories, such as the Fried parameters, the coherence time, and the isoplanetic angle.

Our analysis of the loop error due to the piston evolution with time is based on atmospheric perturbations only. We know that real interferometers, in particular the \textit{VLTI} with \textit{UTs}, are affected by a substantial level of vibration. The \textit{\textit{GRAVITY}+} plan to upgrade the \textit{VLTI} includes passive and active devices that damp these vibrations. A next step of our work
should be to include the vibration power spectrum in the loop error. This would allow a more realistic error budget and consolidate the specifications that the vibration correction should achieve. As the frame time and performances actually achieved by the GFT with the \textit{ATs} are in quite good agreement with our computation, we tend to believe that the combination of a vibration control that would reduce the \textit{UT} vibration levels to those of the \textit{ATs} with Kalman filtering, which allow reducing the impact of vibration on fringe tracking performance, should allow fringe tracking quality on \textit{UTs} to approach or even slightly exceed the atmospheric limit described here. In spite of average coupling efficiency \citep{2019A&A...624A..99L} comparable on \textit{ATs} and \textit{UTs}, the current performance of the GFT on \textit{UTs} displays a deficit of more than 2.5 magnitudes with regard to these theoretical predictions, which are quite well confirmed by the measured performances on \textit{ATs}. The SNR analysis given here suggests that most of this loss should be attributed to very fast instabilities of the AO. This is an additional strong argument for the radical upgrade of this subsystem. However, the effective flux imbalance needed to explain the GFT magnitude deficit on \textit{UTs} looks surprisingly high and should be analysed in more details.

Our analysis of the anisopistonic error is based on atmospheric models and on two campaigns of outer scale measurements at \textit{Paranal} described in \cite{2016MNRAS.458.4044Z}. Although these atmospheric models are relatively well supported by experimental evidence of AO systems, the specific anisopistonic error has not been measured extensively. A coordinated campaign, combining anisopistonic measurements thanks to the prototype \textit{\textit{GRAVITY}}+ wide-field combiner tested at
the \textit{VLTI} with a seeing sensor including outer scale measurements would be very valuable to check the predictions of this
paper and give the parameters to actually evaluate the performance of off-axis tracking. As the current GSMs are expected to allow a 2 h forecast of all the observing parameters \citep[]{2021MNRAS.504.1927G}, the formalism developed here might lead to a forecast of off-axis performance and hence a major optimization of \textit{VLTI} large programs. In Section \ref{sec:QSO-list}, we have seen that moving from the median seeing to the best 20 per cent seeing can change the number of targets that can be fringe tracked off-axis by 40 per cent. A proper evaluation of the off-axis fringe tracking performances from the standard seeing parameters would therefore be highly valuable.

This paper underlines the very large potential of off-axis fringe tracking in the near-IR for observations in the mid-IR. The anisopistonic angle in the $ N $ band with the \textit{UTs} can approach three arcminutes, and the design of the wide-field system that feeds sources with large angular separations in the \textit{VLTI} instrument(s) and tracker chain should take that parameter in consideration. We have set the tracking requirement in $ K $ band at \SI{0.2}{\micro\meter}. This
requirement could be substantially relaxed in terms of FT variance. However, for tracking in near-IR and for observations in the mid-IR, it is crucial to analyze the probability of one $\lambda$ fringe jumps in the tracker that would impact the contrast and ruin the calibration in the science instrument. With a $\lambda/10$ specifications, the fringe jumps should be extremely rare.
Investigating the tracking accuracy that would allow an acceptable fringe jump rate will be a key element to optimize the off-axis tracking for the mid-IR.

Finally, we have compared the potential sky coverage with our best estimate for the current GFT as it could be improved by the \textit{GRAVITY}+ upgrades and an improved FT combining a larger spectral band, a slightly improved transmission, and an optimized concept.
The sky coverage with the GFT appears quite sufficient for the large AGN program that is one of the roots of the \textit{GRAVITY}+ proposal because there is a large enough number of AGNs all over the sky to provide more than enough candidates. For others, more specific and rarer targets, the very substantial gain in sky coverage provided by the HFT would be very valuable. The HFT would be decisive for efficient off-axis tracking with the \textit{ATs}.

\section{Conclusion}
We have developed a complete analysis of the error budget of off-axis fringe tracking in the near-IR for observations in the near-IR and the mid-IR. We have applied it to the \textit{VLTI} with the \textit{ATs} almost as they are and with the \textit{UTs} as they should be after the \textit{GRAVITY}+ upgrade project. We have considered two FTs: the GFT with its current transmission, detector, and $ K $ band operation as well as a new generation FT with a larger bandpass covering partially the $ J $, $ H $ and $ K $ bands with a slightly increased transmission and a new concept of HFT. The contribution of the variation of the atmospheric piston with time and angular separation to the off-axis tracking error has been based on analytic parameters based on the standard seeing sensor measures. The chromatic dependence of the piston, when the tracking and observations are performed at different wavelengths has been evaluated from a Keck Interferometer model updated by recent and ongoing \textit{MATISSE} and \textit{GRAVITY} measurements. We have combined our estimates of the off-axis tracking error budget and parameters with a database of more than \SI{8e8}{} stars, extracted from the \textit{Gaia} second data release, to compute sky coverage maps for the \textit{VLTI} with \textit{UTs} and \textit{ATs} for the GFT and the
HFT in different seeing conditions. With the GFT, the sky coverage with \textit{UTs} is better than 10 per cent in a domain extending from $\pm\SI{20}{\degree}$ to \SI{30}{\degree} away from the Galactic Plane, depending on seeing conditions. With the HFT the sky coverage with \textit{UTs} is larger than 10 per cent in most of the sky observable from \textit{Paranal}. To give a more specific example of a large AGN program on the \textit{VLTI}, we have selected a sample of more than $ 15\,000 $ QSOs from the VHS and we have evaluated the number of QSOs accessible to observations in all bands from $ J $ to $ N $. The first conclusion is that the GFT in the \textit{GRAVITY}+ context will give access to more than enough targets for its large AGN program in the $ K $ band. The second conclusion is that off-axis fringe tracking has an enormous potential for observations in the mid-IR with \textit{MATISSE} that should be considered in \textit{GRAVITY}+ and \textit{MATISSE} with \textit{GRAVITY}+ science programs. To be more specific about the AGN program of \textit{MATISSE}, we have analysed the potential performances of \textit{MATISSE} on a list of 331 nearby AGNs for which we
have detailed information. For 10 per cent of these targets with the GFT and 30 per cent of these targets with the HFT, the SNR and the resolution of \textit{MATISSE} should be sufficient to obtain detailed images of the dust torus in N. The third conclusion is that the HFT would offer a very substantial increase in sky-coverage and tracking precision that would be decisive for many
programs including AGN observations with the \textit{ATs}. We have also evaluated the potential of a new generation interferometer, like the proposed PFI, for off-axis tracking. We show that with an HFT, whose tracking performance does not decrease with the number of apertures, an array of 1.8 m telescopes at a site of \textit{Paranal} quality, the PFI would be able to track off-axis for fine imaging in the $ N $ band of many YSOs.

%% file: 6-Acknowledgements.tex
\section*{Acknowledgements}
\begin{itemize}

	\item[(i)] This work has been supported by the French government
	through the Université Côte d’Azur Jedi (UCAJEDI) Investments
	in the Future project managed by the National Research Agency
	(ANR) with the reference number ANR-15-IDEX-01.
	\item[(ii)] This work has been supported by the Chinese Academy of
	Sciences, through a grant to the Chinese Academy of Sciences South
	America Center for Astronomy (CASSACA) in Santiago, Chile.
	\item[(iii)] This research has made use of the SIMBAD database, operated at Centre de Données de Strasbourg (CDS), Strasbourg, France	\cite[\gmleft{The SIMBAD astronomical database\gmright}][]{2000A&AS..143....9W}.

\end{itemize}

%% file: 7-Data-Availability.tex
\section*{Data Availability}
This paper is using publicly available data from the SIMBAD astronimical database and the \textit{GAIA} second Data Release.